\documentclass[pre,twocolumn,nobalancelastpage,amsmath,amssymb,showpacs,
nofootinbib]{revtex4-1}

\usepackage{graphics}      
\usepackage{graphicx}      
\usepackage{longtable}     
\usepackage{url}           
\usepackage{bm}            
\usepackage{xcolor}
\usepackage{amsmath}
\usepackage{dcolumn}

\begin{document}
\title{Compression Driven Jamming of Athermal Frictionless Spherocylinders in Two Dimensions}
\author{Theodore Marschall}
\author{S. Teitel}
\affiliation{Department of Physics and Astronomy, University of Rochester, Rochester, NY 14627}
\date{\today}

\begin{abstract}
We simulate numerically the compression driven jamming of athermal, frictionless, soft-core spherocylinders in two dimensions, for a range of particle aspect ratios $\alpha$.  We find the critical packing fraction $\phi_J(\alpha)$ for the jamming transition, and the average number of contacts per particle $z_J(\alpha)$ at jamming.  We find that both are nonmonotonic, with a peak at $\alpha\approx 1$.  We find that configurations at the compression driven jamming point are always hypostatic for all $\alpha$, with $z_J<z_\mathrm{iso}=2d_f=6$ the isostatic value.  
We show that, for moderately elongated spherocylinders, there is no orientational ordering upon athermal compression through jamming. We analyze in detail the eigenmodes of the dynamical matrix close to the jamming point for a few different values of the aspect ratio, from nearly circular to moderately elongated.  We find that there are low frequency bands containing $N(z_\mathrm{iso}-z_J)/2$ modes, such that the frequencies of these modes vanish as $\phi\to\phi_J$.  We consider the extended vs localized nature of these low frequency modes, and the extent to which they involve translational or rotational motion, and find many low frequency sliding modes where particles can move with little rotation.
We highlight the importance of treating side-to-side contacts, along flat sides of the spherocylinder, properly for the correct determination of $z_J$.  We note the singular nature of taking the $\alpha\to 0$ limit.
We discuss  the similarities and differences with previous work  on jammed ellipses and ellipsoids, to illustrate the effects that different particle shape have on configurations at jamming.

\end{abstract}
\maketitle


\section{Introduction}
\label{secIntro}

In a system of athermal ($T = 0$) granular particles with only contact interactions, as the particle packing fraction $\phi$ increases, the system will undergo a sharp transition from a liquid-like state to a rigid but disordered solid state.  In the liquid-like state, particles have sufficient room to avoid each other and so there are no contacts, and so no stress, in the system.  As $\phi$ increases, particles come into mutual contact.  In the disordered solid state, force chains percolate across the system giving  it a finite elastic rigidity, and the system  supports a finite stress.  This transition from a  stress-free liquid-like state to a stress-supporting solid state is known as the jamming transition \cite{OHern.PRE.2003,liu2010jamming}.
For particles without intergranular friction, and for a given protocol for compacting the system, this jamming transition occurs at a well defined $\phi_J$  and the transition is  continuous; stress increases continuously from zero as $\phi$ increases above $\phi_J$ \cite{OHern.PRE.2003}.

Much of the work that has been done to analyze behavior near the jamming transition has been for the simple case of perfectly spherical particles.  It is therefore interesting to ask how the jamming transition may be modified if the particles have shapes with a lower rotational symmetry.  Recently, several works have considered the cases of non-spherical particles, in particular monodisperse distributions of aspherical ellipsoids \cite{Donev.PRL.2004,Donev.Science.2004,Man.PRL.2005,Donev.PRE.2007}, oblate ellipsoids \cite{Donev.Science.2004,Man.PRL.2005,Donev.PRE.2007}, and prolate ellipsoids \cite{Donev.Science.2004,Man.PRL.2005,Donev.PRE.2007,Sacanna.JPhysC.2007,Zeravcic.EPL.2009,Schreck.PRE.2012} in three dimensions (3D), and bidisperse distributions of ellipses \cite{Donev.PRE.2007,Mailman.PRL.2009,Schreck.PRE.2012} in two dimensions (2D).
Spherocylinders \cite{Williams2003PRE,Wouterse.JPCM.2007,Azema2010,Azema2012} have been used to model rod-shaped particles, and other work has considered cut spheres \cite{Wouterse.JPCM.2007} in 3D.  For a review, see Ref.~\cite{Borzsonyi.Soft.2013}.

In this work we consider in detail the compression driven jamming of athermal, frictionless, soft-core  2D spherocylinders.  By ``compression driven" we mean a protocol in which we start with a dilute system of non-overlapping particles, then isotropically shrink the system box to increase the density,  passing through the point at which the system jams.  A spherocylinder in 2D consists of a rectangle with two circular end caps.  We will study the behavior of such spherocylinders as a function of the aspect ratio of rectangular length to end cap diameter.  Spherocylinders are unlike ellipses and ellipsoids in that they have parallel flat sides that could in principle lead to  configurations in which particles  stack in parallel layers.  In that sense they share a similarity with the cut spheres in 3D considered by Wouterse et al. \cite{Wouterse.JPCM.2007}.  We will pay particular attention to the effect of these parallel sides on the nature of elastic vibrational modes at jamming.
We will focus on two issues that arise when considering particles with shape anisotropy: (i) do spherocylinders show any orientational ordering as they are compressed through the jamming transition; and (ii) are configurations at jamming isostatic, and if not, what are the characteristics of the unconstrained (to quadratic order in the energy) modes?

{\bf Orientational ordering:} It has long been known, since the work of Onsager \cite{Onsager.NYAS.1949}, that hard-core (no overlaps allowed) rod shaped particles in thermal equilibrium will undergo a  liquid to nematic phase transition  as the density is increased.  Despite the absence of any globally preferred direction in the system, there will be a spontaneous symmetry-breaking transition in which the rods will show a macroscopic alignment in a particular direction.  Bolhuis and Frenkel \cite{Bolhuis.JChemPhys.1997} mapped out the phase diagram for thermalized hard-core spherocylinders in 3D, finding the dependence of the nematic transition as a function of the spherocylinder aspect ratio $\phi_c(\alpha)$ (they also found smectic and crystalline transitions).
Monte Carlo simulations for thermalized hard-core spherocylinders in 2D  \cite{Bates.JChemPhys.2000,Lagomarsino.JChemPhys.2003}  similarly observed a transition, upon increasing density, from an isotropic liquid to a nematic phase with algebraically decaying (rather than long range) orientational correlations.
Experimental studies of vibrated, but otherwise athermal,  elongated  grains in 2D have observed several different types of ordered states with both nematic and tetratic order \cite{Narayan.JStatMech.2006,Galanis.PRL.2006,Galanis.PRL.2010}.  

In contrast to the above, one can ask if a system of athermally compressed rod shaped particles, in which there is neither thermal nor mechanical random agitation of the particles, will show any orientational ordering.  Will the elastic forces that act between particles as they are compressed into mutual contact cause a spontaneous alignment so that the particles can pack more densely, or will they jam into an orientationally disordered state as the packing fraction $\phi$ increases?
Using a configurational statistical mechanics for athermal granular systems with elongated grains, Mounfield and Edwards \cite{Mounfield.PhysicaA.1994} argued that the grains need not order nematically to be minimally compact.  Experimental studies of athermal 3D oblate ellipsoids \cite{Donev.Science.2004} and aspherical ellipsoids \cite{Donev.PRE.2007} at jamming  reported only a  small nematic order parameter that was interpreted as consistent with no orientational ordering. Experiments on prolate ellipsoids \cite{Sacanna.JPhysC.2007} similarly reported no orientational ordering.  Simulations on cut spheres in 3D \cite{Wouterse.JPCM.2007}, however, did find nematic ordering when the aspect ratio was sufficiently large (i.e., thin disk shaped particles).  In this work we will present a detailed investigation as to whether moderately elongated  spherocylinders in 2D show any orientational ordering when athermally and isotropically compressed.  We will find that they do not.


{\bf Isostaticity:}
An important concept for the jamming transition is the notion of isostaticity.  As first discussed by Maxwell \cite{Maxwell.PhilMag.1864},  
a collection of $N$ randomly positioned particles can only be in a mechanically stable rigid state when 
the total number of force constraints  $N_c$ is equal to or greater than the total number of degrees of freedom $Nd_f$, with $d_f$ the number of degrees of freedom per particle.  When equality holds, the system is said to be isostatic.  For frictionless particles, where contact forces are always normal to the particle surface, the number of force constraints is just equal to the number of particle pair contacts $Nz/2$, where $z$ is the average number of contacts per particle and the factor of $1/2$ is because each contact is shared by two particles.  Hence the isostatic condition is $z_\mathrm{iso}=2d_f$.  For spherical particles, where rotations do not change the state of the system and only translational degrees of freedom are important, we have $d_f=d$, the spatial dimension of the system, and so $z_\mathrm{iso}=4$ and $6$ for 2D disks and 3D spheres respectively.  For non-spherical particles, rotational degrees of freedom must be considered, and $d_f$ depends on the rotational symmetries of the particle.  For aspherical ellipsoids in 3D there is no rotational symmetry and so $d_f=6$ with $z_\mathrm{iso}=12$.  For prolate or oblate ellipsoids in 3D, $d_f=5$ and $z_\mathrm{iso}=10$. For spherocylinders or ellipses in 2D, $d_f=3$ and so  $z_\mathrm{iso}=6$.

For frictionless spherical particles, the isostatic condition is found to hold exactly at the jamming transition \cite{OHern.PRE.2003}; at jamming the system is marginally stable -- the average contact 
number at jamming $z_J$ satisfies $z_J=z_\mathrm{iso}$ and  removing one contact causes the system to go unstable  \cite{Wyart.EPL.2005,Wyart.PRE.2005}.  However if the particle shape is infinitesimally perturbed from sphericity, the isostatic value of $z$ would necessarily jump discontinuously from $2d$ to $2d_f$, since even an infinitesimal perturbation will change the rotational symmetry of the particle.  However it seems unlikely that such an infinitesimal perturbation from sphericity should result in a discontinuous structural change in the state of the system at jamming, with an accompanying discontinuous change in either the jamming $\phi_J$ or the number of particle contacts  $z_J$.  Numerical simulations and experiments have  confirmed that  mechanically stable configurations of ellipsoids in 3D \cite{Donev.PRL.2004,Donev.Science.2004,Man.PRL.2005,Donev.PRE.2007,Donev.PRE.2007,Zeravcic.EPL.2009,Schreck.PRE.2012},
and ellipses in 2D \cite{Donev.PRE.2007,Mailman.PRL.2009,Schreck.PRE.2012}, are in general hypostatic at jamming with $z_J<z_\mathrm{iso}$, but with $z_J$ approaching $z_\mathrm{iso}$ as the particles get increasingly elongated.
Spherocylinders and cut spheres in 3D are similarly hypostatic at jamming, but seem to remain so even as the aspect ratio gets very large  \cite{Williams2003PRE,Wouterse.JPCM.2007}, and one study of 2D spherocylinders \cite{Azema2010} suggests that they may remain hypostatic as well.

Donev et al. \cite{Donev.PRE.2007} proposed that the difference $z_\mathrm{iso}-z_J$ at jamming results from  modes of the system that are unconstrained to quadratic order in the expansion of the elastic energy in small particle displacements, and that such modes primarily involve rotations of the particles; such modes were proposed to increase the energy at quartic order, hence enforcing mechanical stability of the system.  Detailed analyses of the eigenmodes of the dynamical matrix of  systems of ellipses and prolate ellipsoids near jamming \cite{Zeravcic.EPL.2009,Mailman.PRL.2009,Schreck.PRE.2012} support this scenario.  In this work we perform a similar analysis of the dynamical matrix for spherocylinders in 2D.  We will find that 2D spherocylinders are similarly hypostatic at jamming, but that as the aspect ratio increases and particles get increasingly elongated,  $z_J$ has a peak near the peak value of $\phi_J$ and then decreases, as opposed to approaching $z_\mathrm{iso}$ as was found for ellipses and ellipsoids.  For moderately elongated spherocylinders we will find that the unconstrained  (to quadratic order) modes are sliding modes, consisting primarily of translations of isolated particles along the spherocylinder axis, rather than rotations.

The remainder of the paper is organized as follows.  In Sec.~\ref{s2} we discuss the details of our model system and our procedure for slowly compressing the system through jamming.  In Sec.~\ref{sorder} we present our results on the lack of orientational ordering of moderately elongated spherocylinders.  In Sec.~\ref{sjam} we present our results for pressure as a function of packing fraction, and determine the  packing fraction at jamming as a function of spherocylinder aspect ratio, $\phi_J(\alpha)$, for compression driven jamming in the quasistatic limit.  In Sec.~\ref{sMS} we describe our energy minimization method for constructing mechanically  stable states, and show that it is necessary to treat side-to-side contacts, where two spherocylinders contact along their flat edges, carefully.  
Doing so, we find that spherocylinders are hypostatic for the entire range of aspect ratios we consider.  
We also note the strong propensity of spherocylinders to have contacts along their flat sides, even for very small $\alpha$, thus suggesting that the $\alpha\to 0$ limit is singular.
In Sec.~\ref{sdos} we analyze the eigenmodes of small vibrations for both nearly circular and moderately elongated spherocylinders near jamming, and relate these results to the hypostaticity of the system.
In Sec.~\ref{sconcl} we summarize our conclusions.


\section{Model and Simulation Method}
\label{s2}

\begin{figure}
\centering
\includegraphics[width=3.5in]{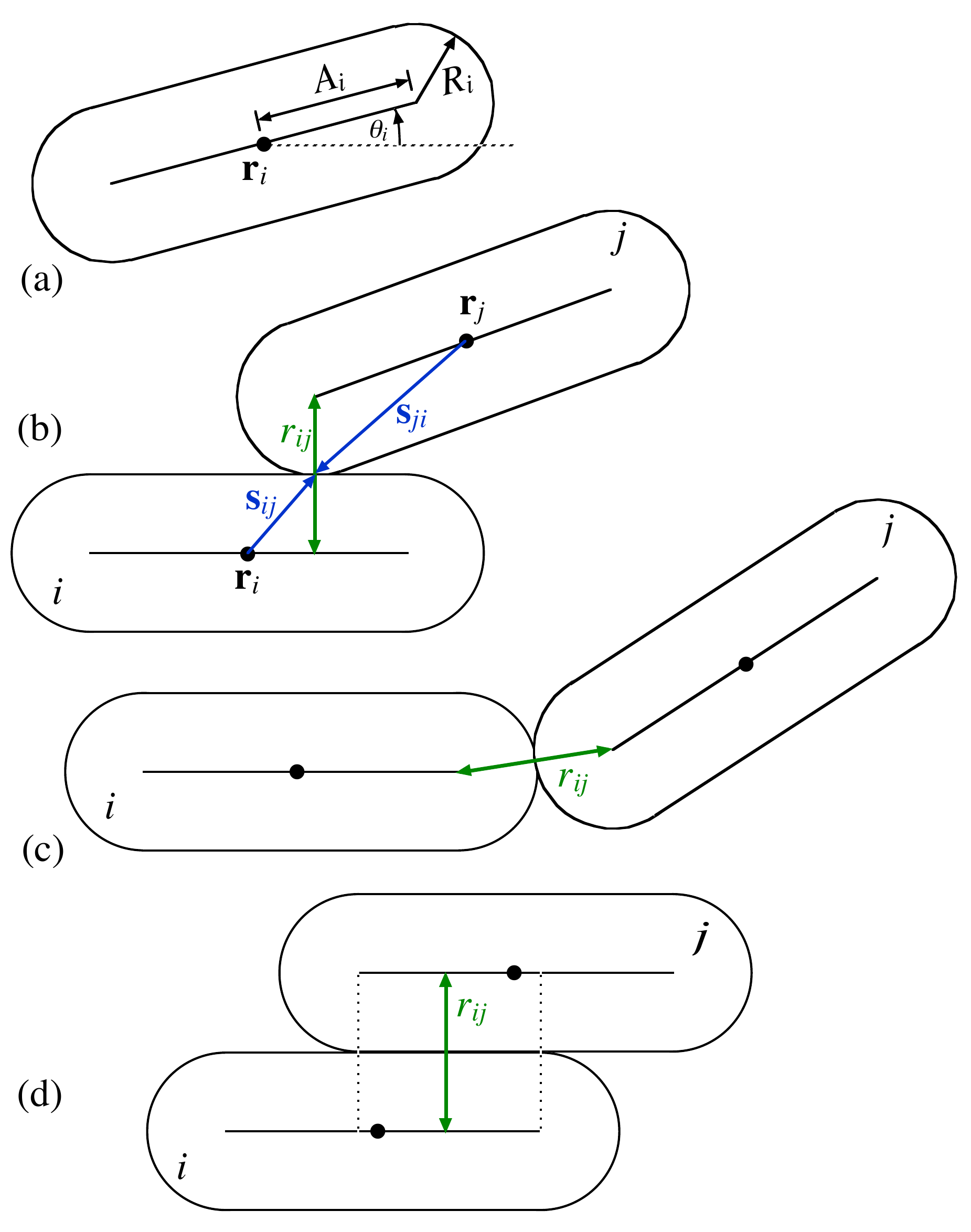}
\caption{ \label{fig:spherocylinder} Geometry of the spherocylinders: (a) An isolated spherocylinder indicating the spine half-length $A_i$, end cap radius $R_i$, center of mass position $\mathbf{r}_i$, and angle of orientation $\theta_i$. (b) Two spherocylinders in tip-to-side contact, indicating the minimal spine separation $r_{ij}$ and the moment arms $\mathbf{s}_{ij}$ and $\mathbf{s}_{ji}$.
(c) Two spherocylinders in tip-to-tip contact. (d) Two spherocylinders in side-to-side contact with the contact point and separation distance $r_{ij}$ taken midway between the spine end points (indicated by the vertical dashed lines).
}
\end{figure}

A two dimensional spherocylinder consists of a rectangle with two circular end caps.  We will denote the half length of the rectangular part of spherocylinder $i$ as $A_i$. The radius of the end cap, which is also the half width of the rectangle, we denote as $R_i$, as illustrated in  Fig.~\ref{fig:spherocylinder}(a).  We will refer to the ``spine" of the spherocylinder as the axis of length $2A_i$ that goes down the center of the rectangle, as indicated by the solid lines in Fig.~\ref{fig:spherocylinder}.  For every point on the perimeter of the spherocylinder, the shortest distance from the spine is $R_i$.  We define the aspect ratio of the spherocylinder as,
\begin{equation}
\alpha_i=A_i/R_i,
\end{equation}
so that $\alpha_i\to 0$ describes a circular particle, and the ratio of the total tip-to-tip length to width is $1+\alpha_i$.
In this work we consider only systems in which all particles have the same aspect ratio $\alpha$.

Our system consists of $N$ such spherocylinders confined within a square box of length $L$.  
We use periodic boundary conditions in both the $\mathbf{\hat x}$ and $\mathbf{\hat y}$ directions.
The packing fraction is,
\begin{equation}
\phi =\frac{1}{L^2} \sum_{i=1}^N \mathcal{A}_{i}, \qquad \mathcal{A}_i = 4 A_i R_i + \pi R_i^2,
\end{equation}
where $\mathcal{A}_i$ is the area of spherocylinder $i$.
Unless otherwise stated, the results in this work are for a bidisperse mixture of spherocylinders, with equal numbers of big and small particles, with $R_b/R_s=1.4$.  However we have also considered a monodisperse system.

We specify the position of a spherocylinder by the location of its center of mass $\mathbf{r}_i=(x_i,y_i)$, which lies at the center of the rectangle. The orientation of the spherocylinder is given by the angle $\theta_i$ that the spine makes with respect to the $\mathbf{\hat x}$ axis, as shown in Fig.~\ref{fig:spherocylinder}(a).  Two spherocylinders $i$ and $j$ come into contact when the shortest distance between their spines, $r_{ij}$, is less than the sum of their radii $d_{ij}=R_i+R_j$.  When $r_{ij}<d_{ij}$ the contact between the spherocylinders may be one of three types, as illustrated in Figs.~\ref{fig:spherocylinder}(b), (c), and (d), respectively: (i) tip-to-side, (ii) tip-to-tip, or (iii) side-to-side.  In order to have a side-to-side contact (iii) rather than a tip-to-side contact (i), in principle it is necessary that the two spherocylinders be perfectly parallel, i.e. $\theta_i=\theta_j$; in practice, due to limitations in the numerical accuracy of our contact detection algorithm \cite{Pournin.GranulMat.2005}, we take two spherocylinders as parallel whenever $|\theta_i-\theta_j|<10^{-8}$.  When this happens, we take the point of contact to be midway between the corresponding endpoints of the spines of $i$ and $j$, as indicated in Fig.~\ref{fig:spherocylinder}(d).

To determine when two spherocylinders are in contact, and if so to then determine the value of $r_{ij}$ and the location of the contact point, we use the efficient algorithm defined in Ref. \cite{Pournin.GranulMat.2005}.
In such a case we model the elastic contact force as a simple one-sided harmonic repulsion which acts at the point of contact only when $r_{ij}<d_{ij}$.  The elastic force on spherocylinder $i$ due to contact with $j$ is thus given by,
\begin{equation}
\mathbf{F}^\mathrm{el}_{ij}=(k_e/d_{ij})(1-r_{ij}/d_{ij})\mathbf{\hat r}_{ij}
\label{eq:Fij}
\end{equation}
where $k_e$ sets the energy scale, and $\mathbf{\hat r}_{ij}$ is the unit normal to the surface at the point of contact, pointing inward to spherocylinder $i$.  
The total elastic force on spherocylinder $i$ is therefore,
\begin{equation}
\mathbf{F}^\mathrm{el}_i={\sum_j}^\prime \mathbf{F}^\mathrm{el}_{ij},
\end{equation}
where the sum is over all spherocylinders $j$ in contact with $i$.  Although the elastic force always acts normal to the surface, there can nevertheless be a torque exerted on the spherocylinder due to the non-circular shape.
The total elastic torque on $i$ is,
\begin{equation}
\tau^\mathrm{el}_i=\mathbf{\hat z}\cdot {\sum_j}^\prime \mathbf{s}_{ij}\times \mathbf{F}^\mathrm{el}_{ij},
\label{eq:taui}
\end{equation}
where $\mathbf{s}_{ij}$ is the moment arm from the center of mass $\mathbf{r}_i$ of spherocylinder $i$  to the point of contact with spherocylinder $j$, as illustrated in Fig.~\ref{fig:spherocylinder}(b).

In this work we are interested in the jamming of the system when it is uniformly compressed from a dilute state.  To model a uniform compression we shrink the box length at a constant rate, $dL/dt = -\kappa L$.  We can then regard the substrate area of the box as undergoing a similar affine contraction, with the local velocity of the substrate at position $\mathbf{r}$ being, 
\begin{equation}
\mathbf{v}_\mathrm{sub}(\mathbf{r})=-\kappa\mathbf{r}.
\end{equation} 
The shrinking box then interacts with the spherocylinders via  a viscous drag force between spherocylinder and substrate.   We denote the local velocity of a position $\mathbf{r}$ on spherocylinder $i$  by,
\begin{equation}
\mathbf{v}_i(\mathbf{r})=\dfrac{d\mathbf{r}_i}{dt} +\dfrac{d\theta_i}{dt}\mathbf{\hat z}\times(\mathbf{r}-\mathbf{r}_i),
\end{equation}
where the first piece is due to the center of mass motion and the second piece is due to the spherocylinder's rotation.
The total dissipative force on spherocylinder $i$ is then,
\begin{equation}
\mathbf{F}^\mathrm{dis}_i = -k_d\int_i d^2r\,[\mathbf{v}_i(\mathbf{r})-\mathbf{v}_\mathrm{sub}(\mathbf{r})],
\end{equation}
where the integral is over the area of spherocylinder $i$.  There is similarly a dissipative torque on the spherocylinder,
\begin{equation}
\tau^\mathrm{dis}_i=-k_d \mathrm{\hat z}\cdot \int_i d^2r\, [\mathbf{r}-\mathbf{r}_i]\times[\mathbf{v}_i(\mathbf{r})-\mathbf{v}_\mathrm{sub}(\mathbf{r})].
\end{equation}
Using $\int_i d^2r\,[\mathbf{r}-\mathbf{r}_i]=0$ by symmetry, taking the area of spherocylinder $i$ as $\int_id^2r=\mathcal{A}_i$, and defining,
\begin{equation}
I=\int_id^2r\,|\mathbf{r}-\mathbf{r}_i|^2/\mathcal{A}_i,
\end{equation}
we can write the dissipative force and torque as,
\begin{align}
\mathbf{F}^\mathrm{dis}_i&=-k_d\mathcal{A}_i\left[\dfrac{d\mathbf{r}_i}{dt}+\kappa\mathbf{r}_i\right],\label{eq:Fdis}\\[10pt]
\tau^\mathrm{dis}_i&=-k_d\mathcal{A}_i I\dfrac{d\theta_i}{dt}.\label{eq:taudis}
\end{align}
For spherocylinders with aspect ratio $\alpha$ we have $I = R^2 (3 \pi + 24 \alpha + 6 \pi \alpha^2 + 8 \alpha^3) / (6 \pi + 24 \alpha)$.  Note, $\tau_i^\mathrm{dis}$ is independent of the compression rate $\kappa$, and serves only to damp out rotations that result from the collisional elastic torque $\tau_i^\mathrm{el}$.

We take the elastic and dissipative forces as the only forces acting on the spherocylinders; there is no interparticle frictional force or collisional dissipation.
Taking an overdamped equation of motion, 
\begin{align}
\mathbf{F}^\mathrm{el}_i+\mathbf{F}^\mathrm{dis}_i&=0,\label{eq:F}\\
\tau^\mathrm{el}_i+\tau^\mathrm{dis}_i&=0,
\label{eq:tau}
\end{align}
Eqs.~(\ref{eq:F}) and (\ref{eq:tau}) can then be numerically integrated to find the motion of the center of mass $\mathbf{r}_i(t)$ and the orientation $\theta_i(t)$.

In the absence of collisions, the above equations of motion become $\mathbf{F}_i^\mathrm{dis}=0$, $\tau_i^\mathrm{dis}=0$, and from Eqs.~(\ref{eq:Fdis}) and (\ref{eq:taudis}) we have $\mathbf{v}_i=-\kappa\mathbf{r}_i$ for the center of mass motion of spherocylinder $i$, while $d\theta_i/dt=0$.  The spherocylinders thus would move according to an isotropic affine compression of the system length $L$ at finite rate $\kappa$, with no rotation.  Collisions then result in the non-affine fluctuations about this affine motion.  Our dynamics is thus a continuous-in-time  analog of the compression protocol used in Refs.~\cite{Mailman.PRL.2009,Schreck.PRE.2012}, in which the system is affinely compressed a small discrete amount $\Delta\phi$, and then energy minimized to reduce the resulting particle overlaps, before another compression step of $\Delta\phi$ is applied.  Rather than compressing in small discrete steps, we compress continuously at a finite, tunable, rate $\kappa$.
Our method of compressing continuously at a finite rate is similar to the Lubachevsky-Stillinger protocol \cite{Donev.PRE.2007,Lubachevsky1990}, except that our dissipative force allows us to  use an overdamped rather than an inertial dynamics, so that configurations at low compression rates are always close to being in energy minimized states.

For our simulations we will take $2R_s=1$ as the unit of distance, $k_e=1$ as the unit of energy, and $t_0=(2R_s)^2k_d/k_e=1$ as the unit of time.  We numerically integrate the equations of motion using a two-stage Heun method with a step size of $\Delta t=0.01$.  With these choices, at each step of integration the packing fraction changes by $\Delta\phi/\phi=2\kappa\Delta t =0.02\kappa $.
We start our compressions from random zero-energy configurations (i.e., no overlapping spherocylinders) at an initial packing $\phi_\mathrm{init}$, and integrate to whatever is the desired final packing fraction $\phi$.  We use  non-overlapping spherocylinders for our initial configurations, rather than completely randomly positioned spherocylinders, since the latter case could result in spherocylinders which completely pass through each other forming a cross; in such a configuration, it becomes ambiguous what is the point of contact at which the repulsive elastic force should be acting.
Our results are typically averaged over $M_s$ separate runs starting from different independent initial configurations.  We will denote each such run as one compression ``sample."

For determining mechanically stable configurations very close to the jamming transition, we will also use a conjugate gradient energy minimization applied to the configurations generated by the above compression protocol.  We defer discussion of this minimization procedure to Sec.~\ref{sMS}.

\section{Results} 

\subsection{Lack of Orientational Ordering}
\label{sorder}

To measure $n-$fold orientational order in two dimensions, the magnitude of the order parameter $S_n$ and its direction of orientation $\theta_n$, for any particular configuration, can be computed as \cite{Donev.PRB.2006},
\begin{equation}
S_n = \max_{\theta_n}\left[\dfrac{1}{N}\sum_{i=1}^N\cos\left(n[\theta_i-\theta_n]\right)\right],
\end{equation}
where the $\theta_n$ that maximizes the sum is the ordering direction.  One can then show that,
\begin{align}
&S_n=\sqrt{\left( \dfrac{1}{N}\sum_{i=1}^N\cos(n\theta_i)\right)^2 + \left( \dfrac{1}{N}\sum_{i=1}^N\sin(n\theta_i)\right)^2},\\[10pt]
&\tan(n\theta_n)={\displaystyle\sum_{i=1}^N\sin(n\theta_i)}\Big/{\displaystyle\sum_{i=1}^N\cos(n\theta_i)}.
\end{align}
Choosing $n = 2$ measures the nematic order while $n = 4$ measures tetratic order.

\begin{figure}
	\centering
	\includegraphics[width=0.9\linewidth]{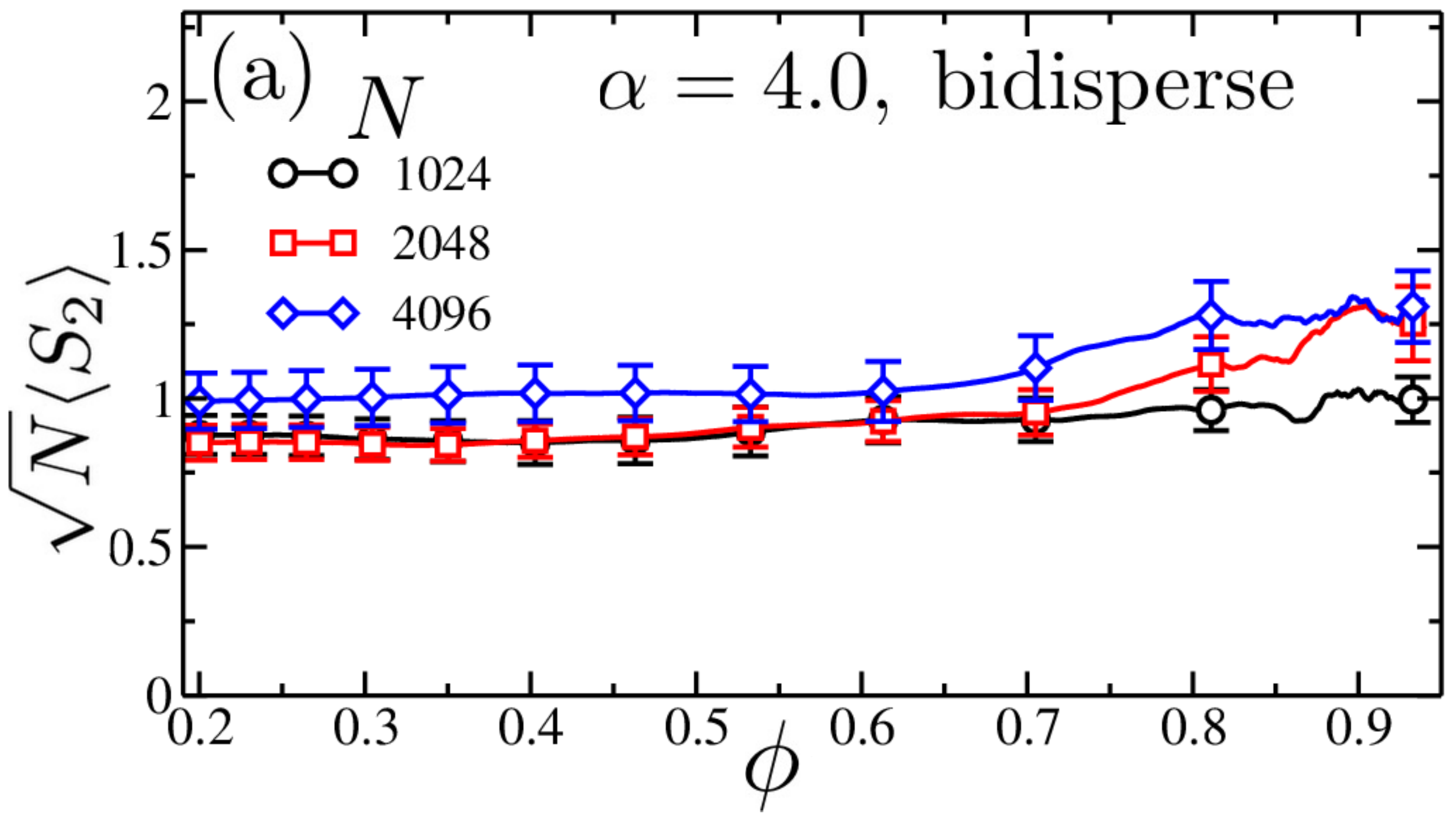}
	\includegraphics[width=0.9\linewidth]{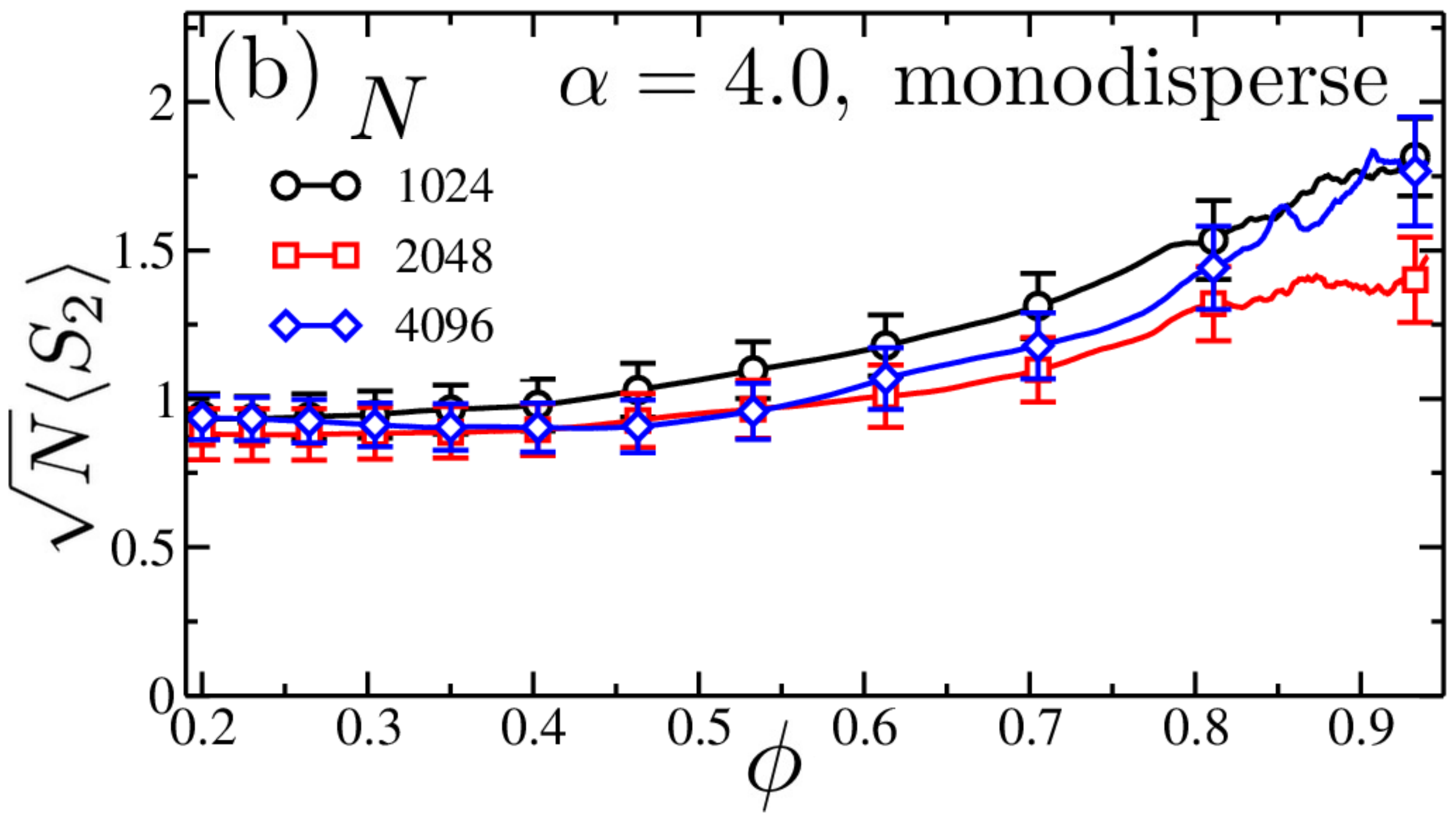}
	\includegraphics[width=0.9\linewidth]{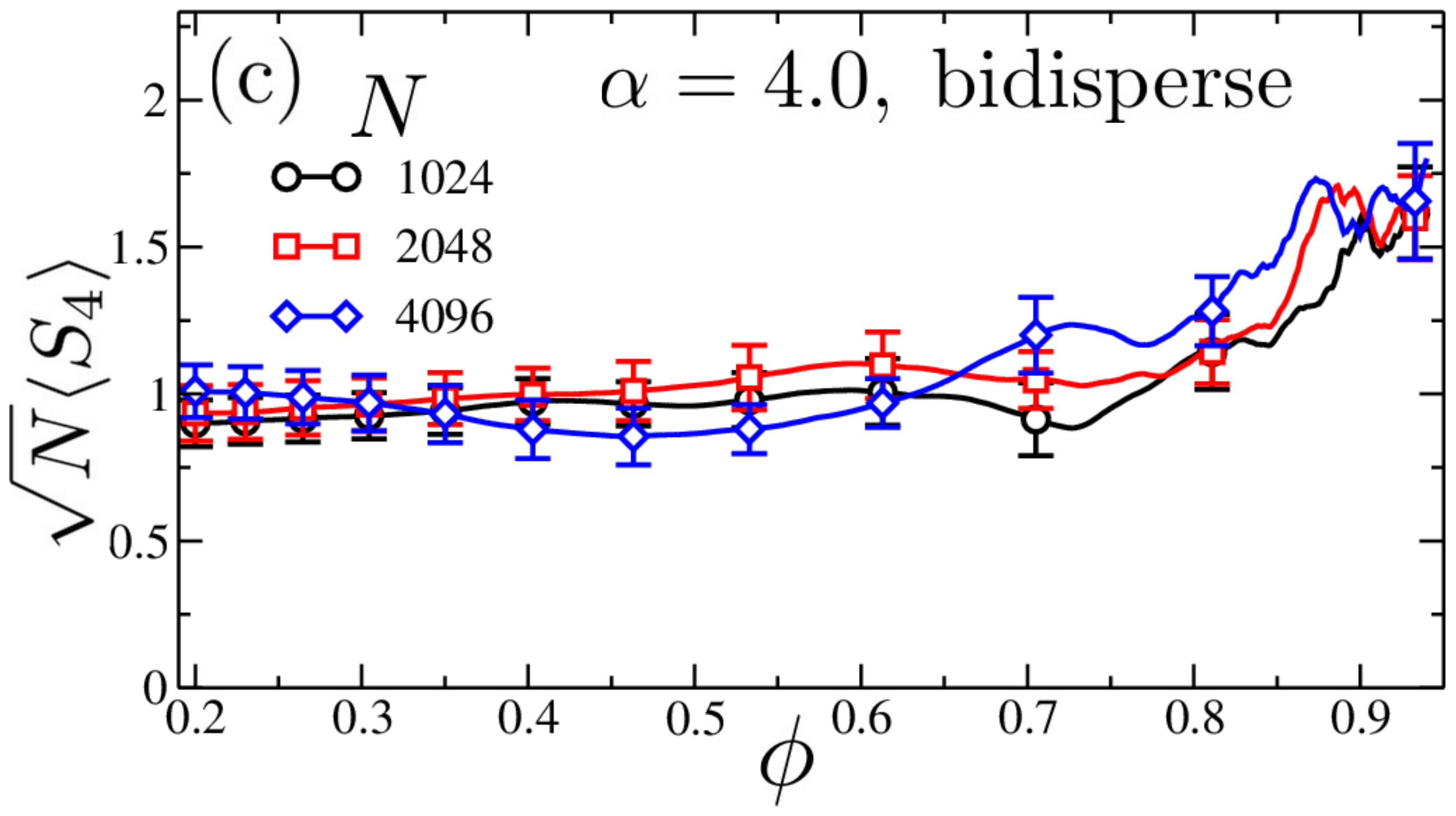}
	\includegraphics[width=0.9\linewidth]{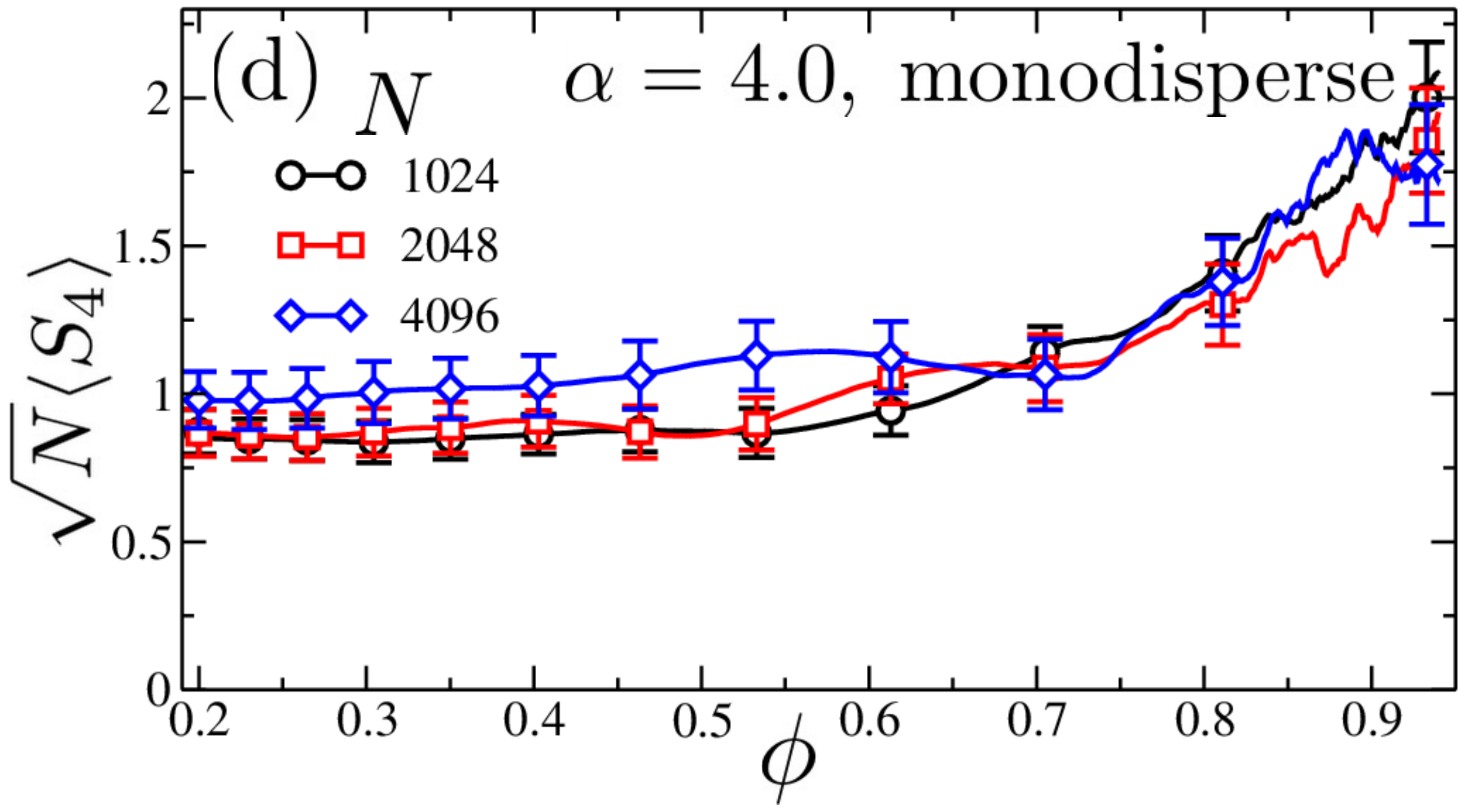}
	\caption{\label{fig:s2sqrtN} For systems with $N = 1024$ (circles) $2048$ (squares), and $4096$ (diamonds) spherocylinders of aspect ratio of $\alpha = 4.0$, scaled nematic order parameter $\sqrt{N}\langle S_2\rangle$ vs $\phi$ for (a) bidisperse size distribution with $R_b/R_s=1.4$ and (b) monodisperse distribution; and scaled tetratic  
	order parameter $\sqrt{N}\langle S_4\rangle$ vs $\phi$ for (c) bidispserse size distribution and (d) monodisperse distribution.
	Averages are calculated over $M_s=20$ to 40 independent samples compressed at a rate $\kappa = 10^{-7}$, starting from random zero-energy configurations at $\phi_\mathrm{init} = 0.2$.  
}
\end{figure}

For an orientationally disordered system we expect to have $S_n=0$.  However, when there are a finite number of particles $N$, even if particles are oriented completely at random,  any individual configuration will in general have a small finite value of $S_n$ as a statistical fluctuation away from the expected average of zero.  For such statistical fluctuations we expect that, for an orientationally disordered state, we would find $S_n\sim 1/\sqrt{N}$. To test for the presence or absence of nematic order, we therefore plot in Fig.~\ref{fig:s2sqrtN} the quantity $\sqrt{N}\langle S_2\rangle$ vs $\phi$, where $\langle S_2\rangle$ is the value of $S_2$ averaged over $M_s$ independent samples.  The error bars shown in Fig.~\ref{fig:s2sqrtN} are the estimated statistical error as obtained from the variance over the independent samples, $\sqrt{N\mathrm{var}[S_2]/(M_s-1)}$.  Error bars for other quantities shown later in this work are computed similarly.

In Fig.~\ref{fig:s2sqrtN}(a) we show results for a bidisperse system with aspect ratio $\alpha=4$ and $N=1024, 2048, 4096$ particles, using $M_s=40, 30$, and $20$ respectively.  Our compression rate is $\kappa=10^{-7}$ and we start from initial random configurations at $\phi_\mathrm{init}=0.2$.
Our results span a range of $\phi$ from below to above the jamming transition (for $\alpha=4$, $\phi_J\approx 0.866$, see Fig.~\ref{fig:p_v_phi}(a)).
We see that the curves for different $N$ are all roughly equal within one or two factors of the estimated statistical error, confirming the $S_2\sim 1/\sqrt{N}$ scaling expected for an orientationally disordered system.  Thus, unlike rods in thermal equilibrium, athermally compressed bidisperse spherocylinders show no nematic ordering transition. In Fig.~\ref{fig:s2sqrtN}(b) we show similar results, but now for a monodisperse system.  Although we see some mild increase in the value of $\sqrt{N}\langle S_2\rangle$ as $\phi$ increases, we again  see behavior consistent with $S_2\sim1/\sqrt{N}$, and so no nematic ordering.  We find similar results at other compression rates, down to our slowest $\kappa=10^{-10}$, but using fewer independent samples $M_s$.
We have also computed the tetratic order parameter $\langle S_4\rangle$, and find that it shows similar behavior, as shown in Figs.~\ref{fig:s2sqrtN}(c) and (d).  
Finally, we have also considered the case of bidisperse, nearly circular, spherocylinders with $\alpha=0.01$, compressed at a rate of $\kappa=10^{-8}$.  Again we find no evidence for orientational ordering.
We thus conclude that athermally compressed states of spherocylinders show no orientational order upon jamming.

\subsection{Jamming Transition}
\label{sjam}

In this section we investigate the jamming transition of a bidisperse mixture of spherocylinders as a function of the aspect ratio $\alpha$.  
Here, and in subsequent sections, we use a system with $N=1024$ spherocylinders.
At sufficiently small packing fraction $\phi$, the spherocylinders are dilute enough that they may avoid all contact with each other and the system is at zero pressure.  As $\phi$ increases, the system will ultimately become so dense that spherocylinders will necessarily come into mutual contact, force chains will percolate across the system, and a finite pressure will develop.  For frictionless particles, the pressure increases continuously from zero at a specific packing fraction $\phi_J$, known as the jamming transition.

To find the density at which systems of different aspect ratios $\alpha$ jam, we compress at finite rates from $\kappa=10^{-6}$ to $\kappa=10^{-10}$. For each aspect ratio and compression rate we start from random zero-energy configurations at $\phi_\mathrm{init}=0.4$.  For each configuration we compute the pressure tensor,
\begin{equation}
\mathbf{p}=-\dfrac{1}{L^2}\sum_{i=1}^N{\sum_{j}}^\prime\mathbf{s}_{ij}\otimes\mathbf{F}^\mathrm{el}_{ij},
\end{equation}
where the primed sum is over all spherocylinders $j$ in contact with spherocylinder $i$, $\mathbf{s}_{ij}$ is the moment arm from the center of mass $\mathbf{r}_i$ of spherocylinder $i$ to the point of 
contact with spherocylinder $j$, as in Eq.~(\ref{eq:taui}), and $\mathbf{F}^\mathrm{el}_{ij}$ is the elastic force on particle $i$ from $j$, as in Eq.~(\ref{eq:Fij}).
The pressure is then defined as,
\begin{equation}
p = \dfrac{1}{2}\mathrm{Tr}[\mathbf{p}]=\dfrac{1}{2}(p_{xx}+p_{yy}).
\end{equation}

\begin{figure}
	\centering
	\includegraphics[width=0.9\linewidth]{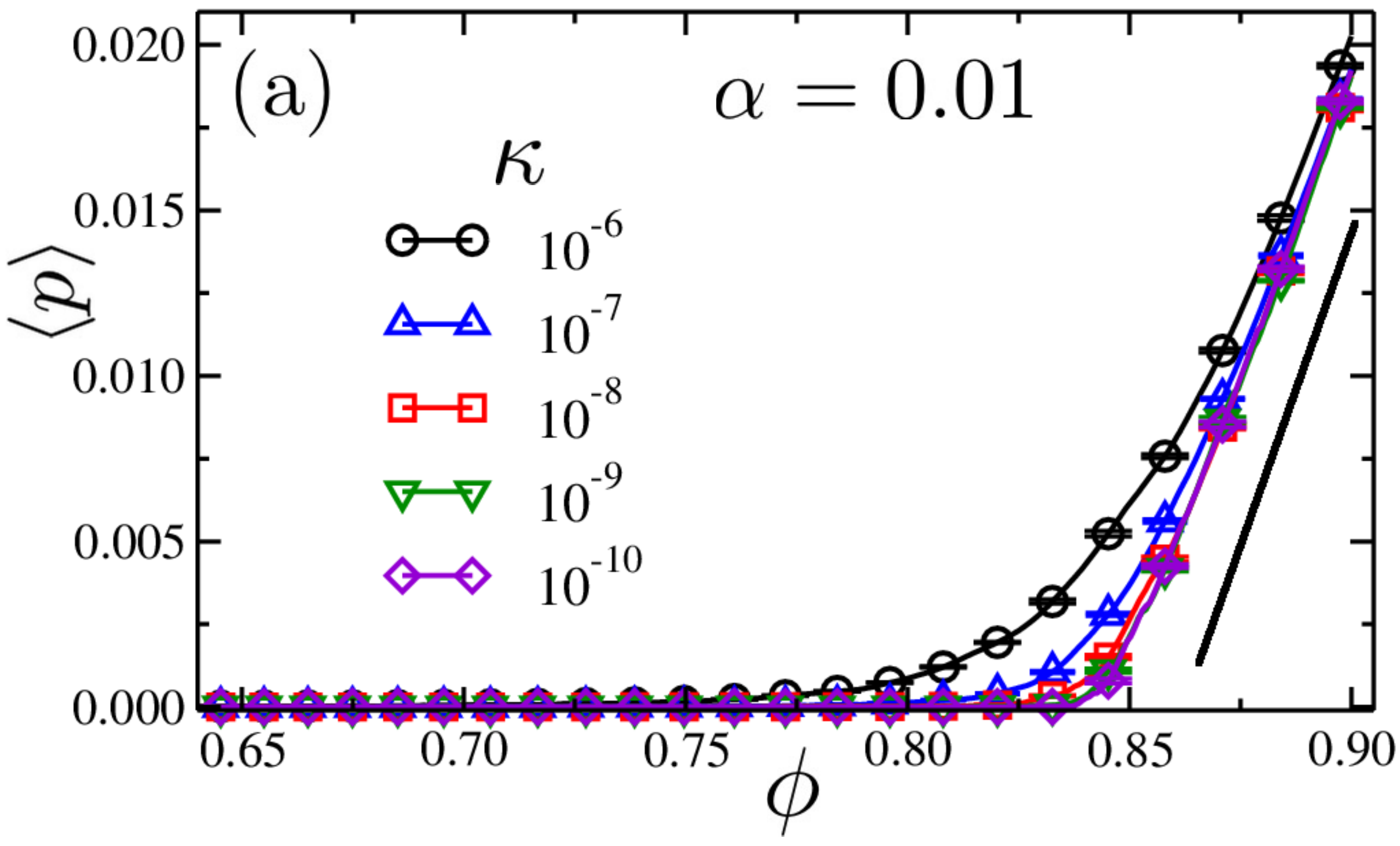}
	\includegraphics[width=0.9\linewidth]{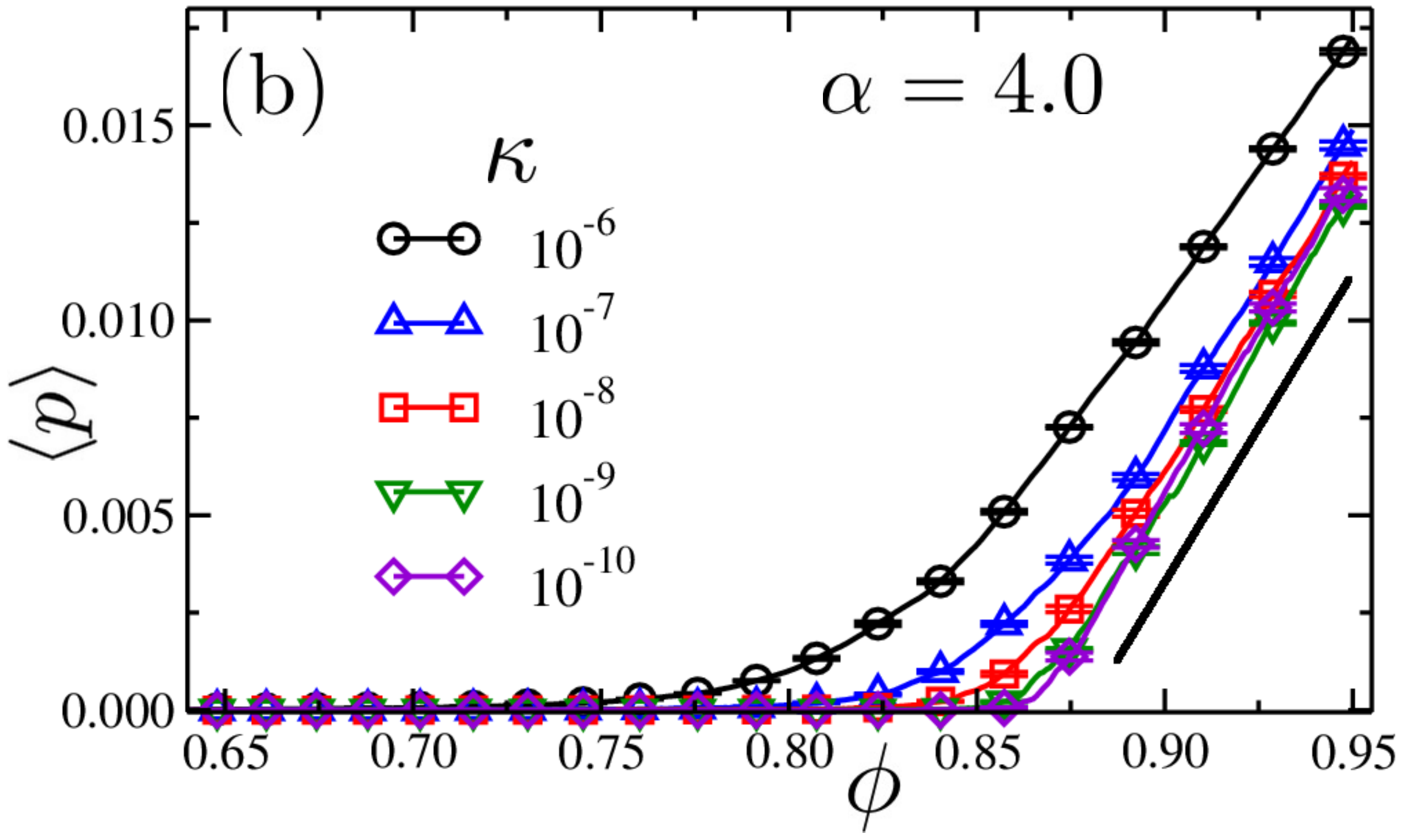}
	\caption{\label{fig:p_v_phi} Average pressure $\langle p\rangle$ vs packing fraction $\phi$ during compression of bidisperse systems of $N=1024$ spherocylinders with aspect ratio (a) $\alpha = 0.01$ and  (b) $\alpha = 4.0$, at compression rates from $\kappa = 10^{-6}$ (circles) to $\kappa = 10^{-10}$ (diamonds).  Each rate is averaged over 6 to 10 independent samples starting from random zero-energy configurations at $\phi_\mathrm{init} = 0.4$.  The straight line to the right serves to guide the eye.}
\end{figure}

We perform such compression runs for a bidisperse system with $N=1024$ spherocylinders, computing the pressure of configurations at regular time intervals
and averaging over the $M_s$ independent samples. In Fig.~\ref{fig:p_v_phi} we plot the resulting average pressure $\langle p\rangle$ vs $\phi$ for the two specific cases of (a) nearly circular disks with $\alpha=0.01$, and (b) moderately elongated spherocylinders with $\alpha=4$.  We use $M_s=6$ to $10$, depending on the compression rate $\kappa$.  

We see that $p=0$ at low $\phi$ and then $p$ increases to finite values as $\phi$ increases above some $\phi_J(\kappa)$. As $\kappa$ decreases, $\phi_J(\kappa)$ increases, the curves sharpen up near $\phi_J(\kappa)$, and $\langle p\rangle$ increases linearly in $\phi$ sufficiently above $\phi_J(\kappa)$, as expected for our harmonic elastic force \cite{OHern.PRE.2003}.    For $\kappa\le 10^{-9}$ we see no change in the $\langle p\rangle$ vs $\phi$ curve, and we have reached the limit of quasistatic compression. The value of $\phi_J$ in this quasistatic limit is the critical packing fraction of the compression-driven jamming transition.   The small tail that is seen near $\phi_J$ in this quasistatic limit is a finite size effect.  For finite $N$, each sample $s$ has a slightly different, sample specific, value of $\phi_{Js}$, as has been observed previously for circular disks \cite{OHern.PRE.2003} and as we confirm for spherocylinders below; as $N\to\infty$, this spread in $\phi_{Js}$ shrinks to zero.

\begin{figure}
	\centering
	\includegraphics[width=0.9\linewidth]{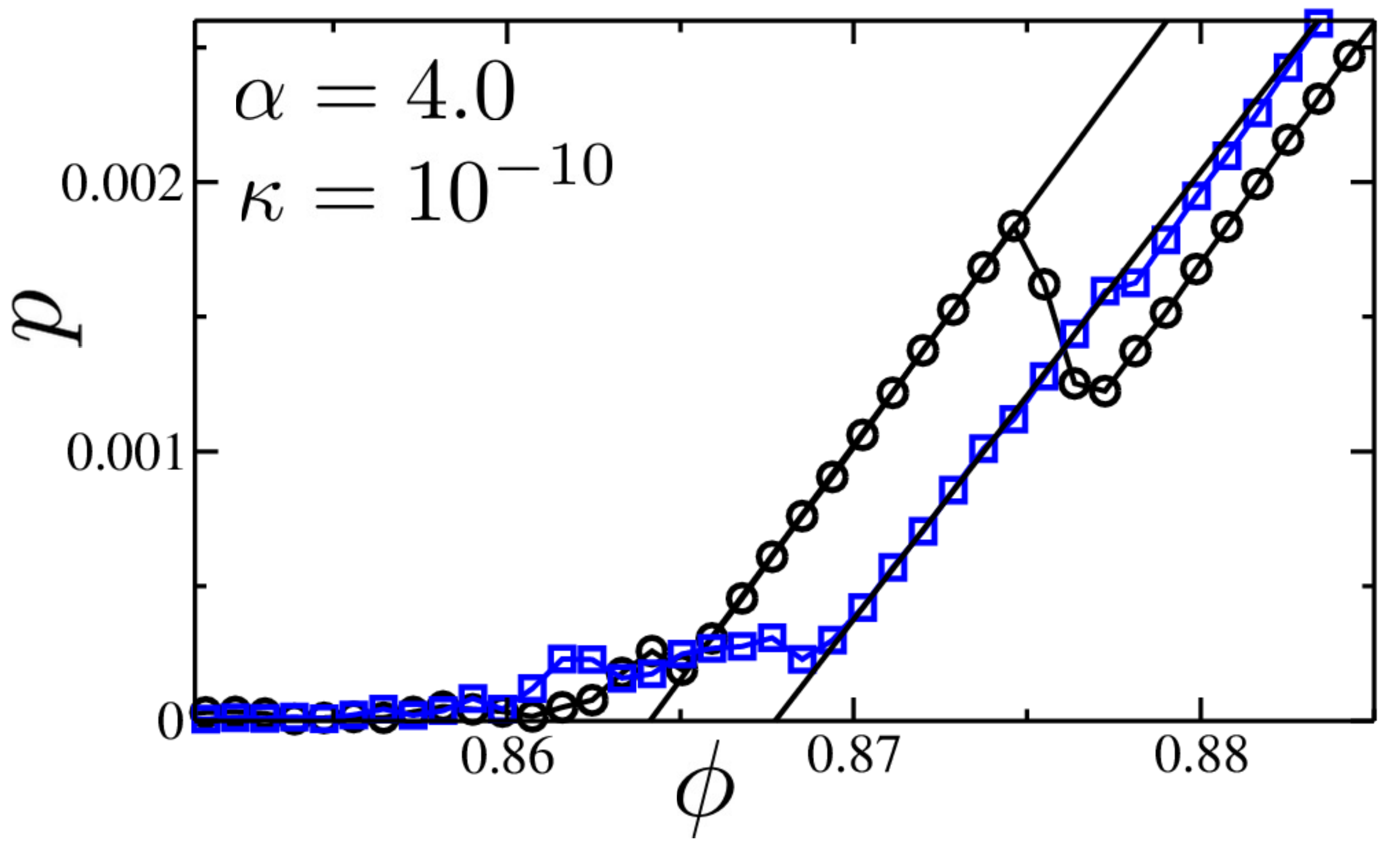}
	\caption{\label{fig:p_v_phi_fits} Pressure $p$ versus packing fraction $\phi$ for two different samples at compression rate $\kappa = 10^{-10}$ in bidisperse systems of $N=1024$ spherocylinders with aspect ratio $\alpha = 4.0$.  Linear fits overlaying the plots of pressure show the extrapolation to the configuration specific jamming density $\phi_{Js}$ for each sample.  There are $5\times10^6$ compression steps between the data points shown in the curves. 
	}
\end{figure}

\begin{figure}
	\centering
	\includegraphics[width=0.9\linewidth]{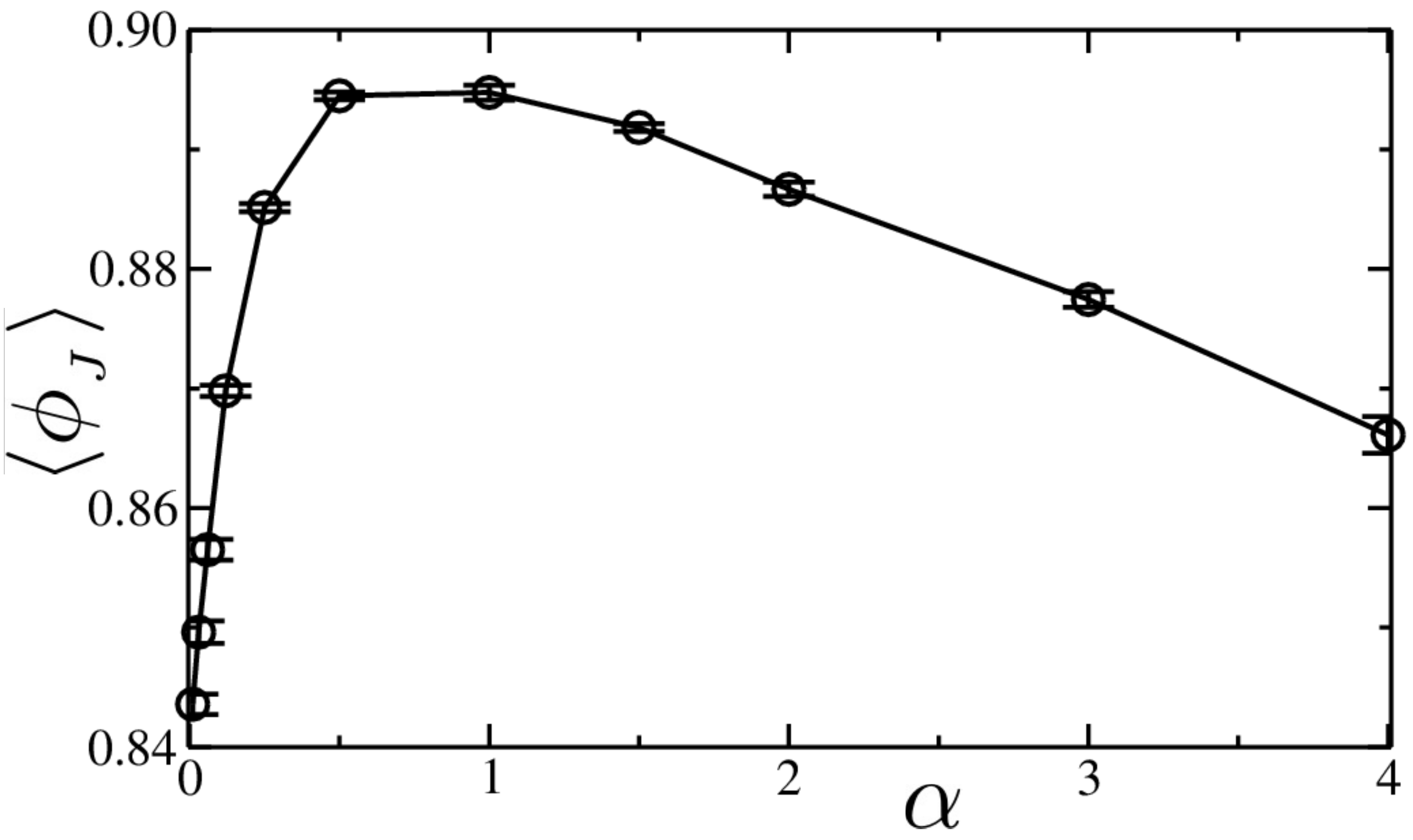}
	\caption{\label{fig:phiJ-zJ_v_aspect} The compression-driven jamming density $\langle\phi_J\rangle$ vs spherocylinder aspect ratios $\alpha$, for a bidispserse system of $N=1024$ spherocylinders.
}
\end{figure}

To estimate the value of $\phi_J$ for each aspect ratio $\alpha$, we consider the runs at $\kappa=10^{-10}$, which are in the quasistatic limit.  We look at each of the $M_s$ samples separately and fit the part of the $p$ vs $\phi$ curve where the pressure first develops a linear behavior upon increasing $\phi$, before there occurs any plastic rearrangements that may lead to discontinuous drops in pressure.  Extrapolating this linear region to $p=0$ then determines $\phi_{Js}$ for this particular sample.  In Fig.~\ref{fig:p_v_phi_fits} we show two examples of such determinations for the case $\alpha=4.0$.  We then average over these $\phi_{Js}$ to determine $\langle\phi_J\rangle$.  In Fig.~\ref{fig:phiJ-zJ_v_aspect} we plot the resulting $\langle\phi_J\rangle$ vs aspect ratio $\alpha$.  At $\alpha=0$ we find $\langle\phi_J\rangle= 0.8412 \pm 0.0005$, consistent with earlier results for circular disks \cite{OHern.PRE.2003,Vagberg.PRE.2011}.  As $\alpha$ increases, $\langle\phi_J\rangle$ increases to a maximum $\langle\phi_J\rangle\approx 0.8875$ around $\alpha\approx 1$, and then decreases.  The results we see here for $\phi_J(\alpha)$ are qualitatively similar to those found in previous simulations of ellipsoids and spherocylinders in 3D
\cite{Donev.PRL.2004,Donev.Science.2004,Man.PRL.2005,Donev.PRE.2007,Sacanna.JPhysC.2007,Williams2003PRE,Wouterse.JPCM.2007}.

\subsection{Mechanically Stable Configurations and Lack of Isostaticity}
\label{sMS}

To investigate the question of isostaticity at jamming in spherocylinders, we will want to consider the density of states for vibrational modes at the jamming transition.  The density of states is found by expanding the energy of the system about a mechanically stable state (i.e., a local energy minimum) to second order in small displacements of the degrees of freedom, and finding the eigenstates of the resulting dynamical matrix.
Since the configurations we obtain from compressing are the result of a dynamical (albeit slow) process, they are not necessarily in exact mechanical equilibrium.  We therefore wish to energy minimize the configurations we obtain from compression.

To get configurations  close to jamming, we will consider configurations in which the total elastic energy $U$ (defined below) is fixed to the value $U_0/L^2=10^{-15}$.  
We choose configurations at a fixed value of $U$, rather than a fixed value of $\phi$, since the jamming point $\phi_{Js}$ varies slightly from sample $s$ to sample $s^\prime$; fixing $U$, rather than $\phi$,  ensures that all our samples will be about the same distance from  their sample specific jamming transition. 
For our harmonic elastic force we have $U/L^2\propto (\phi-\phi_{Js})^2$, and we find  that $U_0/L^2=10^{-15}$ corresponds to $(\phi-\phi_{Js})\lesssim 10^{-7}$.

To locate configurations with the desired $U_0$, we start with a configuration with $U>U_0$ obtained from our continuous compression runs at a fixed rate $\kappa$, and energy minimize it using  a conjugate gradient algorithm (see Appendix A for details).  
Depending on whether the resulting minimized energy $U$ is greater or less than $U_0$, we carry out an affine decompression or compression of the box,
\begin{equation}
L\to (1\pm\lambda)L,\quad\mathbf{r}_i\to (1\pm\lambda)\mathbf{r_i},
\end{equation}
and then energy minimize the resulting configuration.  We continue such decompression or compression steps until $U$ crosses the value $U_0$.  We then reduce $\lambda$ by half, and reverse direction, i.e., if we had been decompressing, we now compress, and vice versa.  We continue in this fashion until we have narrowed in on the desired value $U=U_0$.  We start this process with a value $\lambda=10^{-6}$ and stop when $\lambda<10^{-16}$, which we find gives and accuracy in the energy of $| U - U_0 | / U_0 \lesssim 10^{-7}$.

To implement the above minimization procedure we must define a global energy function consistent with the elastic forces of Eq.~(\ref{eq:Fij}).  We use,
\begin{equation}
U(\{\mathbf{r}_k,\theta_k\})=\sum_{(i,j)}V_{ij}(r_{ij}), \quad
\end{equation}
with
\begin{equation}
V_{ij}(r_{ij})=
\dfrac{1}{2}k_e\left( 1-\dfrac{r_{ij}}{d_{ij}}\right)^2,
\label{eVij}
\end{equation}
where the sum is over all pairs of contacts between spherocylinders $i$ and $j$, $r_{ij}$ is the shortest distance between their two spines, and $d_{ij}=R_i+R_j$ is the sum of their end cap radii.  

Once we have found energy minimized states sufficiently close to jamming, we will then wish to construct the dynamical matrix. We find it convenient to convert the orientation angle $\theta_i$ into a length, and thus we take the coordinates of a given spherocylinder $i$ to be written as $\boldsymbol{\zeta}_i=(x_i,y_i,A_i\theta_i)$, where $\zeta_{i1}=x_i$, $\zeta_{i2}=y_i$ and $\zeta_{i3}=A_i\theta_i$.
The dynamical matrix is then the $3N\times3N$ matrix, 
\begin{equation}
M_{ia,jb}=\left.\dfrac{\partial^2 U}{\partial \zeta_{ia}\partial\zeta_{jb}}\right|_\mathrm{min},
\label{eq:dm}
\end{equation}
where $i,j=1,2\dots, N$, and $a,b=1,2,3$, and  the derivatives are evaluated at the energy minimized configuration.

To evaluate $M_{ia,jb}$ we need to know how $r_{ij}$ depends on the coordinates $\boldsymbol{\zeta}_i$ and $\boldsymbol{\zeta}_j$ of the two spherocylinders in contact, since we have,
\begin{equation}
\dfrac{\partial V_{ij}(r_{ij})}{\partial \zeta_{ia}} 
=-\dfrac{k_e}{d_{ij}}\left(1-\dfrac{r_{ij}}{d_{ij}}\right)
\dfrac{\partial r_{ij}}{\partial \zeta_{ia}}.
\end{equation}
The dependence of $r_{ij}$ on the spherocylinder coordinates depends on which of the three types of contacts of Fig.~\ref{fig:spherocylinder}(b), (c), (d) that one is considering.  For the tip-to-tip contact of Fig.~\ref{fig:spherocylinder}(c), a small displacement of any of the two spherocylinder's coordinates will keep the contact tip-to-tip.  We can therefore write,
\begin{equation}
r_{ij}=\sqrt{(\Delta x_{ij})^2+(\Delta y_{ij})^2}
\end{equation}
where
\begin{align}
\Delta x_{ij}&=[x_i\pm A_i\cos\theta_i]-[x_j\pm A_j\cos\theta_j],\\
\Delta y_{ij}&=[y_i\pm A_i\sin\theta_i]-[y_j\pm A_j\sin\theta_j],
\end{align}
with the appropriate signs taken so as to minimize $|\Delta x_{ij}|$ and $|\Delta y_{ij}|$.

For a tip-to-side contact, as in Fig.~\ref{fig:spherocylinder}(b), a motion of either spherocylinder parallel to the side with the contact, or a rotation of the spherocylinder with the tip contact, will result in a sliding of the location of the side contact.  The calculation of $r_{ij}$ must be done more carefully.  If $i$ is the spherocylinder with the side contact and $j$ is the spherocylinder with the tip contact, then,
\begin{equation}
\begin{array}{c}
r_{ij}=|(y_j-y_i)\cos\theta_i-(x_j-x_i)\sin\theta_i\\[10pt]
\pm A_j\sin(\theta_j-\theta_i)|,
\end{array}
\label{eq:tipside}
\end{equation}
where the sign is taken so as to minimize $r_{ij}$.

\begin{figure}
	\centering
	\includegraphics[width=0.9\linewidth]{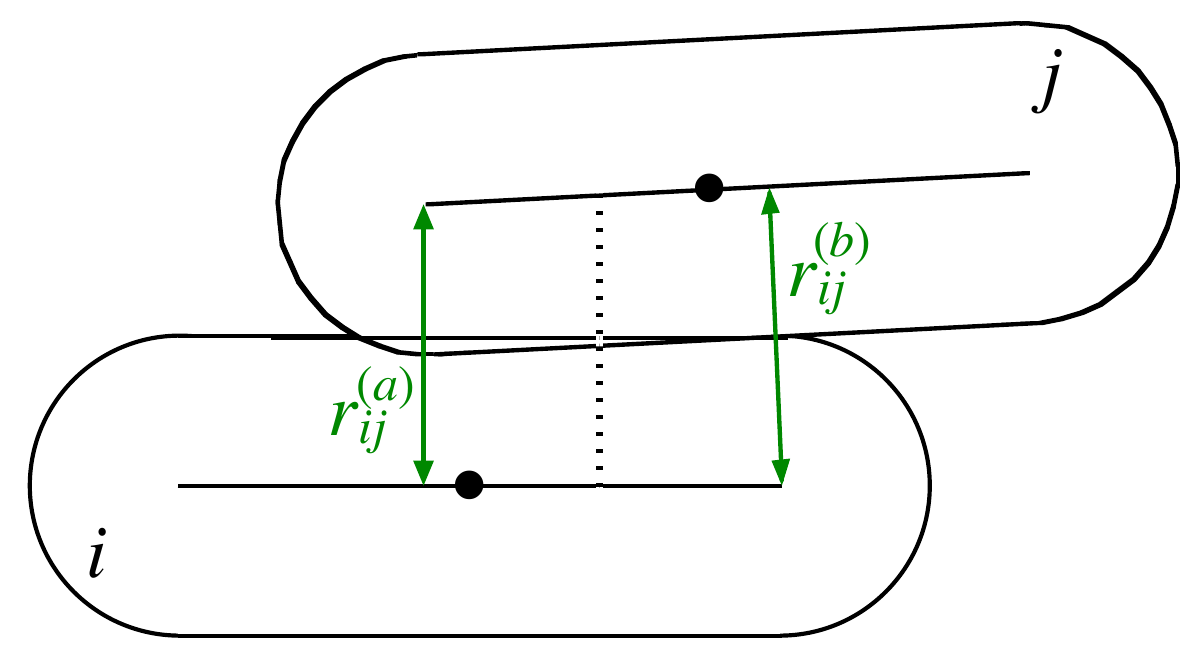}
	\caption{\label{fig:rodConfigs2} Configuration of two spherocylinders that are in approximate side-to-side contact.  For energy minimization and for computing the dynamical matrix, we consider this case as if there were two contacts at the points indicated by the separations $r_{ij}^{(a)}$ and $r_{ij}^{(b)}$.
	}
\end{figure}

For a side-to-side contact, if we take the location of the contact bond as illustrated in Fig.~\ref{fig:spherocylinder}(d), then a small rotation of either spherocylinder changes the configuration from a side-to-side contact to a tip-to-side contact, with a resulting discontinuous jump in the location of the contact point, and hence in the torques on the spherocylinders.  This discontinuity makes the derivatives needed for the dynamical matrix ill-defined, and moreover also causes difficulties carrying out the conjugate gradient minimization procedure.  We therefore modify the contact energy for this case as illustrated in Fig.~\ref{fig:rodConfigs2}.  Instead of a single contact located midway between the ends of the opposing spines (dotted line in Fig.~\ref{fig:rodConfigs2}) we now model the side-to-side contact as two bonds located at the corresponding ends of the spines (solid lines labeled ${r}_{ij}^{(a)}$ and ${r}_{ij}^{(b)}$ in Fig.~\ref{fig:rodConfigs2}).  We use the same convention when doing our conjugate gradient minimization of the energy $U$, provided both $r_{ij}^{(a)}$ and $r_{ij}^{(b)}$ are points of spherocylinder overlap, i.e. $r_{ij}^{(a)},r_{ij}^{(b)}<d_{ij}$.  Taking spherocylinder $j$ as the one whose tip comes closest to the spine of spherocylinder $i$, then the bond where the spherocylinder overlap is larger (i.e., $r_{ij}^{(a)}$ in Fig.~\ref{fig:rodConfigs2}), is given by the same relation as Eq.~(\ref{eq:tipside}).  The bond where the overlap is smaller   (i.e., $r_{ij}^{(b)}$ in Fig.~\ref{fig:rodConfigs2}) is given by,
\begin{equation}
\begin{array}{c}
r_{ij}^{(b)}=|(y_j-y_i)\cos\theta_j - (x_j-x_i)\cos\theta_j \\[10pt]
\pm A_i\sin(\theta_j-\theta_i)|,
\end{array}
\end{equation}
where the sign is taken so as to maximize $r_{ij}^{(b)}$.  

Modeling a side-to-side contact by two contact bonds as described above, rather than one, is also physically reasonable since in the hard-core limit a side-to-side contact will constrain two degrees of freedom:  translational motion perpendicular to the spherocylinder spine, as well as rotational motion.\footnote{in Ref.~\cite{Wouterse.JPCM.2007} a similar effect was noted for plane-to-plane contacts in cut spheres, where each such planar contact constrains three degrees of freedom.  However the authors of that work continued to count such planar contacts as a single contact.  Their result for the number of contacts as a function of aspect ratio is therefore  an underestimate of the correct constraint counting that should be done to test for isostaticity.}
  In contrast, tip-to-side and tip-to-tip contacts constrain only one degree of freedom.  Hence when counting the number of contact bonds per particle $z$, we count each side-to-side contact as two bonds.  A similar observation was made in Ref.~\cite{Azema2013} for polyhedral shaped particles.

Using the above procedure, we construct mechanically stable configurations with the desired $U_0/L^2=10^{-15}$, very close to jamming.  
Having obtained the mechanically stable configurations, we then remove any ``rattler" particles.  We take a rattler to be any particle which has only zero or one contact with another particle (here, and here only, a side-to-side contact is counted as one contact).  For a particle with only two contacts, we also take it to be a rattler unless the two contacts are oriented on  opposite sides parallel to the spherocylinder spine; in this case the spherocylinder may still be important for the stability of the contact network, even if it has a zero-energy sliding mode in the direction parallel to the spine.  Passing through the configuration to remove such rattlers, we then iterate the process until no further rattlers are found.  

\begin{figure}[h!]
	\centering
	\includegraphics[width=0.8\linewidth]{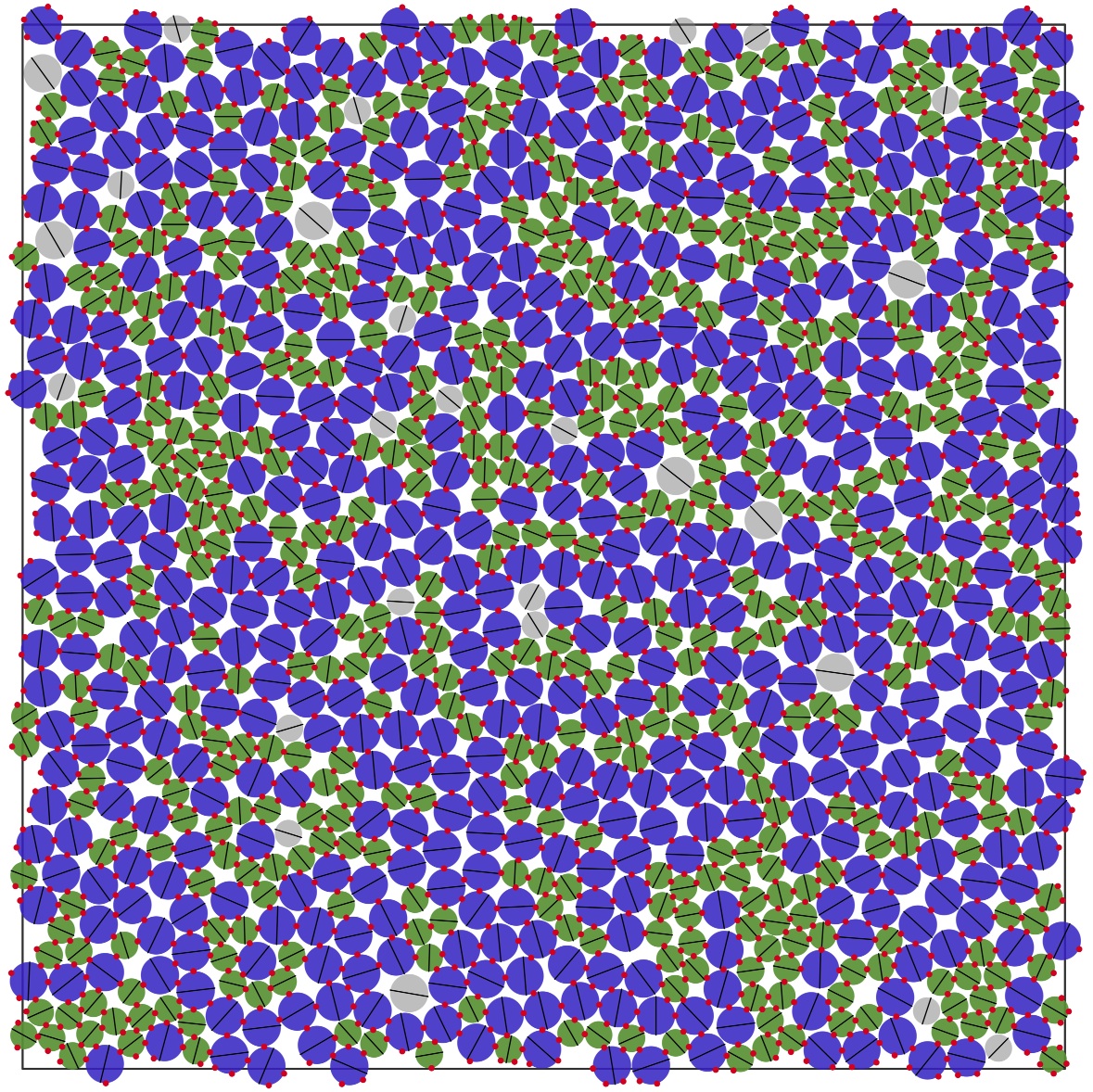}\\	
	\includegraphics[width=0.8\linewidth]{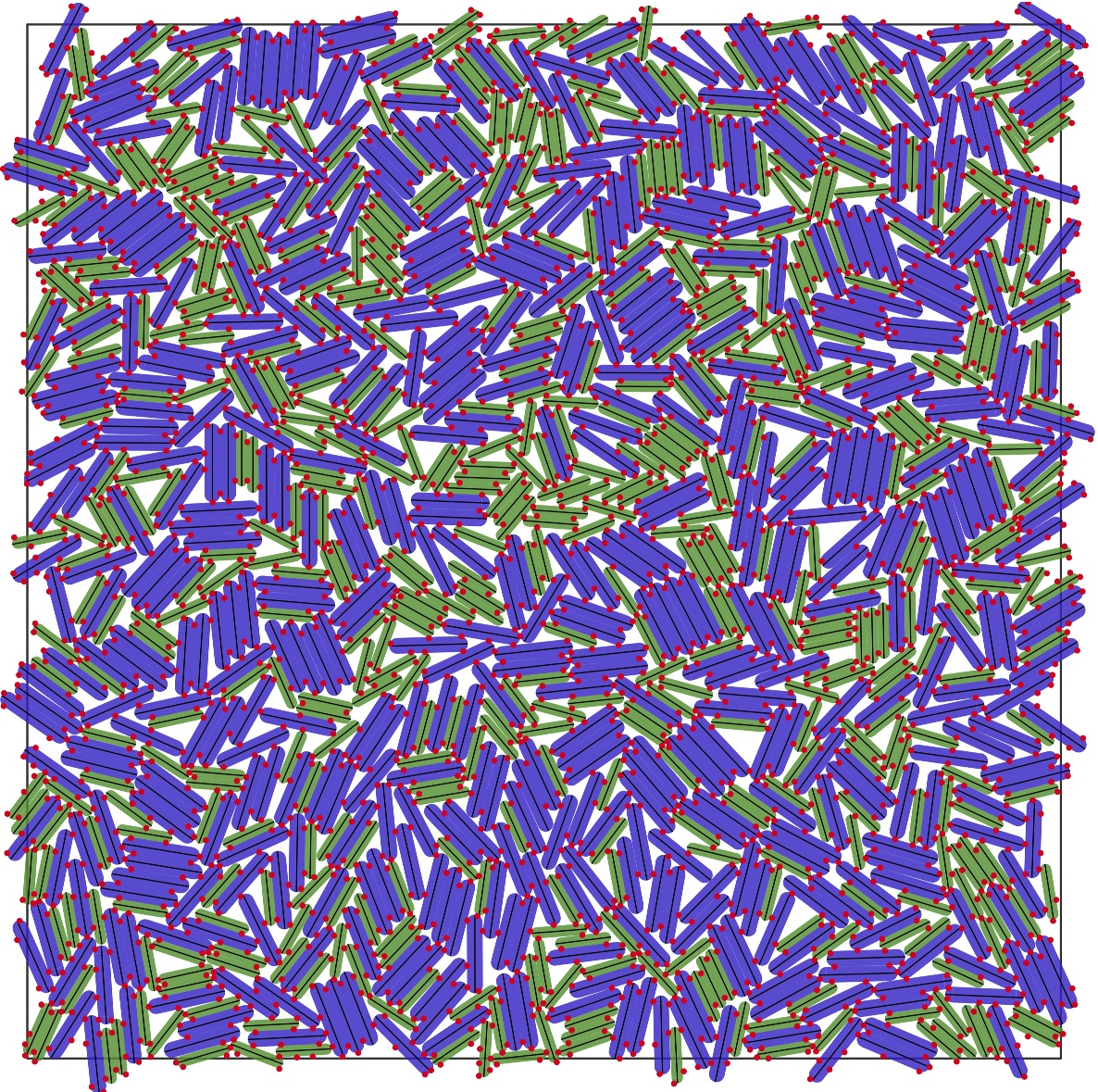}\\	
	\caption{\label{fig:snapshots}  Snapshots of energy minimized configurations of $N=1024$ bidisperse spherocylinders, at $U/L^2=10^{-15}$, very close to the jamming transition.  Results are shown for   aspect ratios (a) $\alpha=0.01$ and (b) $\alpha=4.0$.  Big and small spherocylinders are colored blue and green respectively, while rattlers are gray.  Solid black lines on each spherocylinder denote the direction of the spine.  Red dots indicate contacts between adjacent spherocylinders; for side-to-side contacts, we show the two contact bonds as illustrated in Fig.~\ref{fig:rodConfigs2}.	  
		}
\end{figure}

In Fig.~\ref{fig:snapshots} we show snapshots of sample configurations resulting from this procedure for $\alpha=0.01$, corresponding to nearly circular spherocylinders, and $\alpha=4.0$, corresponding to moderately elongated spherocylinders. 
In our systems with $N=1024$ spherocylinders, we find roughly $N_r=34$ rattlers ($3.3\%$) for $\alpha=0.01$,  while  $N_r=1.2$ ($0.1\%$) for $\alpha=4.0$.  For $\alpha=1$, which we will see gives the maximum number of contacts per particle, we find $N_r=0$.
For the remaining $\mathcal{N}=N-N_r$ particles forming the rigid backbone of the system, we can then compute $z_J$, the average number of contacts per particle, counting each side-to-side contact twice as discussed above, to determine if the system is isostatic at jamming or not.
We find $\langle z_J\rangle  = 4.43\pm0.03$ for $\alpha=0.01$ and $\langle z_J\rangle=5.64\pm 0.01$ for $\alpha=4.0$, both smaller than $z_\mathrm{iso}=6$. Hence both cases are hypostatic.

\begin{figure}
	\centering
	\includegraphics[width=0.9\linewidth]{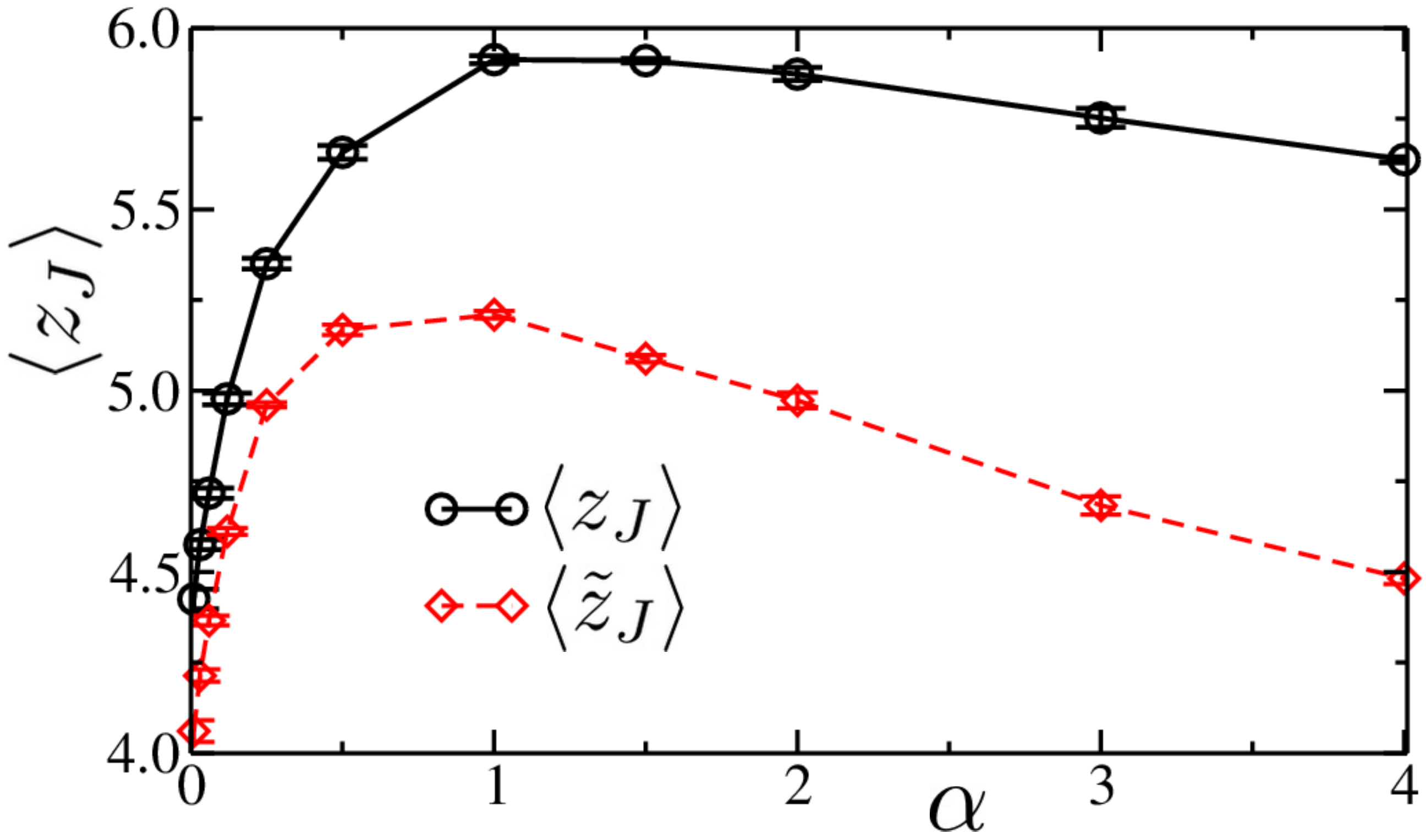}
	\includegraphics[width=0.9\linewidth]{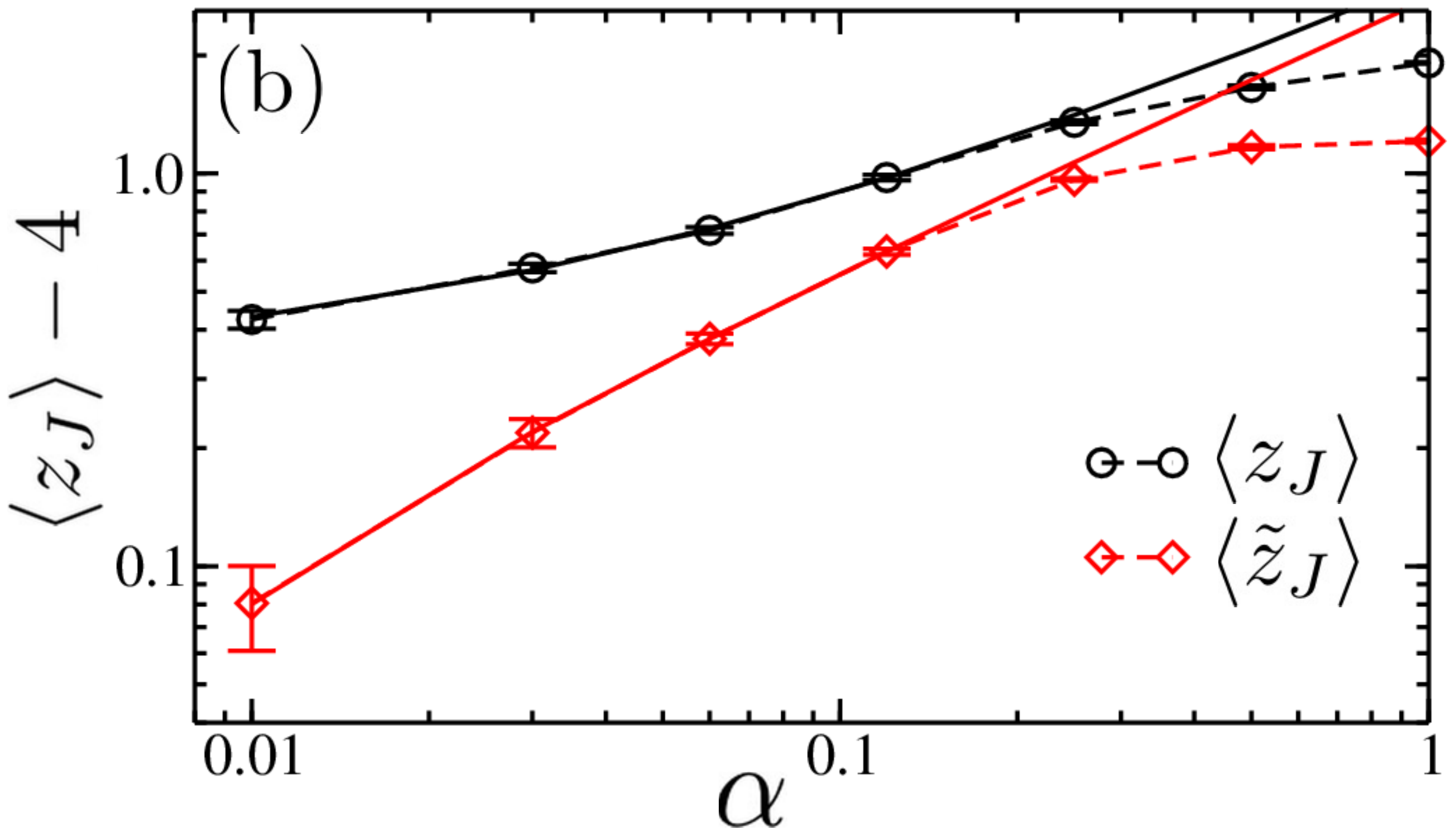}
	\caption{\label{fig:zJ_v_a} (a) Average number of contacts per particle at jamming $\langle z_J\rangle$ vs spherocylinder aspect ratio $\alpha$.  Rattler particles do not enter the average.  Circular data points give the values of $\langle z_J\rangle$ when counting each side-to-side contact as two bonds (see discussion in the main text).  For comparison, diamond shaped data points give the values of $\langle \tilde z_J\rangle$ when counting each side-to-side contact as only a single bond.  (b) $\langle z_J\rangle -4$ and $\langle \tilde z\rangle -4$ vs $\alpha$ on a log-log scale, so as to highlight the behavior as $\alpha\to 0$; for perfectly circular particles, with $\alpha=0$,  $\langle z_J\rangle=4$.  Solid lines are fits to $a+b\alpha^x$.    For $\langle z_J\rangle$ we find $x=0.68\pm 0.13$  and $a=0.31\pm 0.08$, demonstrating that the $\alpha\to 0$ limit of $\langle z_J\rangle$ is discontinuous.  For $\langle \tilde z_J\rangle$ we find $x=0.67\pm 0.11$ and $a=-0.048\pm 0.065$, consistent with $a=0$, and demonstrating that the discontinuity in $\langle z_J\rangle$ at $\alpha=0$ is due to the side-to-side contacts.
The system has $N=1024$ spherocylinders.}
\end{figure}

Having such mechanically stable configurations, obtained by energy minimization as discussed above, will be essential for our analysis of the  eigenmodes of the small elastic vibrations of the system, to be discussed in the next section.
However, we find in practice (checking explicitly for $\alpha=0.01, 1.0$ and 4.0) that $\langle z_J\rangle$ changes negligibly if we compare the value computed in these energy minimized configurations with the value computed in the quasistatically compressed configurations from which we start the minimization procedure.  Rather than carry out energy minimization at all values of $\alpha$, we therefore use the values of $\langle z_J\rangle$ found from our quasistatically compressed configurations. In Fig.~\ref{fig:zJ_v_a}(a) we plot the resulting $\langle z_J\rangle$ vs $\alpha$. In this figure the circular data points give the values of $\langle z_J\rangle$ when we count each side-to-side contact as two bonds, as illustrated in Fig.~\ref{fig:rodConfigs2}; we believe this is the correct approach to properly count the number of constraints, and all cited numerical values  for $\langle z_J\rangle$ represent  values computed in this way.  
We see that $\langle z_J\rangle$ has a peak near the same value of $\alpha\approx 1$ that gives the peak in $\langle\phi_J\rangle$, and that it decreases as $\alpha$ increases further.  Thus, unlike 2D ellipses and 3D ellipsoids \cite{Donev.PRL.2004,Donev.Science.2004,Man.PRL.2005,Donev.PRE.2007,Zeravcic.EPL.2009,Mailman.PRL.2009,Schreck.PRE.2012}, spherocylinders in 2D are not  approaching the isostatic limit as they get increasingly elongated.  At the peak value, $\langle z_J\rangle\approx 5.91\pm 0.01$, close to the isostatic value of 6, but still smaller, so the system is always hypostatic.

For comparison, we also consider the average contact number at jamming when we count each side-to-side contact as only as single bond.  We denote this alternative value as $\langle \tilde z_J\rangle$ and plot it as the diamond shaped data points in Fig.~\ref{fig:zJ_v_a}(a). The noticeable difference between $\langle z_J\rangle$ and $\langle\tilde z_J\rangle$ is therefore a measure of the  fraction of contacts which are side-to-side, and we see that this increases as $\alpha$ increases.
For perfectly circular particles, i.e. $\alpha=0$, the jamming transition is isostatic   \cite{OHern.PRE.2003} with $\langle z_J\rangle=4$.  
In Fig.~\ref{fig:zJ_v_a}(b) we therefore plot $\langle z_J\rangle-4$ and $\langle\tilde z_J\rangle-4$ vs $\alpha$ on a log-log plot, so as to highlight the behavior as $\alpha\to 0$.  The solid lines in Fig.~\ref{fig:zJ_v_a}(b)  are fits of the four smallest $\alpha$ data points to the form $a+b\alpha^x$.  For $\langle z_J\rangle$ we find $x=0.68\pm 0.13$  and $a=0.31\pm 0.08$, while for $\langle \tilde z_J\rangle$ we find $x=0.67\pm 0.11$ and $a=-0.048\pm 0.065$, consistent with $a=0$.  This demonstrates that the contact number $\langle z_J\rangle$ for spherocylinders is discontinuous as $\alpha\to 0$, taking a jump from $4$ at $\alpha=0$ to $4.31$ as $\alpha$ becomes finite, and that this discontinuity is due to the persistence of a finite fraction of side-to-side contacts even as $\alpha\to 0$.  This discontinuity in $\langle z_J\rangle$ is thus specifically a consequence of the flat sides of the spherocylinders, which constrain two degrees of freedom whenever there is a side-to-side contact.  We would expect no such discontinuity in $\langle z_J\rangle$ for ellipses or other shapes without flat sides.

\begin{figure}
	\centering
	\includegraphics[width=0.9\linewidth]{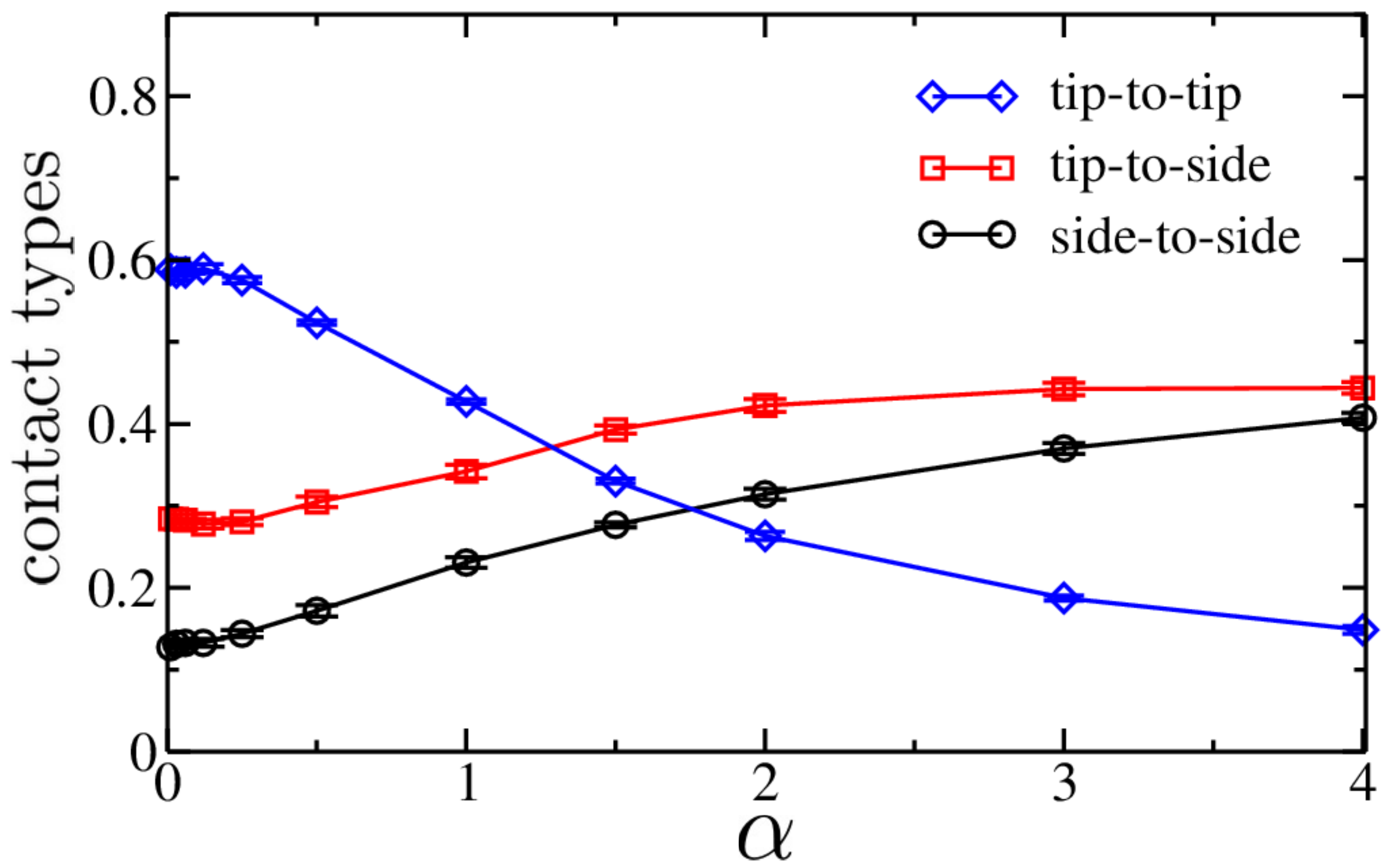}
	\caption{\label{fig:contacts} Fraction of side-to-side, tip-to-side, and tip-to-tip contact bonds in configurations of $N=1024$ bidisperse spherocylinders at the jamming transition, as a function of spherocylinder aspect ratio $\alpha$.  	Each side-to-side contact is counted as two contact bonds, as explained in the main text.	}
\end{figure}

In Fig.~\ref{fig:contacts} we show the fraction of contact bonds   of each of the three different types (i.e., side-to-side, tip-to-side, tip-to-tip) as a function of the aspect ratio $\alpha$.  
As we do in computing $z$, each side-to-side contact is counted as two bonds.
Not surprisingly, the fraction of side-to-side contacts increases  as $\alpha$ increases.  However, consistent with our preceding arguments  concerning $\langle z_J\rangle$, we find that the fraction of side-to-side  contacts remains finite as $\alpha\to 0$.  The fraction of  tip-to-side contacts similarly stays finite as $\alpha\to 0$.  Indeed, for $\alpha=0.01$, we find that virtually all the particles ($96.3\%$ of them) have a contact on at least one of their two flat sides, even though the flat sides represent only $0.63\%$ of the perimeter length.  This is readily seen in Fig.~\ref{fig:snapshots}(a).  

\begin{figure}
	\centering
	\includegraphics[width=0.9\linewidth]{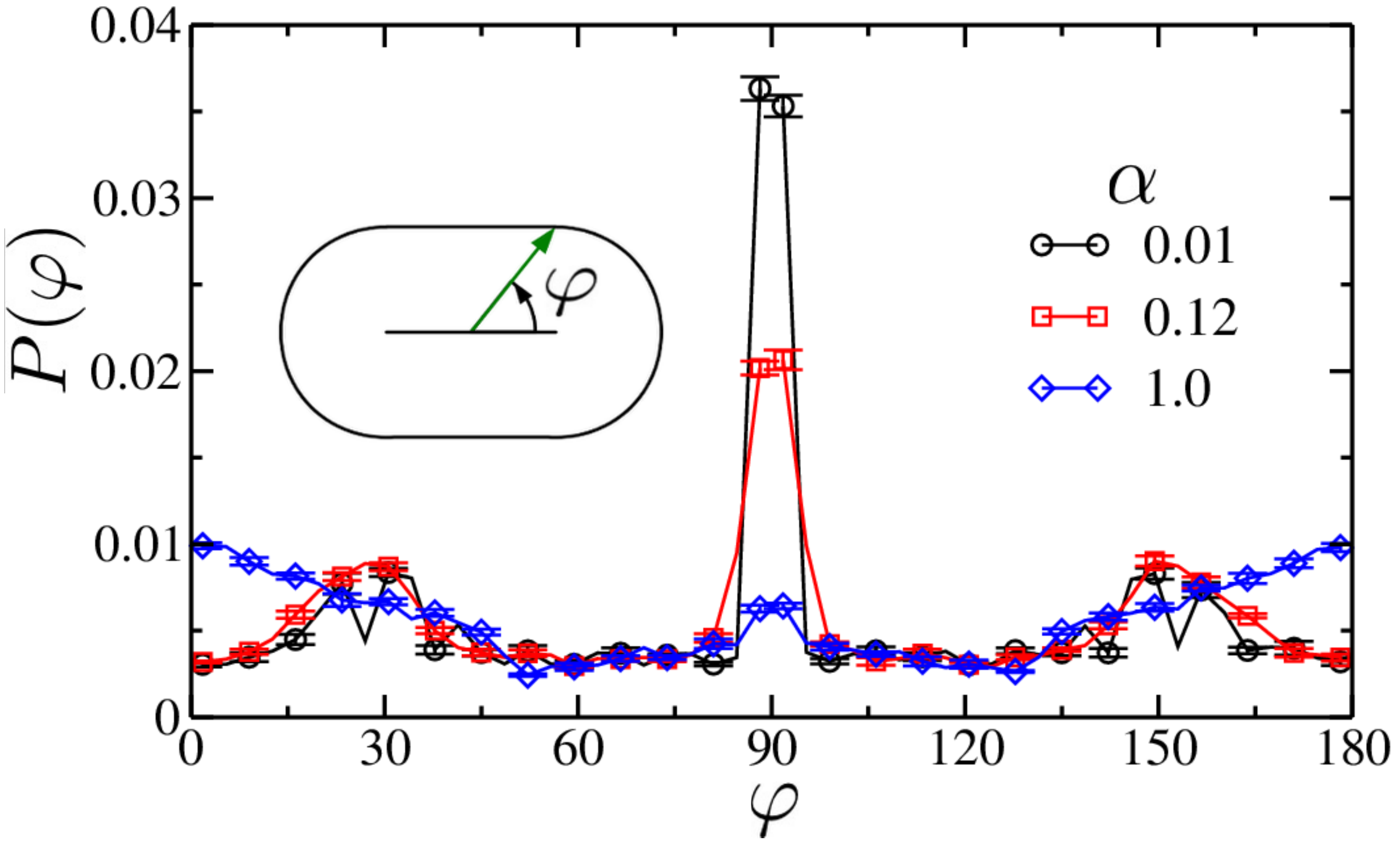}
	\caption{\label{fig:Pvarphi} Probability distribution $P(\varphi)$ for there to be a contact on the surface of a spherocylinder at an angle $\varphi$ relative to the direction of the spherocylinder spine, as illustrated in the inset, for mechanically stable configurations at $U/L^2=10^{-15}$ very close to jamming. Results are shown for aspect ratios $\alpha=0.01$, 0.12, and 1.0, and represent  averages over both big and small spherocylinders.
	}
\end{figure}

To examine this propensity for spherocylinders at small $\alpha$ to have contacts on their flat sides, we  measure the probability for a spherocylinder to have a contact at a particular point on its surface.
Defining $\varphi$ as the polar angle that a given point on the spherocylinder surface makes with respect to the spine (see inset to Fig.~\ref{fig:Pvarphi}), we measure the probability density $P(\varphi)$  to have a contact at angle $\varphi$.  We average over both big and small particles.  In calculating this distribution, we will count side-to-side contacts as only a single bond, located along the flat side as in Fig.~\ref{fig:spherocylinder}(b), since we are more interested in the geometry of the contacts rather than counting constraints.  In Fig.~\ref{fig:Pvarphi} we plot $P(\varphi)$ vs $\varphi$ for values of $\alpha=1.0, 0.12$ and $0.01$.  We see clearly that as $\alpha$ decreases, a  sharp peak grows at $\varphi=90^\circ$, i.e. along the flat side.  In contrast, for a circular disk this distribution would be flat. The smaller, broader, side peaks observed near $\varphi=30^\circ$ and $150^\circ$ may be interpreted as a shadow effect; if a contact exists at an angle $\varphi$, then a neighboring contact is generally no closer than $\varphi\pm 60^\circ$.  

The prevalence of contacts along the flat sides of the spherocylinders, even as $\alpha \to 0$ and the length of these flat sides becomes a negligible fraction of the total spherocylinder surface, suggests that the presence of flat sides makes the $\alpha\to 0$ limit in some sense singular.  As the length $2L$ of the spherocylinder spine shrinks to zero, the system nevertheless seems to remember what direction that spine is in.  This conclusion appears to be robust, as we have demonstrated by the following check.  Rather than starting our energy minimization to obtain mechanically stable states from our quaistatically compressed configurations, we start from a jammed configuration of perfectly circular disks ($\alpha=0$).  We then choose a random spine direction for each particle and distort it into a spherocylinder with $\alpha=0.01$.  We then follow the procedure discussed above to vary the system box size, and energy minimize, so as to obtain a new mechanically stable state of the spherocylinders at $U/L^2=10^{-15}$, close to jamming.  The resulting $P(\varphi)$ for these configurations is found to be the same as in Fig.~\ref{fig:Pvarphi}. 

In response to our above observation, Vanderwerf et al. \cite{Vanderwerf} have recently computed the analogous $P(\varphi)$ for a bidisperse distribution of 2D elliptical particles with minor to major axis ratio $b/a$.  Although the effect is not as dramatic as we find for spherocylinders, they similarly find an increasing probability for contacts along the minor axis of the ellipse, as one takes the limit $b/a\to 1$. This suggests that the effect may hold generally for barely aspherical particles, rather than be specifically due to the flat sides of the spherocylinders.

\subsection{Density of States and Eigenmodes}
\label{sdos}

Having obtained mechanically stable configurations and eliminated rattler particles, 
in this section we analyze in detail the spectrum of the eigenmodes of the $3\mathcal{N}\times 3\mathcal{N}$ dynamical matrix $M_{ia,jb}$ of Eq.~(\ref{eq:dm}), determining the matrix eigenvalues $\lambda_m$ and the corresponding normalized eigenvectors,
\begin{equation}
\mathbf{\hat u}_m=(\mathbf{u}_{1m}, \mathbf{u}_{2m}, \dots), 
\label{eq:eigenvector1}
\end{equation}
where
\begin{equation}
\mathbf{u}_{im}=(\delta x_{im},\delta y_{im}, A_i\delta\theta_{im}) 
\label{eq:eigenvector2}
\end{equation}
gives the small displacement of spherocylinder $i$ in eigenmode $m$.  The frequencies of elastic vibration are then $\omega_m=\sqrt{\lambda_m}$.  We  define the 
resulting density of vibrational states $D(\omega)$ by counting the number of modes within bins of equal relative widths $\Delta\omega/\omega$.  We normalize the density of states so that $\int d\omega D(\omega)=1$.

We  will also wish to characterize the nature of the eigenvectors, in particular whether they correspond to localized or extended modes, and the extent to which they involve translational or rotational motion of the particles.  To measure the extent to which the modes are extended, involving the correlated motion of large groups of spherocylinders, or localized,  involving only a few spherocylinders, we compute the participation ratio $P_m$, defined as \cite{Zeravcic.EPL.2009},
\begin{equation}
P_m=\dfrac{\left[\sum_{i=1}^\mathcal{N}\left[(\delta x_{im})^2+(\delta y_{im})^2 + (A_i\delta\theta_{im})^2\right]\right]^2}
{\mathcal{N}\sum_{i=1}^\mathcal{N}\left[(\delta x_{im})^4+(\delta y_{im})^4 + (A_i\delta\theta_{im})^4\right]},
\end{equation}
where  $\delta x_{im}$, $\delta y_{im}$, $A_i\delta\theta_{im}$ give the components of the eigenvector $\mathbf{\hat u}_m$ as in Eq.~(\ref{eq:eigenvector2}).
For an extended mode in which each degree of freedom is excited equally, we have $P_m=3$; for a localized mode in which only a single degree of freedom is excited, we have $P_m=1/\mathcal{N}$. Taking the average of $P_m$ over all eigenmodes with frequencies $\omega_m$ within bins of equal width $\Delta\omega/\omega$, we then define the participation ratio $P(\omega)$ for modes at frequency $\omega$.

To measure the extent to which a given eigenvector involves translational motion parallel to the spherocylinder's spine, translational motion perpendicular to the spine, or rotational motion about the center of mass, 
we define \cite{Zeravcic.EPL.2009,Mailman.PRL.2009,Schreck.PRE.2012} quantities $u^2_{\parallel m}$, $u^2_{\perp m}$ and $u^2_{\theta m}$,
\begin{align}
u^2_{\parallel m}&=\sum_{i=1}^\mathcal{N} \left[\delta x_{im}\cos\theta_{i} +\delta y_{im}\sin\theta_{i}\right]^2,\\
u^2_{\perp m}&=\sum_{i=1}^\mathcal{N}\left[\delta y_{im}\cos\theta_{i}-\delta x_{im}\sin\theta_{i})\right]^2,\\
u^2_{\theta m}&=\sum_{i=1}^\mathcal{N} \left[A_i\delta\theta_{im}\right]^2.
\end{align}
Because each eigenvector is normalized to unity, we have $u^2_{\parallel m}+u^2_{\perp m}+u^2_{\theta m}=1$.  Taking the average of these quantities over all eigenmodes with frequencies $\omega_m$ within bins of equal width $\Delta\omega/\omega$, we then define $u^2_\parallel(\omega)$, $u^2_\perp(\omega)$ and $u^2_\theta(\omega)$ to describe the average behavior of modes at frequency $\omega$.

\subsubsection{Nearly circular spherocylinders: $\alpha=0.01$}

\begin{figure}
	\centering
	\includegraphics[width=0.9\linewidth]{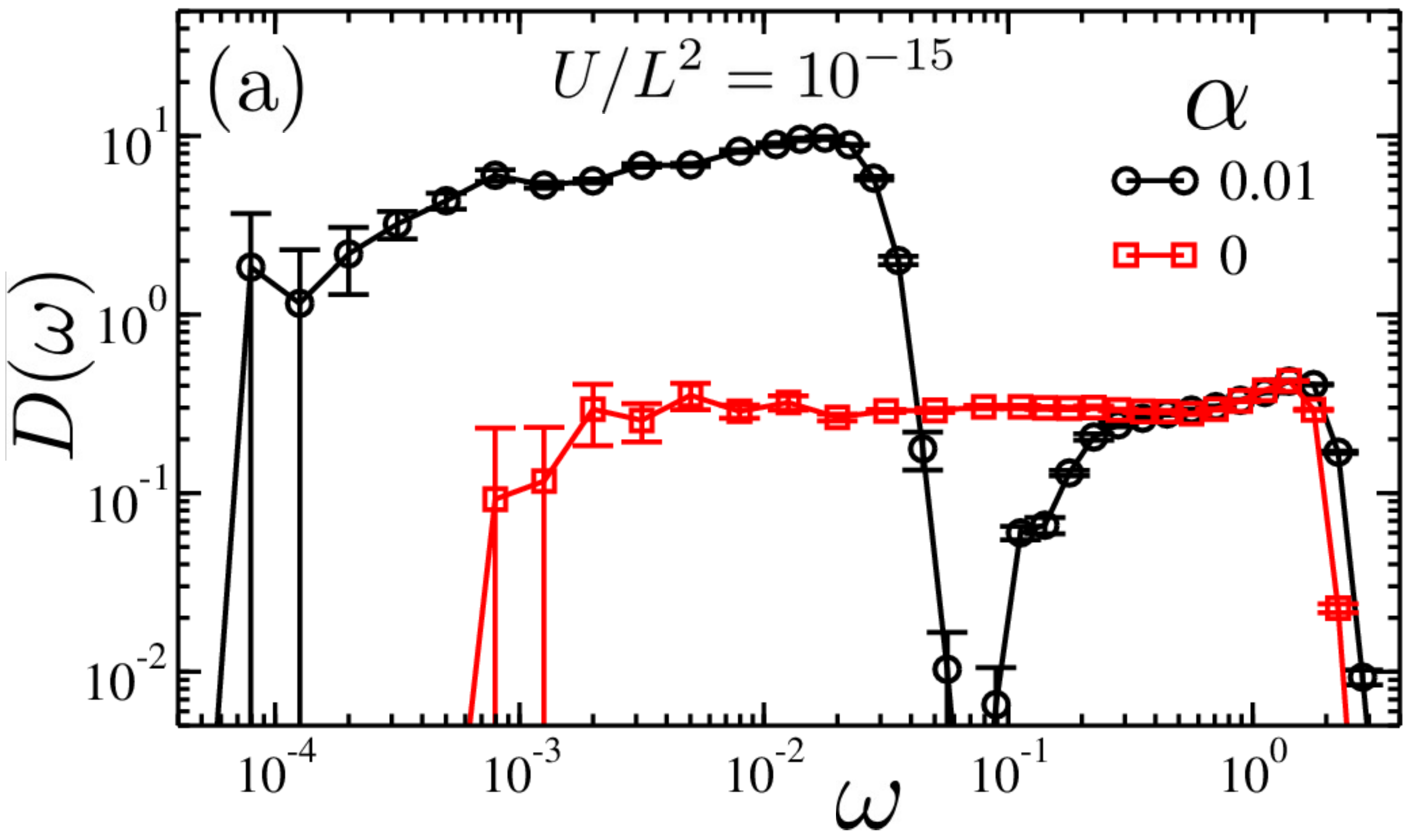}
	\includegraphics[width=0.9\linewidth]{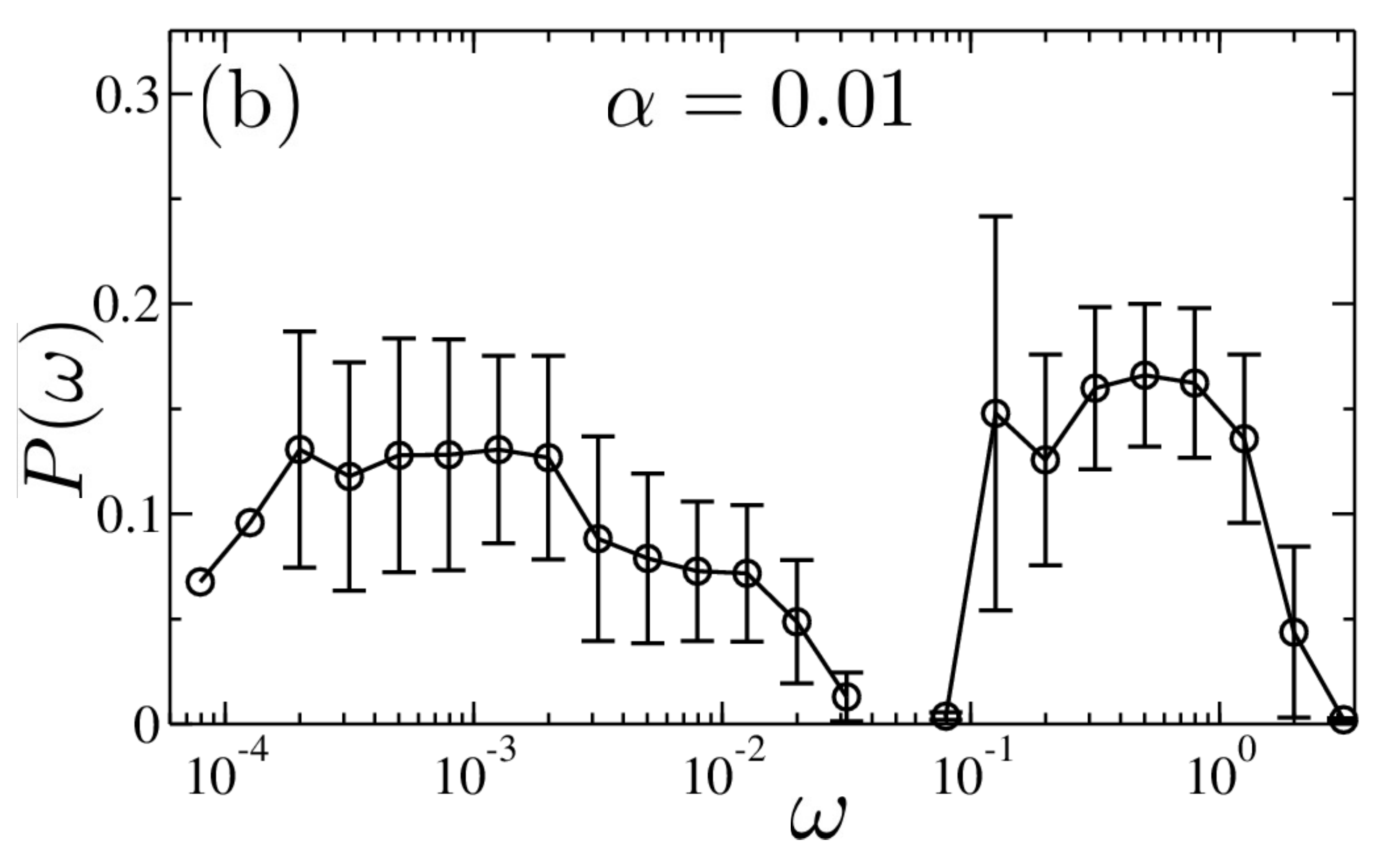}
	\includegraphics[width=0.9\linewidth]{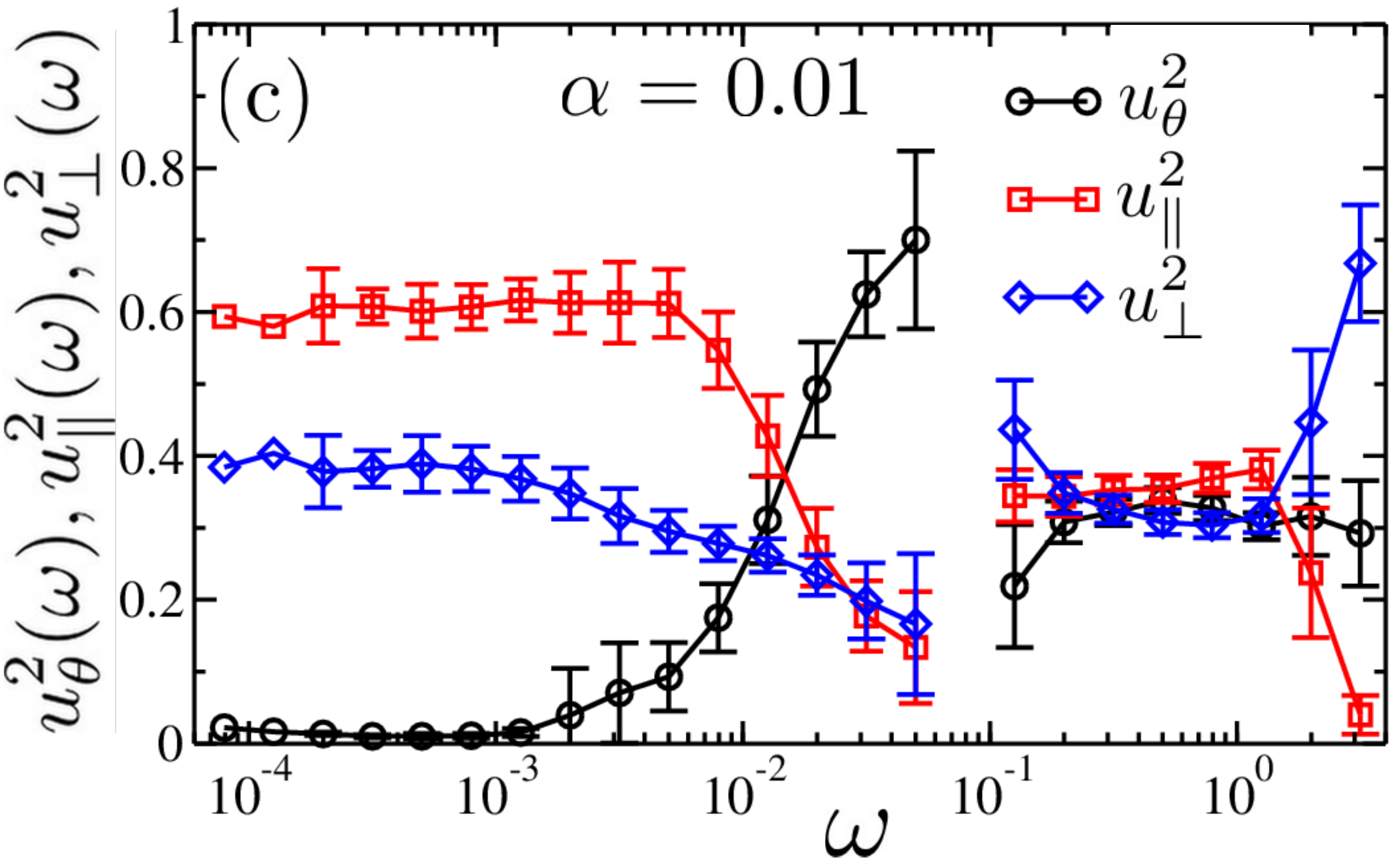}
	\caption{\label{fig:participation_ratio-a001} For a bidisperse system of $N=1024$ nearly circular spherocylinders with   aspect ratio $\alpha = 0.01$, at energy $U/L^2=10^{-15}$ close to the jamming transition, 
	(a) the density of vibrational modes $D(\omega)$ vs $\omega$, and also for comparison  $D(\omega)$ for circular disks with $\alpha=0$;
	(b) the participation ratio $P(\omega)$ vs $\omega$;
	(c) the average rotational motion $ u_\theta^2(\omega) $ (circles),  parallel translational motion $ u_{\parallel}^2(\omega) $ (squares) and perpendicular translational motion $ u_\perp^2(\omega) $ (diamonds) vs $\omega$. ``Parallel" and ``transverse" refer to directions relative to the direction of the spherocylinder spine.
	Each point is averaged over all of the modes in the same frequency bin for six different independent samples.
	}
\end{figure}

In this work we will treat  in detail three specific cases:  aspect ratio $\alpha=0.01$, corresponding to  nearly circular particles,   $\alpha=4.0$, corresponding to moderately elongated particles, and $\alpha=1.0$, corresponding to the peak value of the packing fraction and also the largest value of $\langle z_J\rangle\approx 5.91$.  We start by considering $\alpha=0.01$.
In Fig.~\ref{fig:participation_ratio-a001}(a) we plot our results for the density of states $D(\omega)$ vs $\omega$ for mechanically stable configurations at energy   $U/L^2=10^{-15}$, very close to the jamming transition.   
For comparison, we also show results for $\alpha=0$, i.e., perfectly circular disks; for $\alpha=0$ there is also a delta function contribution (not shown) at $\omega=0$ that represents the $\mathcal{N}$ non-interacting rotational modes of the $\mathcal{N}$ disks.
Our results here, and for other quantities in this section, are averaged over six independent samples.

For  $\alpha=0.01$ we find behavior qualitatively similar to that found previously for elliptical and ellipsoidal particles \cite{Zeravcic.EPL.2009,Mailman.PRL.2009,Schreck.PRE.2012}.  $D(\omega)$ shows two distinct bands of frequencies, separated by a clear gap.  The upper frequency band consists of the finite energy modes usually associated with disordered granular solids near jamming.  Rather than the $D(\omega)\sim \omega^2$ behavior found in uniform elastic solids, there is a proliferation of floppy modes (the ``boson peak" \cite{OHern.PRE.2003}) as $\omega$ decreases, causing $D(\omega)$ to drop sharply as one goes to the low frequency edge of this upper frequency band.  
We find that the total number of modes in this upper frequency band is precisely $\mathcal{N}\langle z\rangle/2$, corresponding to the $\mathcal{N}\langle z\rangle$ contacts that serve to  constrain any large-length scale motion, so that the system is jammed and can support a finite pressure.  

Note, we see a substantial difference in the frequency of the lower cutoff for the upper frequency band when comparing $\alpha=0.01$ with $\alpha=0$.  Although we have not systematically explored the dependence of this cutoff on the spherocylinder aspect ratio $\alpha$, Schreck et al. \cite{Schreck.PRE.2012} have done such calculations for ellipses, and find that as the asphericity of the ellipse $\alpha = a/b-1$ vanishes (with $a/b$ the ratio of the major to minor axis lengths), the lower cutoff frequency of the upper frequency band vanishes as $\omega_\mathrm{cut}\sim\alpha^{1/2}$.  For $\alpha=0.01$ they find this cutoff to be $\omega_\mathrm{cut}\sim 0.1$, similar to what we find here for the spherocylinders.

The total number of modes in the lower frequency band is $\mathcal{N}(z_\mathrm{iso}-\langle z\rangle)/2$, with $z_\mathrm{iso}=2d_f=6$ the isostatic value for non-circular particles, so that the total number of modes in both bands is simply $\mathcal{N}d_f$.
The low frequency band may be thought of as the modes that evolve from the  non-interacting rotational modes of pure disks ($\alpha=0$) once the particle shape is perturbed away from perfect circularity.  As we will demonstrate later in this section, these low band modes 
are in principle zero-frequency modes exactly at jamming, consisting of  cooperative small  displacements of the particles that are unconstrained at the level of a quadratic approximation to the energy minimum.  The frequency of these modes is small but finite in Fig.~\ref{fig:participation_ratio-a001}(a) because we are at a finite energy $U/L^2=10^{-15}$ slightly above jamming.
Note, the equality we find between the number of modes in the upper band and $\mathcal{N}\langle z\rangle/2$ holds only at the jamming point; above the jamming point, the formation of additional contacts does not necessarily act to constrain  previously unconstrained modes of the lower band, but rather may act to overconstrain  modes in the upper band.


In Fig.~\ref{fig:participation_ratio-a001}(b) we show the participation ratio $P(\omega)$.  We see that modes at the edges of either the high frequency band or the low frequency band are localized, with small values of $P(\omega)$.  But modes in the center of either band are fairly extended.  In Fig.~\ref{fig:participation_ratio-a001}(c) we plot the quantities $u^2_\theta(\omega)$, $u^2_\parallel(\omega)$ and $u^2_\perp(\omega)$. We see that modes in the high frequency band are mixed in nature, with roughly equal participation in each of the rotational and translational degrees of freedom.  For the lower frequency band, however the situation is different.  Modes in the upper part of this band involve primarily rotational motions, similar to what was found for the entire low frequency band for ellipses and ellipsoids \cite{Zeravcic.EPL.2009, Mailman.PRL.2009, Schreck.PRE.2012}.  However, unlike with ellipses and ellipsoids, modes towards the lower edge of  this band involve only translational motion, with motion parallel to the spherocylinder spine somewhat greater than motion perpendicular to the spine.  That the lowest energy modes involve translational motion  is presumably a reflection of the flat surfaces that exist on the sides of the spherocylinders.

\begin{figure}
	\centering
	\includegraphics[width=0.9\linewidth]{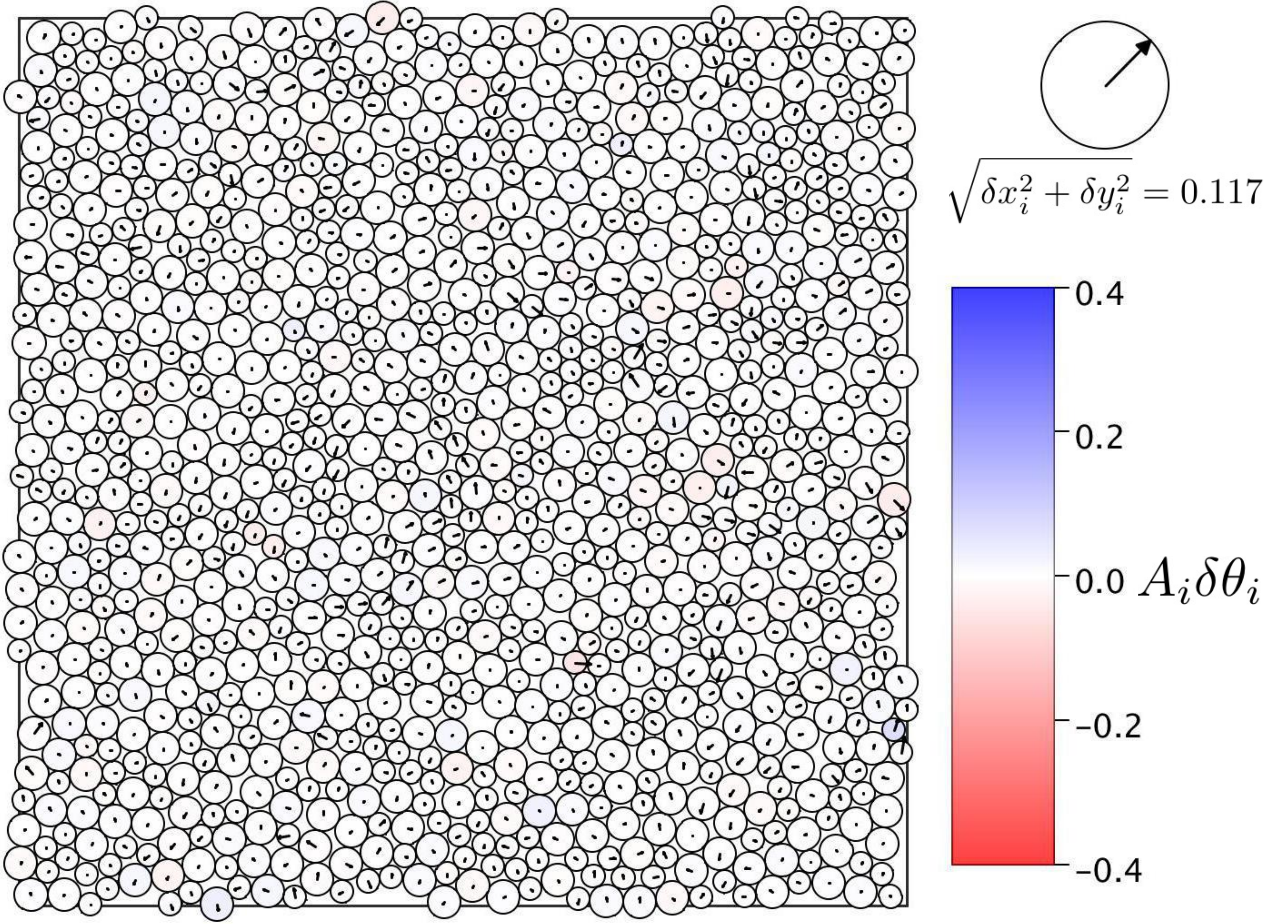}\\
	\large{(a) $\omega_m=3\times 10^{-3}\qquad\qquad$}
	\includegraphics[width=0.9\linewidth]{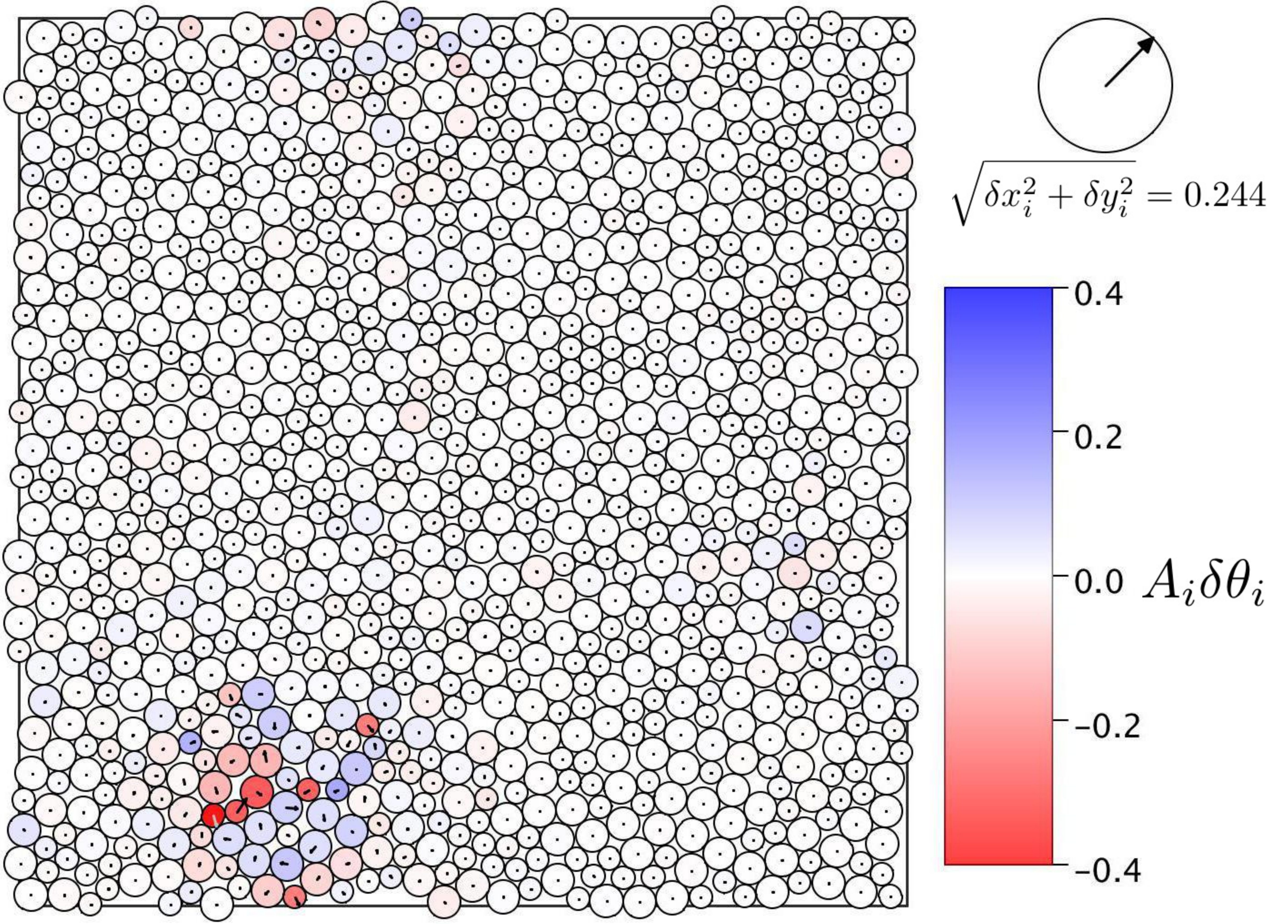}\\
	\large{(b) $\omega_m=3\times 10^{-2}\qquad\qquad$}
	\caption{\label{fig:mode001} For $N=1024$ nearly circular spherocylinders with aspect ratio $\alpha=0.01$, at energy   $U/L^2=10^{-15}$ very close to the jamming transition,
	(a) eigenmode at $\omega_m=3\times 10^{-3}$ near the middle of the low frequency band, and 
	(b) eignemode at $\omega_m=3\times 10^{-2}$ near the upper edge of the low frequency band.  Arrows on the spherocylinders indicate the relative translational motion, and color the relative rotational motion, according to the legend on the right hand side.
		}
\end{figure}

In Fig.~\ref{fig:mode001} we illustrate graphically two examples of eigenmodes in the low frequency band.  
In these figures, arrows on each spherocylinder are proportional to the translational displacement of the spherocylinder, while the color of each spherocylinder indicates the degree of rotation: blue is a counterclockwise rotation, while red is clockwise, with the darkness of the color proportional to the amount of the rotation.  
Fig.~\ref{fig:mode001}(a) is for a mode at $\omega_m=3\times 10^{-3}$, somewhat in the middle of the band.  As expected from Fig.~\ref{fig:participation_ratio-a001}, one clearly sees that this mode is extended throughout the system and is mixed between translational and rotational motion.  In contrast, the mode in Fig.~\ref{fig:mode001}(b)  at $\omega_m=3\times 10^{-2}$, near the upper edge of the band, is clearly seen to be more localized and consists primarily of rotational motion.

\begin{figure}
	\centering
	\includegraphics[width=0.9\linewidth]{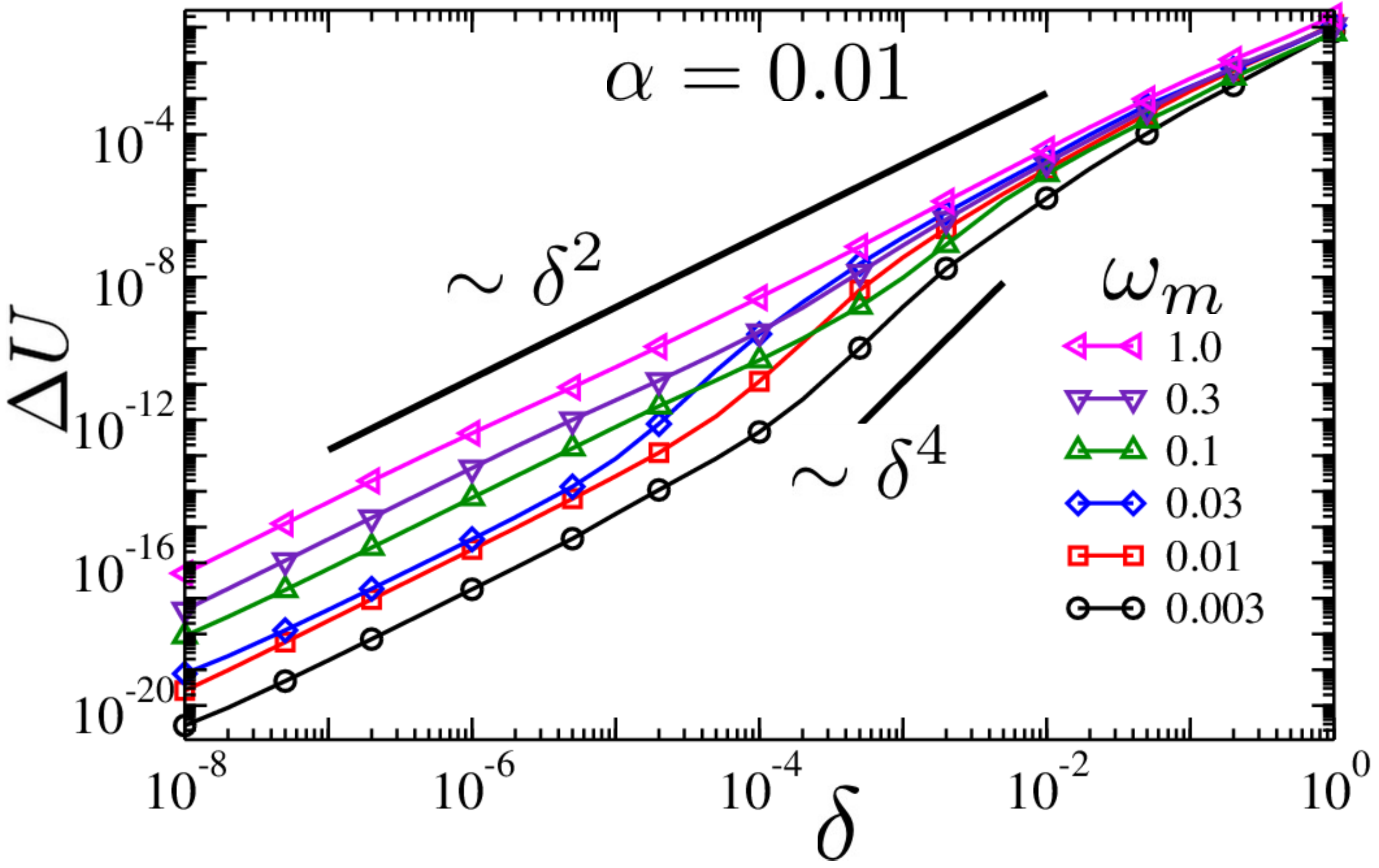}
	\caption{\label{fig:energy_basin_001} Change in energy $\Delta U$ vs displacement $\delta$ for perturbations of spherocylinder positions $\delta\mathbf{\hat u}_m$ along different eigenmode directions $\mathbf{\hat u}_m$ at frequencies $\omega_m$ as indicated.  The system of $N=1024$ bidisperse spherocylinders is at energy $U/L^2=10^{-15}$, very close to jamming, and the aspect ratio is $\alpha=0.01$. Solid black lines indicate the behaviors $\Delta U\sim \delta^2$ and $\Delta U\sim \delta^4$.}
\end{figure}

When displacing the spherocylinders an amount $\delta$ in the direction of a given eigenmode $\mathbf{\hat u}_m$, the energy of the system will increase by $\Delta U=\omega_m^2\delta^2$, to lowest order in $\delta$.  It is interesting to see how $\Delta U(\delta)$ behaves as one increases $\delta$ away from the small $\delta$ limit.
Note, for a highly localized mode we expect that $\delta\sim 1$ corresponds to the displacement of a particle on the order of a particle diameter $2R_s$; for a highly delocalized mode, we expect that $\delta\sim 1$ corresponds to the displacement of particles on the order of $2R_s/\sqrt{3N}\sim 2R_s/50$.  In Fig.~\ref{fig:energy_basin_001} we plot $\Delta U(\delta)$ vs $\delta$ 
for several typical modes $m$ at the different frequencies $\omega_m$ as shown.  For $\omega_m$ in the high frequency band, we see that $\Delta U \sim \delta^2$ for the entire range $0\le\delta\le 1$.  For $\omega_m$ in the low frequency band, we see that $\Delta U$ crosses over from a form $A\delta^2$ at small $\delta$ to a form $B\delta^2$ at larger $\delta$, with $A\ll B$.  The small $A$ corresponds to the small eigenvalue $\lambda_m=\omega_m^2$ of the lower band; this $\lambda_m$ would be zero if the system were exactly at $\phi_J$.  The cross over region to the larger $B$ suggests the quartic nature of these low band modes  (i.e., $\Delta U\sim \delta^4$) that has been predicted  \cite{Donev.PRE.2007} to hold exactly at the jamming $\phi_J$.  As $\delta$ increases further we find that the contact network starts to change significantly; the ``grazing" particle overlaps (overlap $\sim \delta^2$) that characterize the quartic nature of the low band modes at small $\delta$ start to break, and new ``ordinary" contacts start to form as particles push into each other with overlaps similar to those found in the modes of the higher band (overlap $\sim\delta$).  This results in the larger value $B$, which is comparable to  similar  values found in the high frequency band.

\begin{figure}
	\centering
	\includegraphics[width=0.9\linewidth]{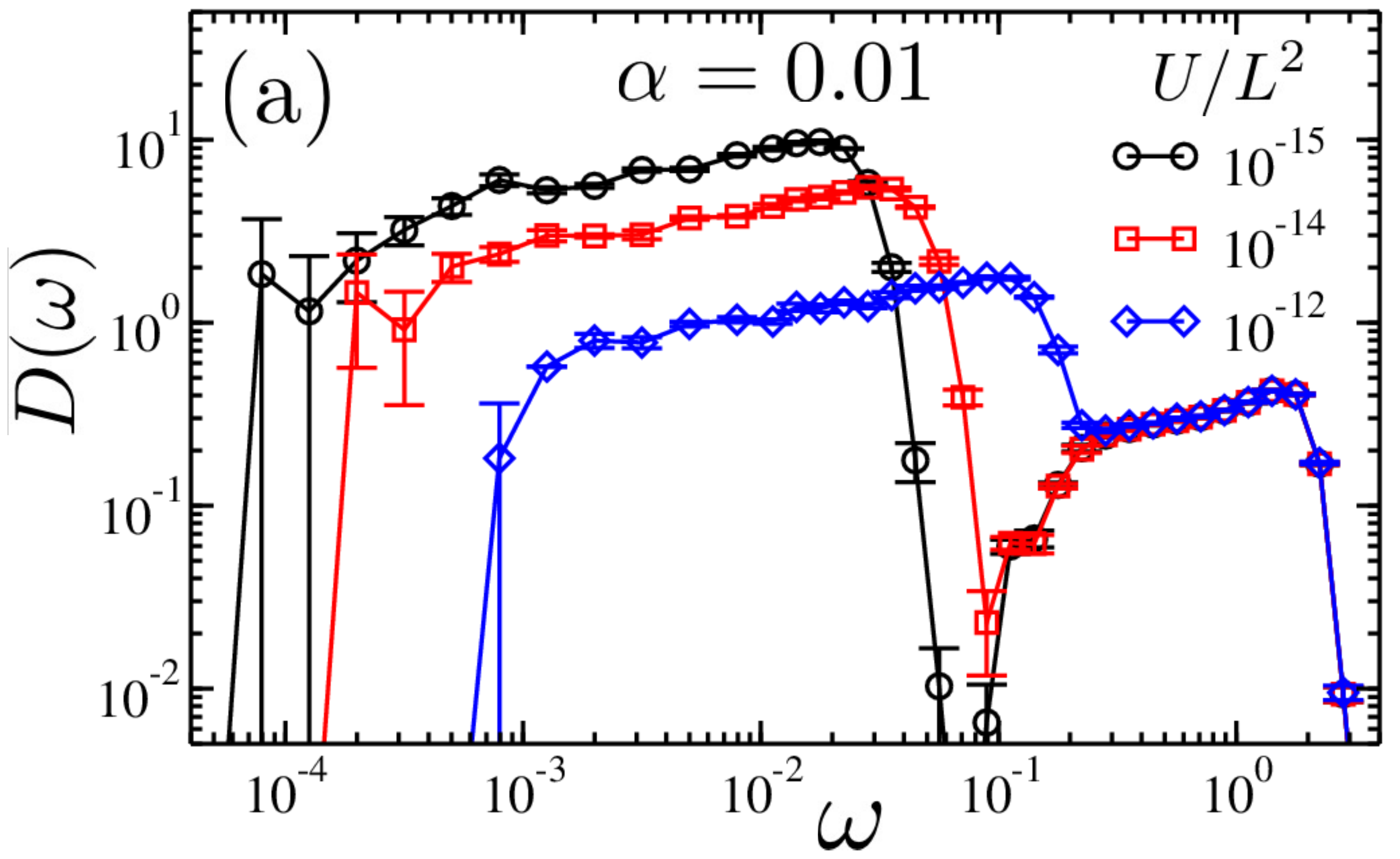}
	\includegraphics[width=0.9\linewidth]{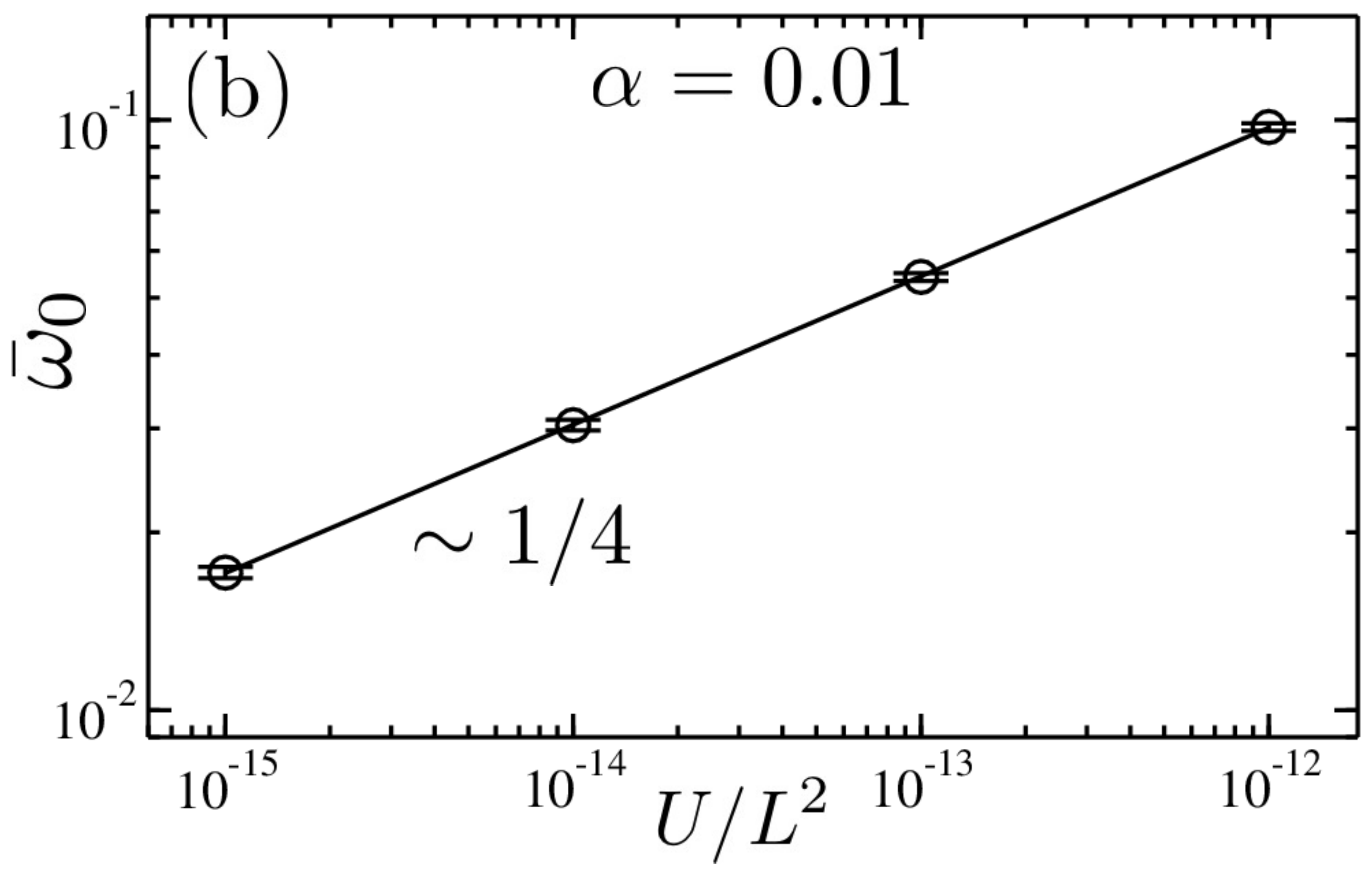}
	\caption{\label{fig:density_states_v_U-01} (a) The density of vibrational states $D(\omega)$ for $N=1024$ bidisperse spherocylinders with aspect ratio $\alpha = 0.01$ at configuration energies  $U/L^2 = 10^{-15}$,  $10^{-14}$, and $10^{-12}$, corresponding to  packing fractions of $\phi-\phi_J \approx 7.4\times10^{-8}$, $2.4\times 10^{-7}$, and $2.4\times10^{-6}$ respectively.
	(b) Average frequency  of modes in the
	lower frequency band $\bar\omega_0$  vs energy $U/L^2$.  We find $\bar\omega_0\sim (U/L^2)^{1/4}$.
	}
\end{figure}

Finally we explore the dependence of the eigenmodes on the energy of the system $U$.
In Fig.~\ref{fig:density_states_v_U-01}(a) we plot $D(\omega)$ vs $\omega$ for mechanically stable configurations at several different values of $U$.  As was found previously for ellipses and ellipsoids \cite{Zeravcic.EPL.2009, Mailman.PRL.2009, Schreck.PRE.2012}, we see that as $U$ increases, the upper frequency band changes relatively little, while the lower frequency band increases, the gap between the two bands narrows, and ultimately the two bands merge.  Defining $\bar\omega_0$ as the average frequency of the modes in the lower band, 
in Fig.~\ref{fig:density_states_v_U-01}(b) we plot $\bar\omega_0$
vs $U/L^2$.  We see a perfect power law dependence,
strongly suggesting that the lower band of modes collapses to $\omega\to 0$ as $U/L^2\to 0$ at jamming.  In particular we find $\bar\omega_0\sim (U/L^2)^{1/4}$.  Since, for our harmonic elastic interaction $U/L^2\sim (\phi-\phi_J)^2$, this gives, $\bar\omega_0\sim(\phi-\phi_J)^{1/2}$, in agreement with results found previously for ellipses and ellipsoids \cite{Schreck.PRE.2012}.

\subsubsection{Moderately elongated spherocylinders: $\alpha=4.0$}

\begin{figure}
	\centering
	\includegraphics[width=0.9\linewidth]{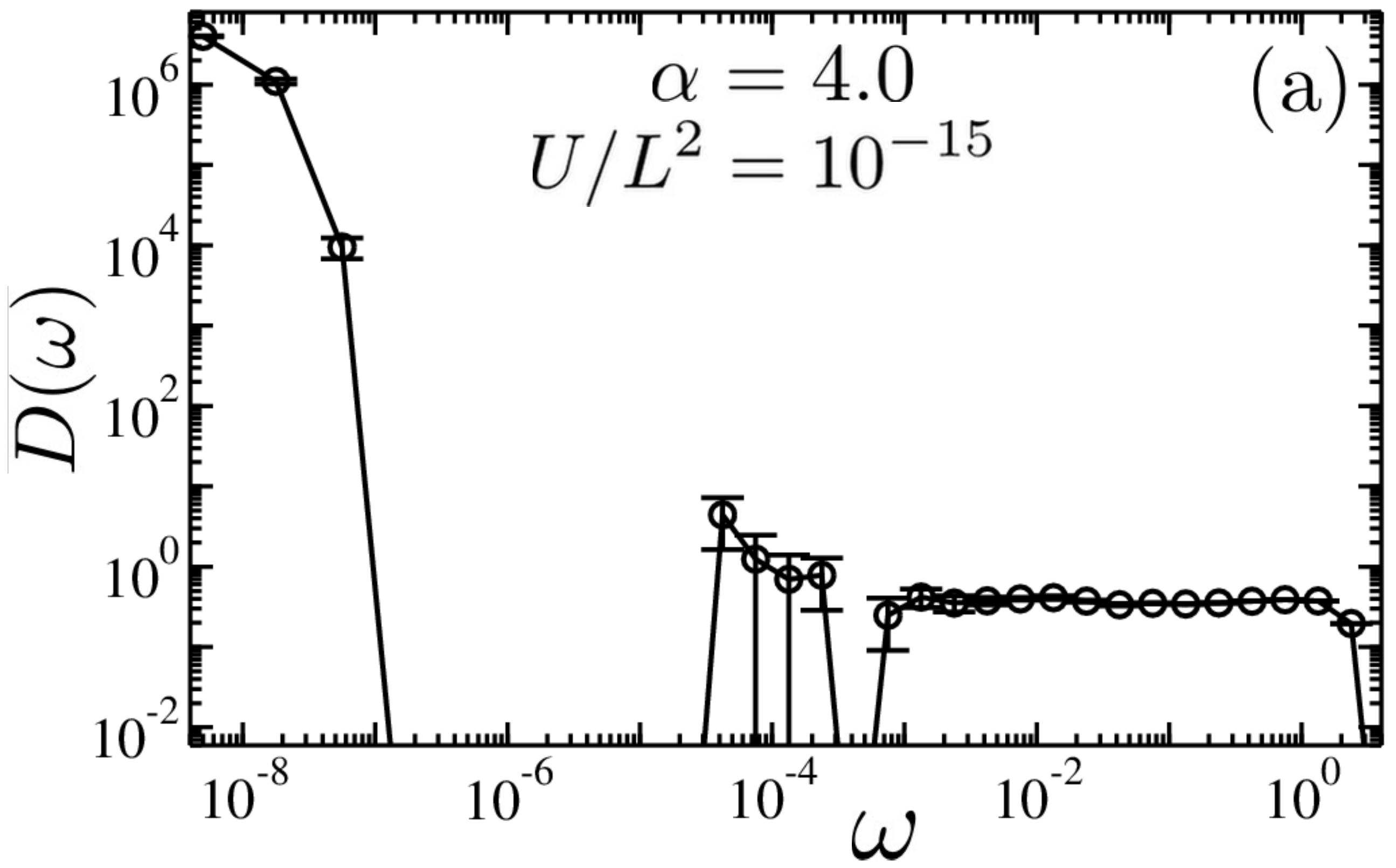}
	\includegraphics[width=0.9\linewidth]{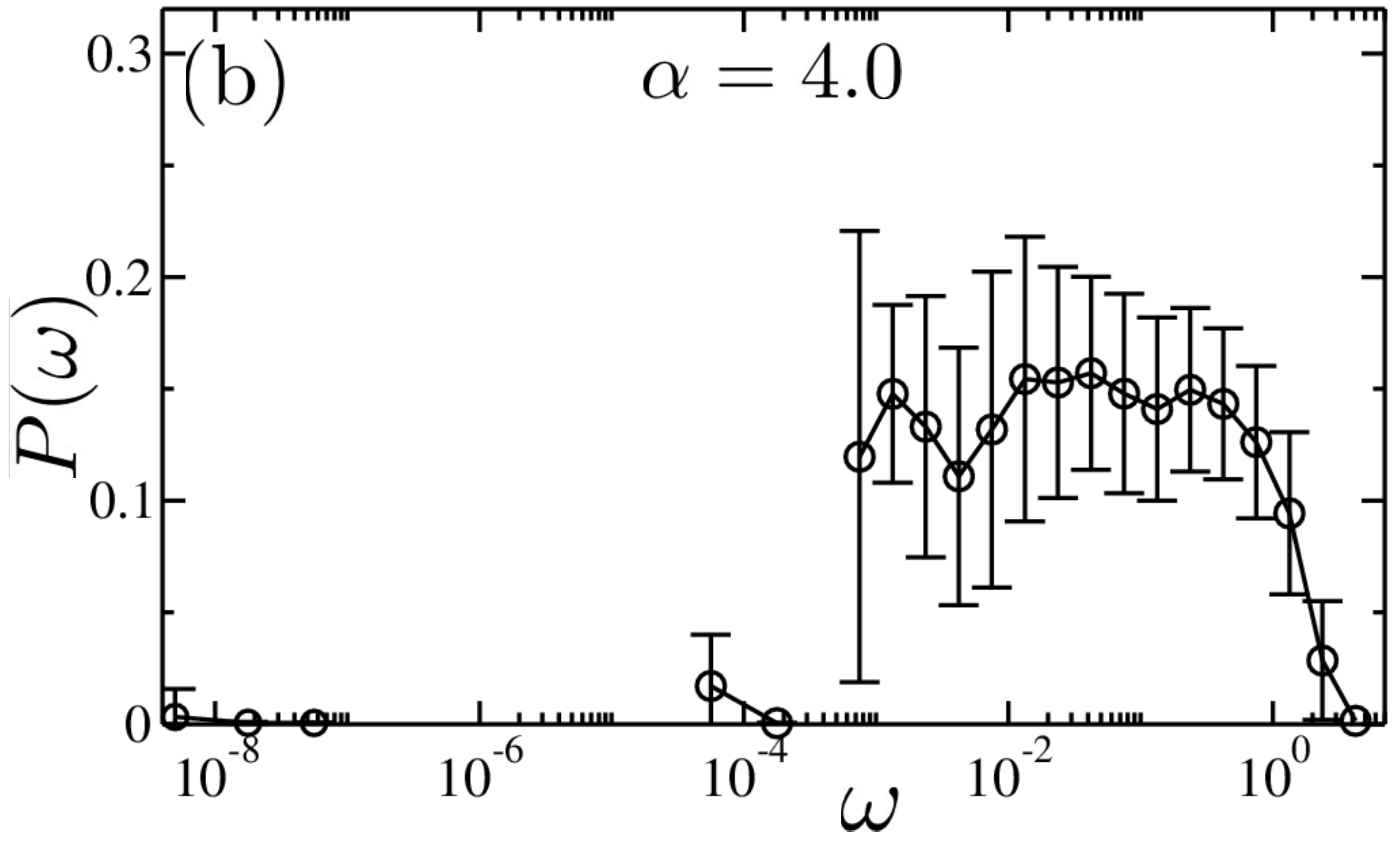}
	\includegraphics[width=0.9\linewidth]{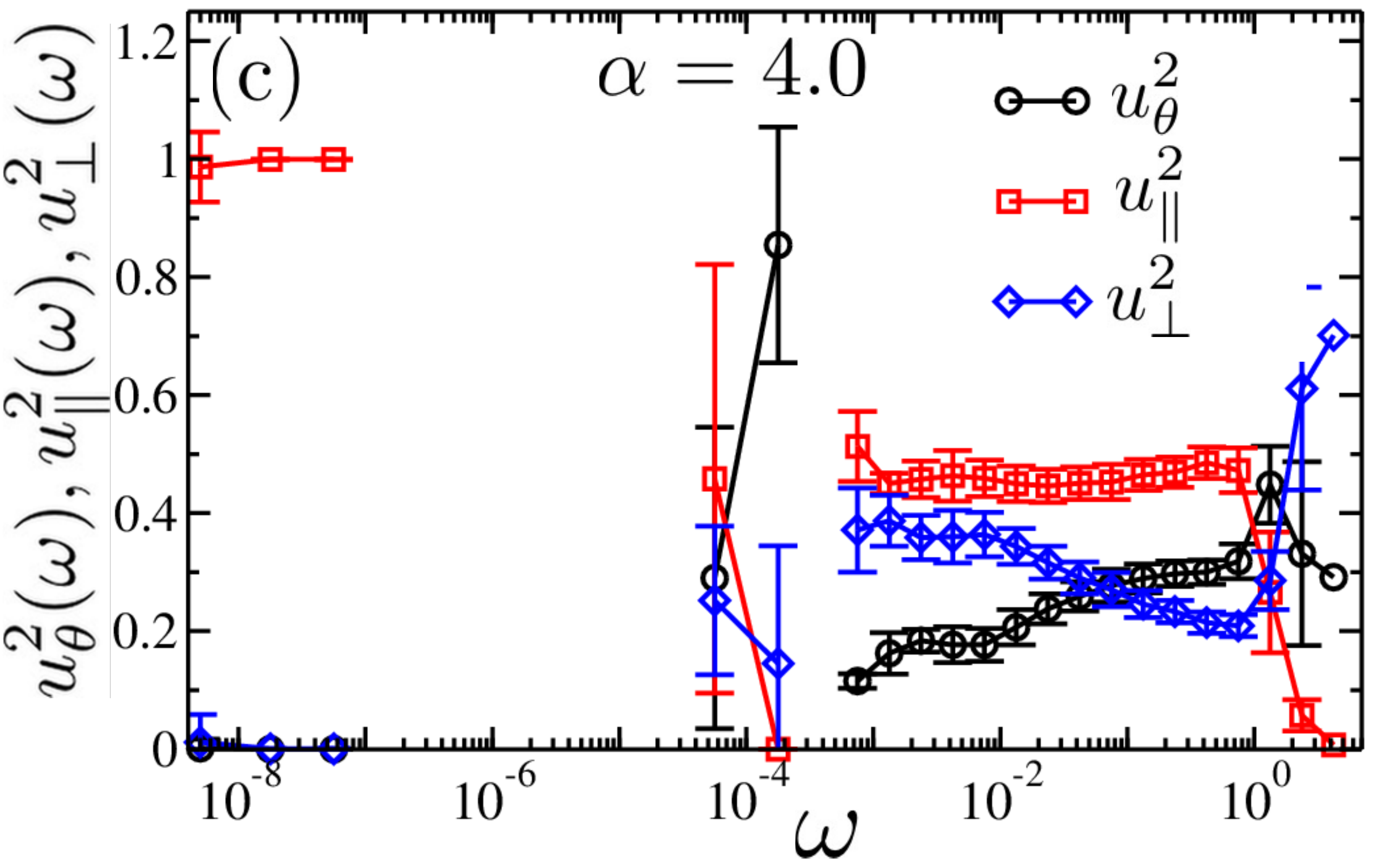}
	\caption{\label{fig:participation_ratio-a4} For a bidisperse system of $N=1024$ moderately elongated spherocylinders with   aspect ratio $\alpha = 4.0$, at energy $U/L^2=10^{-15}$ close to the jamming transition, 
	(a) the density of vibrational modes $D(\omega)$ vs $\omega$;
	(b) the participation ratio $P(\omega)$ vs $\omega$;
	(c) the average rotational motion $ u_\theta^2(\omega) $ (circles),  parallel translational motion $ u_{\parallel}^2(\omega) $ (squares) and perpendicular translational motion $ u_\perp^2(\omega) $ (diamonds) vs $\omega$. ``Parallel" and ``transverse" refer to directions relative to the direction of the spherocylinder spine.
	Each point is averaged over all of the modes in the same frequency bin for six different independent samples.
	}
\end{figure}

We now consider the case of moderately elongated spherocylinders with aspect ratio $\alpha=4.0$.
In Fig.~\ref{fig:participation_ratio-a4}(a) we plot our results for the density of states $D(\omega)$ vs $\omega$ for mechanically stable configurations at energy  $U/L^2=10^{-15}$, very close to the jamming transition.   
In this case we see three bands of eigenmodes: a high frequency band similar to that found for $\alpha=0.01$, a low frequency band, and a narrow middle band.  
We note that the frequencies in the lowest band are so exceedingly small that it was  necessary to adopt a perturbative approach to compute them accurately (see Appendix B). 
We will show below that both the low and middle frequency bands scale to zero as $U\to 0$ and the system approaches the jamming transition; hence these lower two bands correspond to the modes which are unconstrained, to quadratic order, exactly at the jamming $\phi_J$.  As with $\alpha=0.01$, we find that the number of modes in the high frequency band is exactly $\mathcal{N}\langle z\rangle/2$, while the total number of modes in the low and middle frequency bands is $\mathcal{N}(z_\mathrm{iso}-\langle z\rangle)/2$.  The fraction of modes in the low, middle, and high frequency bands is 0.0638, 0.00033, and 0.936 respectively.   

In Fig.~\ref{fig:participation_ratio-a4}(b) we plot the participation ratio $P(\omega)$, and in Fig.~\ref{fig:participation_ratio-a4}(c)  we plot the 
quantities $u_\theta^2(\omega)$, $u_\parallel^2(\omega)$ and $u_\perp^2(\omega)$.
We see that the high frequency band consists of a set of mostly extended modes with mixed rotational and translational motion, similar to what was found for $\alpha=0.01$.  However we see that all the modes in the low and middle bands are  strongly localized.  In the low band these modes are  entirely translational in the direction parallel to the spine of the spherocylinder.  In the middle band these modes are primarily rotational.

\begin{figure}
	\centering
	\includegraphics[width=0.9\linewidth]{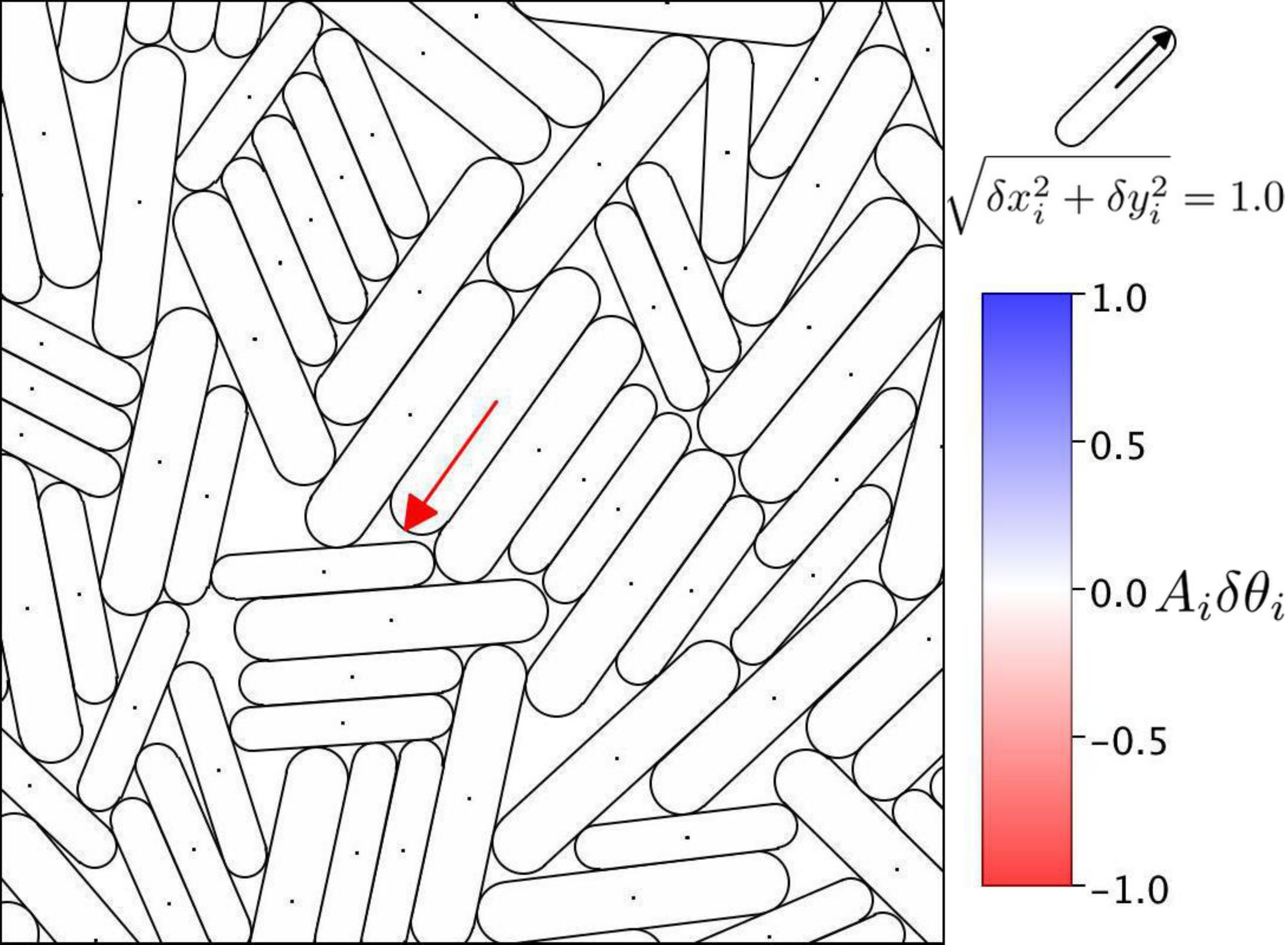}\\
	\large{(a) $\omega_m=10^{-8}\qquad\qquad$}\\
	\includegraphics[width=0.9\linewidth]{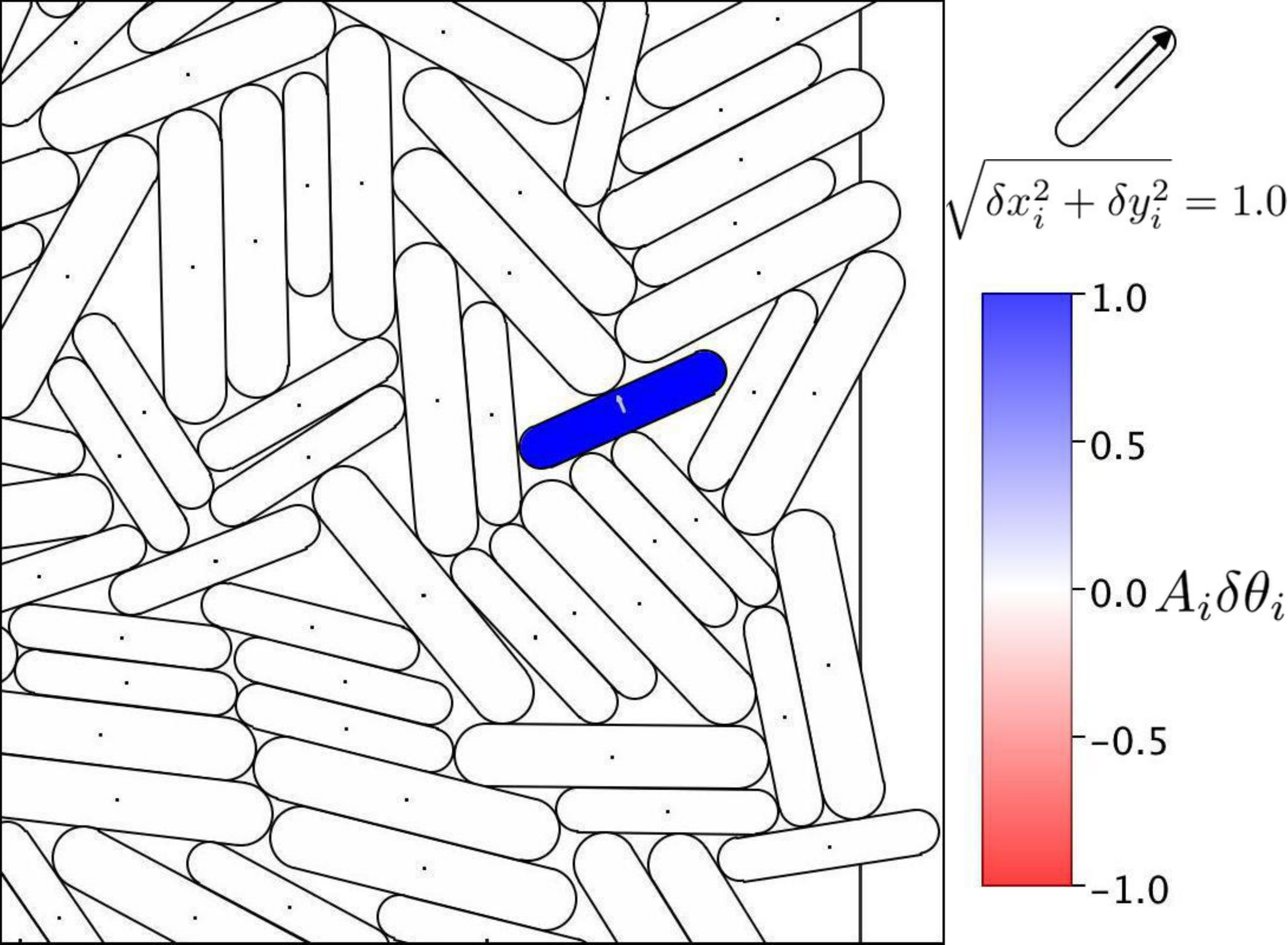}\\
	\large{(b) $\omega_m=10^{-4}\qquad\qquad$}
	\caption{\label{fig:mode4} For $N=1024$ moderately elongated spherocylinders with aspect ratio $\alpha=4.0$, at energy   $U/L^2=10^{-15}$ very close to the jamming transition:
	(a) eigenmode at $\omega_m=10^{-8}$ in the low frequency band; the red arrow denotes the spherocylinder with the largest motion. And 
	(b) eignemode at $\omega_m=10^{-4}$ in the middle band; the dark blue  spherocylinder denotes the one with the largest motion.  Arrows on the spherocylinders indicate the relative translational motion, and color the relative rotational motion, according to the legend on the right hand side.  In each case we show only a subregion of the entire system
		}
\end{figure}

In Fig.~\ref{fig:mode4} we illustrate graphically two examples of eigenmodes, one in the low frequency band and one in the middle frequency band.  Because these modes are highly localized, we show only a subregion of the system containing the spherocylinder that moves, rather than the entire system.
Fig.~\ref{fig:mode4}(a) is for a mode at $\omega_m=10^{-8}$, in the low band.  It is clear that this mode consists of only a single spherocylinder that slides parallel to its spine.  Fig.~\ref{fig:mode4}(b)  is for a mode at $\omega_m=10^{-4}$, in the middle band.  Again this mode consists of only a single spherocylinder, but now the motion is primarily rotational.
In both cases, although the isolated spherocylinder may move along one degree of freedom with  low cost in energy, its presence is nevertheless clearly important for the global rigidity of the system.

\begin{figure}
	\centering
	\includegraphics[width=0.9\linewidth]{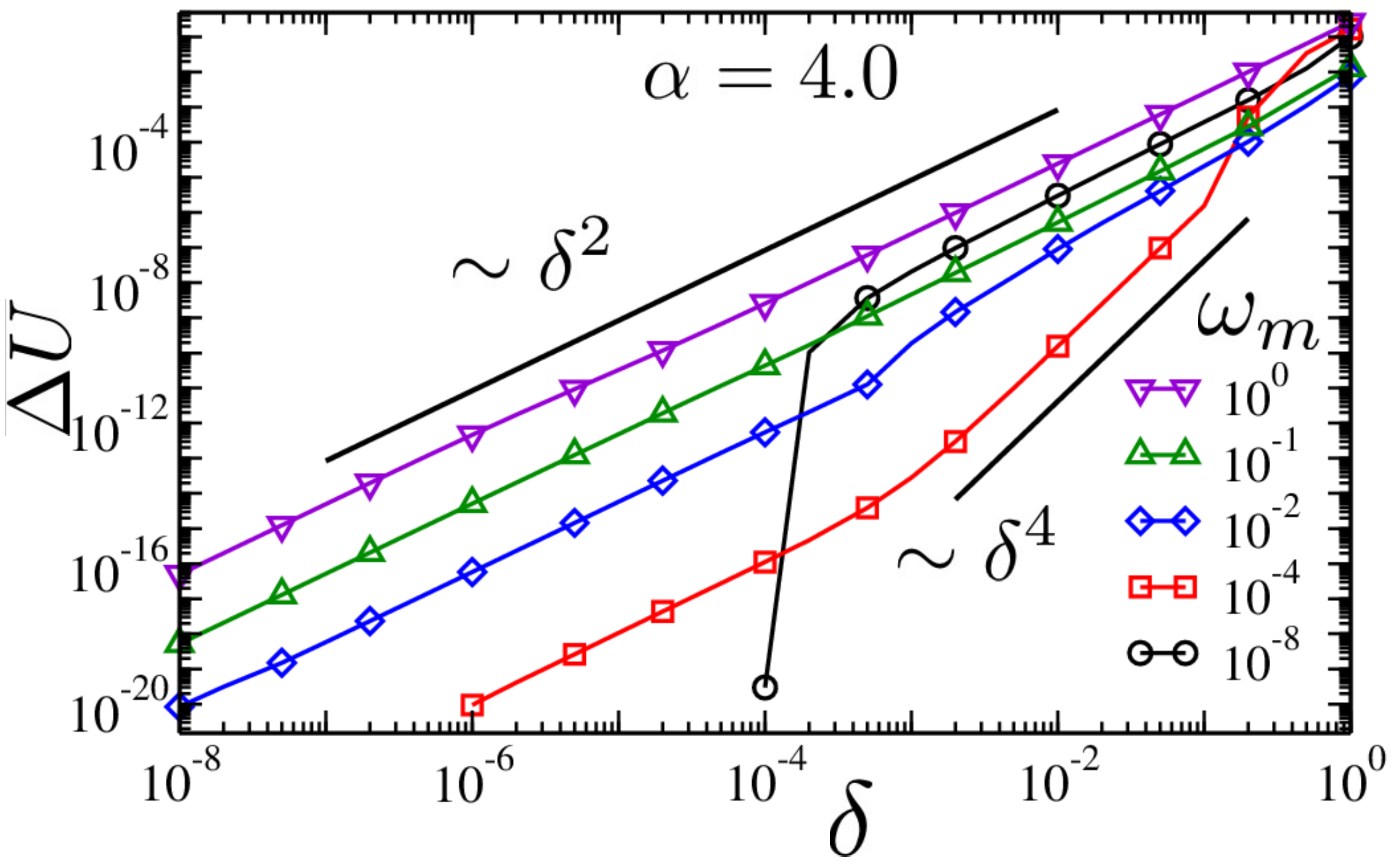}
	\caption{\label{fig:energy_basin_4} Change in energy $\Delta U$ vs displacement $\delta$ for perturbations of spherocylinder positions $\delta\mathbf{\hat u}_m$ along different eigenmode directions $\mathbf{\hat u}_m$ at frequencies $\omega_m$ as indicated.  The system of $N=1024$ bidisperse spherocylinders is at energy $U/L^2=10^{-15}$, very close to jamming, and the aspect ratio is $\alpha=4.0$. Solid black lines indicate the behaviors $\Delta U\sim \delta^2$ and $\Delta U\sim \delta^4$.}
\end{figure}

In Fig.~\ref{fig:energy_basin_4} we show the change in energy $\Delta U(\delta)$ as the spherocylinders are displaced a distance $\delta$ in the direction of several given eigenmodes $\mathbf{\hat u}_m$.  For the modes in the high frequency band, we see $\Delta U \sim \delta^2$ for most of the range of $\delta$.  
For the two lowest modes shown, at $\omega_m=10^{-4}$ and $10^{-8}$ in the middle and low band respectively (these are the same two modes illustrated in Fig.~\ref{fig:mode4}), 
we do not have sufficient numerical accuracy to compute $\Delta U$ at the smallest values of $\delta$.  For the middle band mode, which is mostly rotational, we see behavior similar to that found for the low band modes of $\alpha=0.01$.  As $\delta$ increases, $\Delta U(\delta)$ transitions from a small $\delta$ behavior of $A\delta^2$, with $A=\omega_m^2$,  to $B\delta^2$ with $A\ll B$; the transition region has a quartic dependence $\sim \delta^4$.
For the translational mode in the low band, the small $\delta$ behavior $\Delta U=\omega_m^2\delta^2$ is too small to be computed accurately and so does not appear in our plot.  We believe this very small energy is due to the fact that the spherocylinder involved in this mode is not exactly parallel with its neighbors, and so the energy in the side-to-side contact changes ever so slightly as the spherocylinder slides parallel to its spine.  The sharp step upwards seen for this mode in Fig.~\ref{fig:energy_basin_4} corresponds to a displacement large enough that the tip of the  sliding spherocylinder starts to contact and overlap a neighbor that it previously did not touch.  We suspect that exactly at $\phi_J$, such sliding modes may be strictly unconstrained (for small enough $\delta$), rather than quartic.

\begin{figure}[h!]
	\centering
	\includegraphics[width=0.9\linewidth]{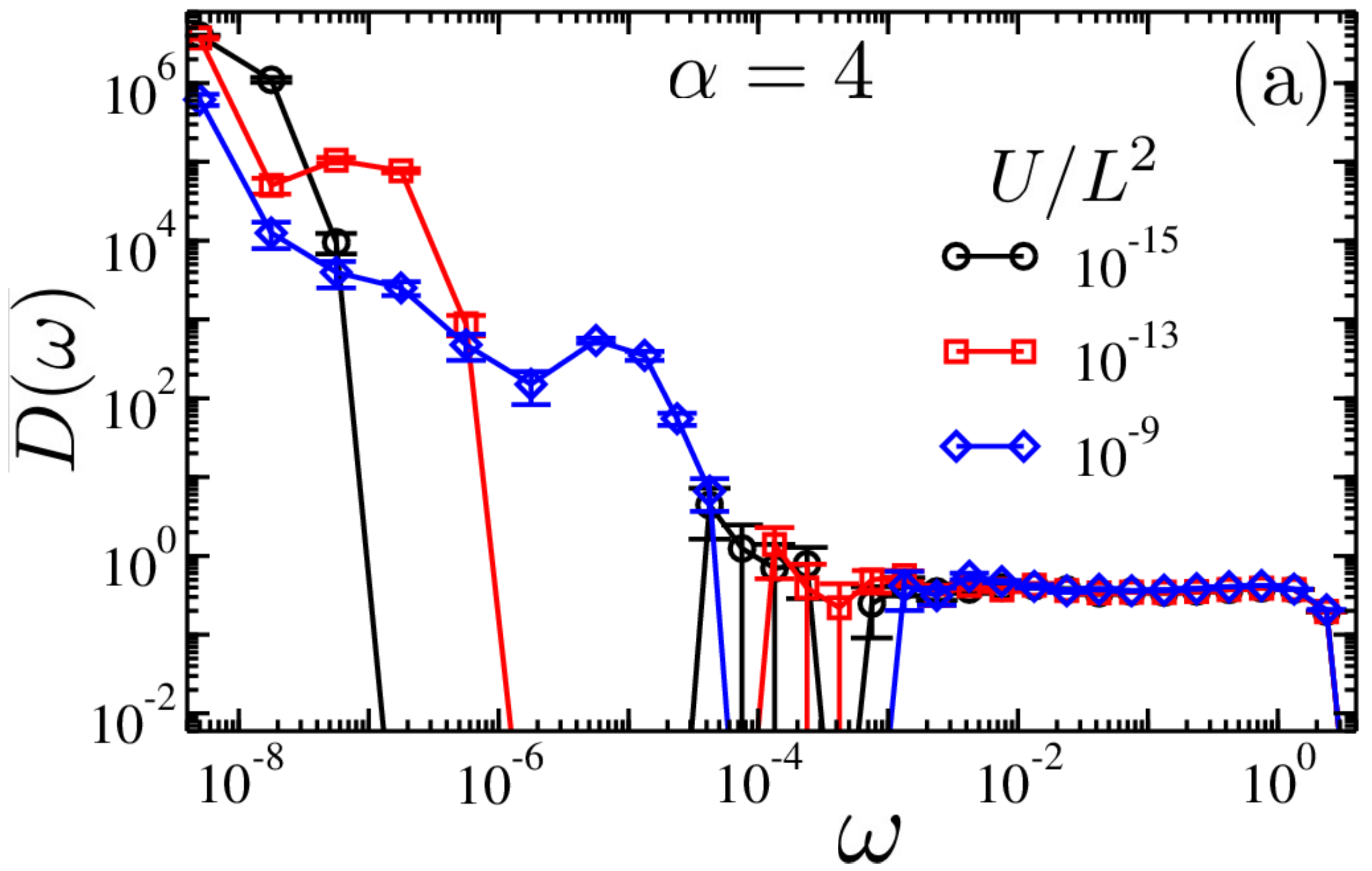}
	\includegraphics[width=0.9\linewidth]{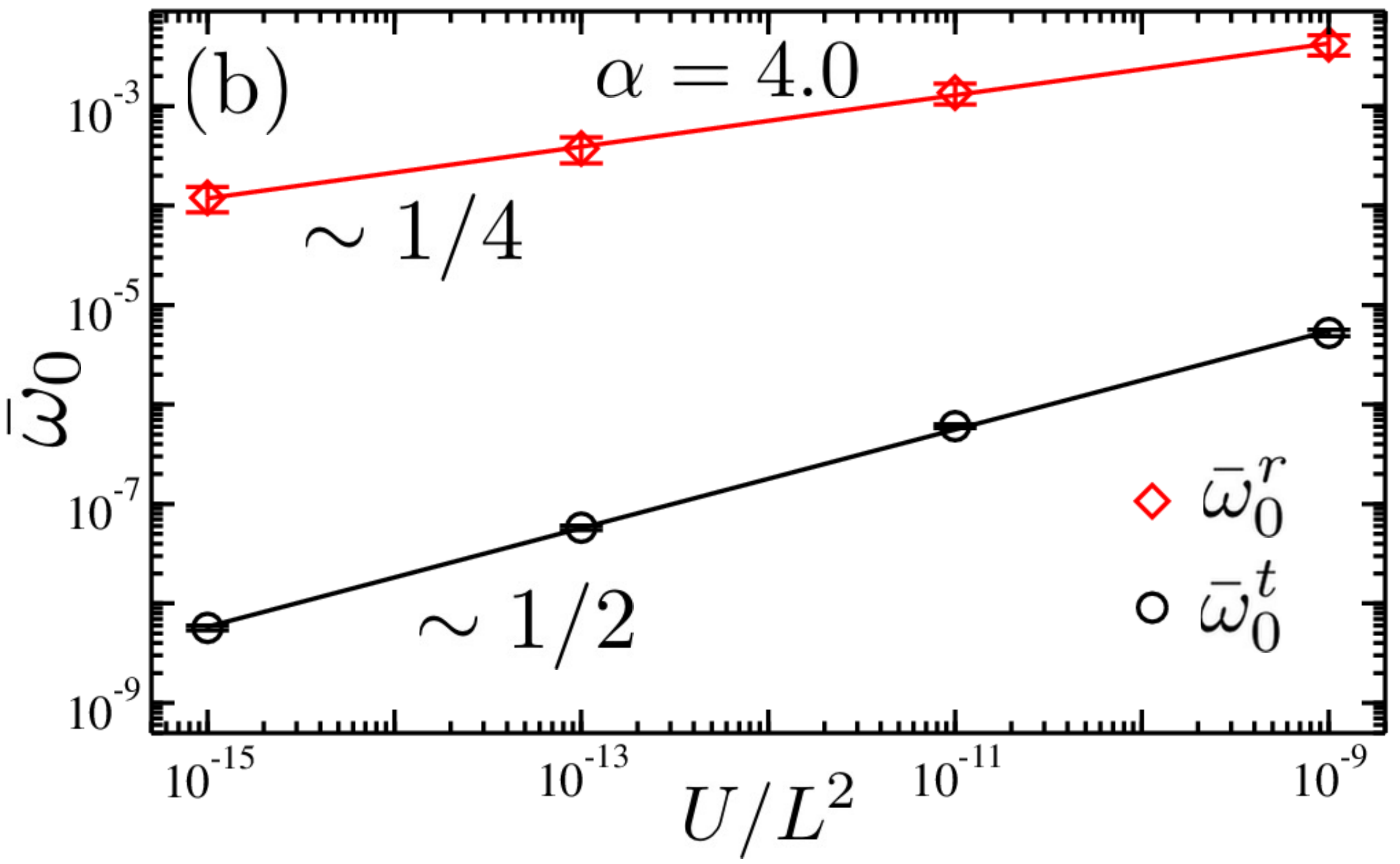}
	\caption{\label{fig:density_states_v_U-4} (a) The density of vibrational states $D(\omega)$ for $N=1024$ bidisperse spherocylinders with aspect ratio $\alpha = 4.0$ at configuration energies  $U/L^2 = 10^{-15}$,  $10^{-11}$, and $10^{-9}$, corresponding to  packing fractions of $\phi-\phi_J \approx 9.1\times 10^{-8}$, $9.3\times 10^{-6}$, and $4.3\times 10^{-4}$ respectively.
	(b) Average frequency $\bar\omega_0^t$ of modes in the low frequency band, and average frequency $\bar\omega_0^r$ of the modes in the middle frequency band,  vs energy $U/L^2$.
	We find $\bar\omega_0^r\sim (U/L^2)^{1/4}$, while $\bar\omega_0^t\sim (U/L^2)^{1/2}$.
	}
\end{figure}

Finally in Fig.~\ref{fig:density_states_v_U-4}(a) we show how the density of states $D(\omega)$ changes as $U$ increases and the system moves further from jamming.  As $U$ increases, the high frequency band changes little.  For sufficiently large $U$, the middle band merges with the upper band.  Defining the average frequency of the modes in the lowest band of localized, translational, modes as $\bar\omega_0^t$, and the average frequency of the modes in the middle band of localized, primarily rotational, modes as $\bar\omega_0^r$, we plot $\bar\omega_0^t$ and $\bar\omega_0^r$ vs $U/L^2$ in Fig.~\ref{fig:density_states_v_U-4}(b).  We see that they both vary as a power law as $U/L^2$ decreases, strongly indicating that the frequencies of the modes in these two bands vanish exactly at $\phi_J$. However we see that they vanish with different power laws.  We find for the middle band modes that  $\bar\omega_0^r\sim (U/L^2)^{1/4}\sim (\phi-\phi_J)^{1/2}$, the same as was found for the low band modes for nearly circular spherocylinders with $\alpha=0.01$, and the same as was found for ellipses and ellipsoids \cite{Schreck.PRE.2012}.  However for the low band modes we find that they vanish more rapidly as $U\to 0$, with  $\bar\omega_0^t\sim(U/L^2)^{1/2}\sim (\phi-\phi_J)$.

\subsubsection{Spherocylinders near the peak packing fraction: $\alpha=1.0$}

Finally in this section we consider spherocylinders with aspect ratio $\alpha=1.0$, thus being a case between those considered in the two previous sections; $\alpha=1$ also corresponds to the aspect ratio that gives the peak packing fraction $\langle\phi_J\rangle\approx 0.8875$ and  which is also closest to being isostatic, with $\langle z_J\rangle= 5.91\pm 0.01$.

\begin{figure}
	\centering
	\includegraphics[width=0.9\linewidth]{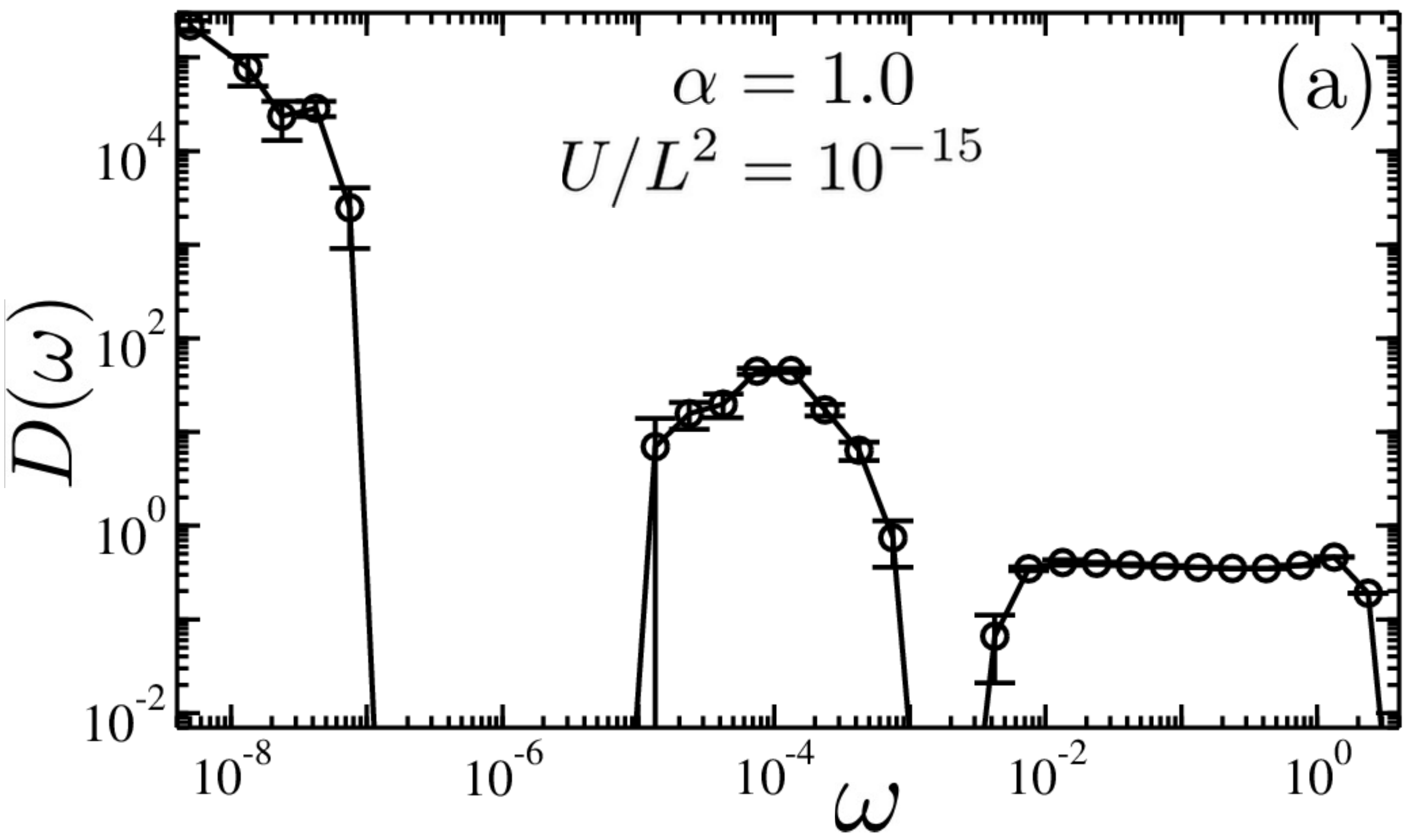}
	\includegraphics[width=0.9\linewidth]{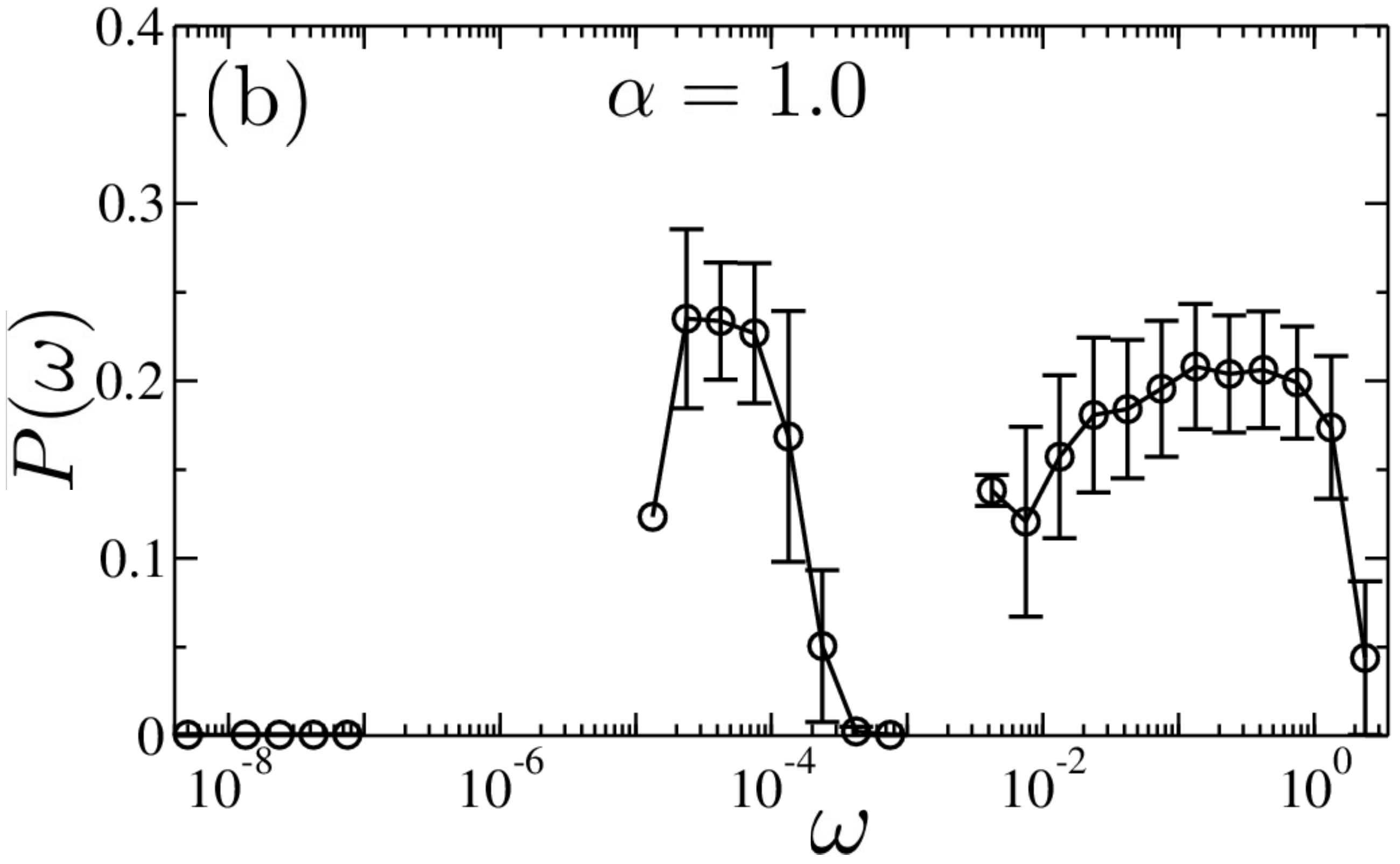}
	\includegraphics[width=0.9\linewidth]{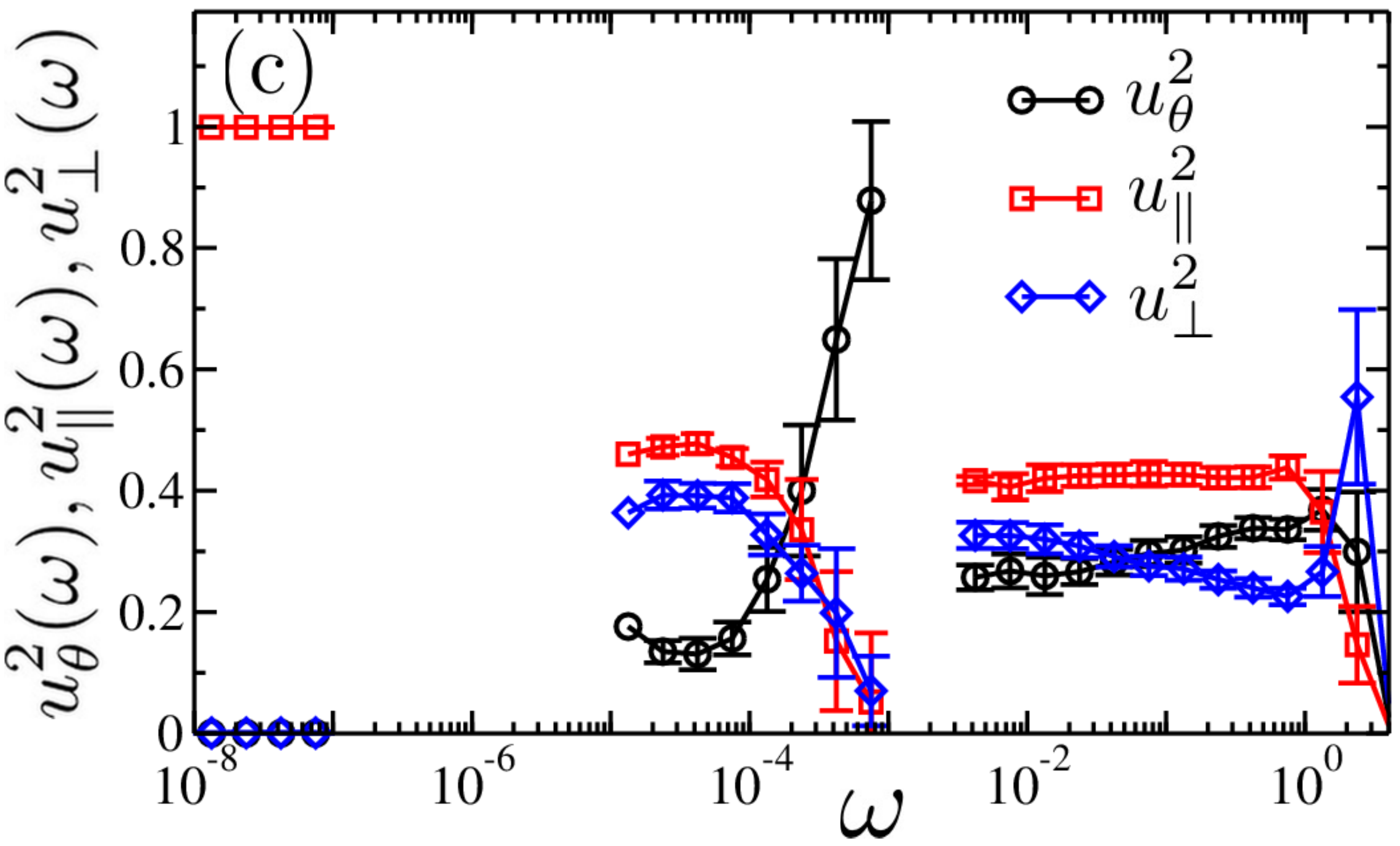}
	\caption{\label{fig:participation_ratio-a1} For a bidisperse system of $N=1024$ spherocylinders with   aspect ratio $\alpha = 1.0$, at energy $U/L^2=10^{-15}$ close to the jamming transition, 
	(a) the density of vibrational modes $D(\omega)$ vs $\omega$;
	(b) the participation ratio $P(\omega)$ vs $\omega$;
	(c) the average rotational motion $ u_\theta^2(\omega) $ (circles),  parallel translational motion $ u_{\parallel}^2(\omega) $ (squares) and perpendicular translational motion $ u_\perp^2(\omega) $ (diamonds) vs $\omega$. ``Parallel" and ``transverse" refer to directions relative to the direction of the spherocylinder spine.
	Each point is averaged over all of the modes in the same frequency bin for 6 different independent samples.
	}
\end{figure}

In Fig.~\ref{fig:participation_ratio-a1}(a) we plot our results for the density of states $D(\omega)$ vs $\omega$ for mechanically stable configurations at energy  $U/L^2=10^{-15}$, very close to the jamming transition.   
In Fig.~\ref{fig:participation_ratio-a4}(b) we plot the participation ratio $P(\omega)$, and in Fig.~\ref{fig:participation_ratio-a4}(c)  we plot the 
quantities $u_\theta^2(\omega)$, $u_\parallel^2(\omega)$ and $u_\perp^2(\omega)$.
We see that the situation at $\alpha=1.0$ is a natural combination of the two previous cases.  There are three distinct frequency bands, with the two upper bands looking essentially the same as was found for $\alpha=0.01$.  States are localized near the edges of these bands but extended in the middle.  The highest frequency band consists of modes that are mostly of mixed translational and rotational character.  The middle frequency band is primarily rotational towards its upper edge, but primarily translational towards its lower edge.  The lowest frequency band is like that found for $\alpha=4.0$, consisting of highly localized sliding modes that are purely translational parallel to the spherocylinder spine.
We may thus speculate that this represents the generic case.  As $\alpha$ decreases from unity, the low frequency band of sliding modes shrinks and disappears while the middle frequency band grows.  As $\alpha$ increases from unity, the middle frequency band shrinks while the low frequency band grows.  For $\alpha=1.0$ we find the fraction of modes in the low, middle, and high frequency bands to be 0.0039, 0.0105, and 0.986 respectively

\begin{figure}[h!]
	\centering
	\includegraphics[width=0.9\linewidth]{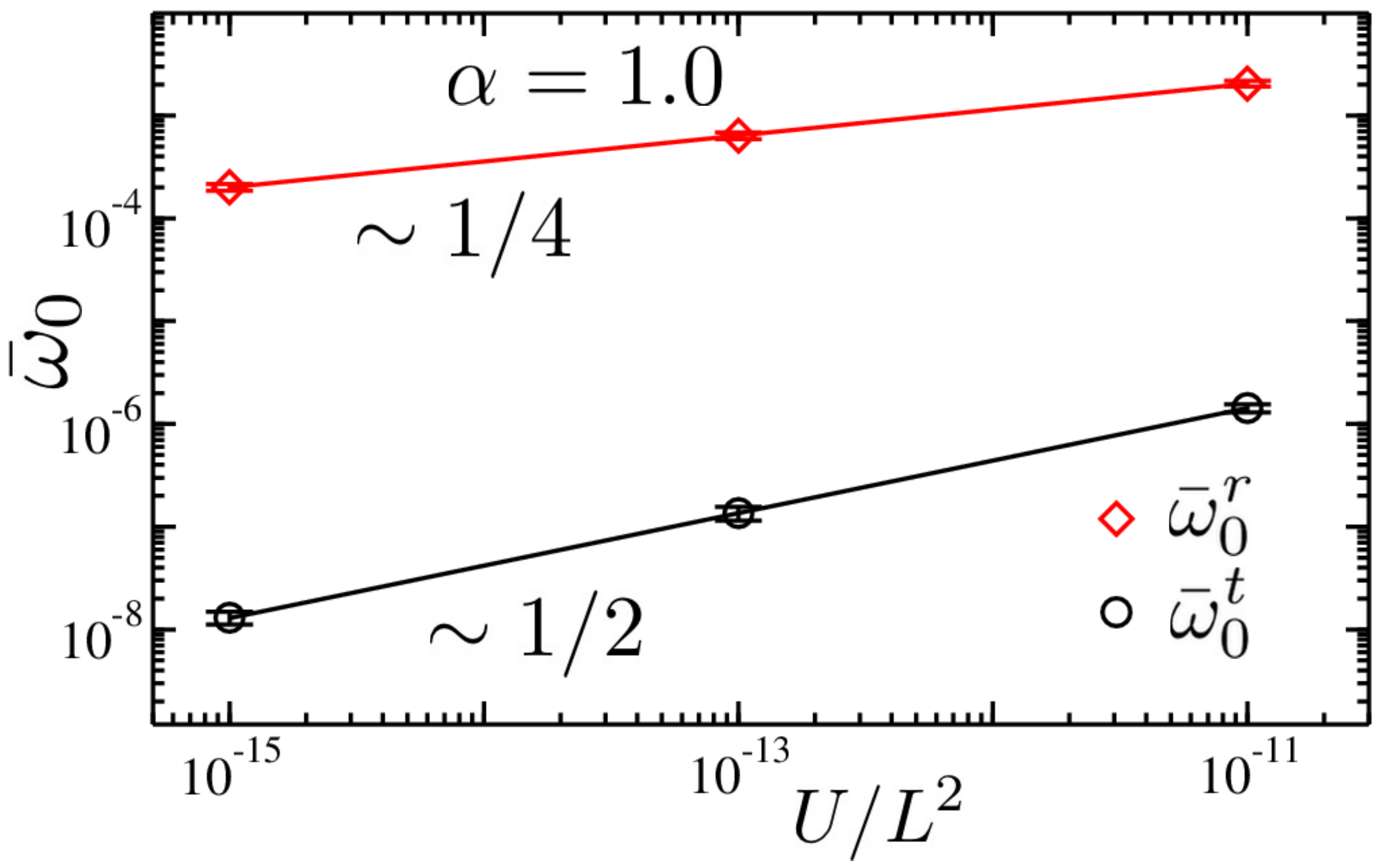}
	\caption{\label{fig:w_v_U_a1} 
	For a bidisperse system of $N=1024$  spherocylinders with   aspect ratio $\alpha = 1.0$, the 
	average frequency $\bar\omega_0^t$ of modes in the low frequency band, and average frequency $\bar\omega_0^r$ of the modes in the middle frequency band,  vs energy $U/L^2$.
	We find $\bar\omega_0^r\sim (U/L^2)^{1/4}$, while $\bar\omega_0^t\sim (U/L^2)^{1/2}$.
	}
\end{figure}

It is interesting to note that three frequency bands were also reported for ellipses and ellipsoids at low aspect ratios \cite{Mailman.PRL.2009,Schreck.PRE.2012}.  In that case the authors argued that it was only their lowest frequency band that represented the quadratically unconstrained states.  In our case of spherocylinders at $\alpha=1.0$, however, it is both the lowest two bands that represent quadratically unconstrained states.  Our high frequency band is found to contain all the $\mathcal{N}\langle z\rangle/2$ modes expected for the quadratically constrained states, while the modes in the lowest two bands scale to zero as $U\to 0$ at the jamming transition.  To demonstrate this, we compute the average frequency $\bar\omega_0^r$ of modes in the middle band, and the average frequency $\bar\omega_0^t$ of modes in the lower band, as a function of the system energy $U/L^2$. As we found in the preceding section for $\alpha=4.0$, we similarly find here that $\bar\omega_0^r\sim (U/L^2)^{1/4}$ while $\bar\omega_0^t\sim (U/L^2)^{1/2}$ (see Fig.~\ref{fig:w_v_U_a1}).

\section{Conclusions}
\label{sconcl}

In this work we have considered the behavior of ensembles of two dimensional soft-core spherocylinders as they are isotropically compressed, under athermal ($T=0$) conditions, from dilute packing fractions $\phi_\mathrm{init}$, to packings above the jamming transition $\phi_J$.  We use particles obeying an overdamped dynamics, in response to an affinely shrinking box, to model the compression.  We then use conjugate gradient energy minimizations to relax these configurations to mechanically stable states at their local energy minimum, so as to explore further details of the small vibrational modes of the system.

We first considered the question of orientational ordering.  Rod-shaped particles in thermal equilibrium are known to have a normal-liquid to nematic-liquid phase transition as the density increases, where the orientational ordering increases from zero as one goes above the transition.  In contrast, we find that,
for moderately elongated spherocylinders with aspect ratio $\alpha=4.0$,  there is no orientational ordering as the system is athermally compressed  to above jamming.  We find this to be true for both monodisperse and bidisperse ensembles.  Thus the fluctuations induced by athermal collisions under compression would seem to be qualitatively different from thermally induced fluctuations.

We  then investigated the dependence of the jamming transition packing fraction $\phi_J$, and the average number of contacts per particle at jamming $z_J$, as a function of the spherocylinder aspect ratio $\alpha$.  As was found previously for ellipses and ellipsoids, we find that $\phi_J$ increases, reaches a maximum, and then decreases as $\alpha$ increases.  The maximum occurs at $\alpha\approx 1$, which gives a maximal packing fraction $\phi_J\approx 0.8875$.  Also as found for ellipses and ellipsoids, we find that for spherocylinders only slightly distorted from circular (i.e., at small $\alpha$) the configurations at the jamming $\phi_J$ are hypostatic, with the average number of contacts per particle $z_J < z_\mathrm{iso}=2d_f=6$ smaller than the isostatic value.  However, unlike ellipses and ellipsoids where $z_J\to z_\mathrm{iso}$ as  the aspect ratio increases, for spherocylinders we find that $z_J$ reaches a maximum value $z_J\approx 5.91 <z_\mathrm{iso}=6$ at $\alpha\approx 1$, and then decreases as $\alpha$ increases further.  2D spherocylinders at jamming are hypostatic for all aspect ratios.  

We then considered the limit $\alpha\to 0$, in which spherocylinders are approaching the limit of perfectly circular disks.  Surprisingly, we found that this limit appears to be singular, with a strong probability developing for the spherocylinders to form contacts along their flat edges, even as those flat edges shrink to a negligible fraction of the particle surface. Similar results have recently been found for 2D ellipses \cite{Vanderwerf}, suggesting that this may be a general behavior for non-spherical particles.

Finally we examined the density of states and the nature of the vibrational eigenmodes.
As with ellipses and ellipsoids, we find that for $\phi$ slightly above $\phi_J$,  the density of states $D(\omega)$ splits into distinct bands.  
In general, as illustrated by the case $\alpha=1.0$, there are three  frequency bands.  The high frequency band is similar to the modes found for circular disks, and consists of $Nz/2$ modes of mixed rotational and translational character; these modes remain at finite frequency even at $\phi_J$.  The lower bands consist of $N(z_\mathrm{iso}-z)/2$ modes that, as $\phi\to\phi_J$, become unconstrained at quadratic order in the expansion of the elastic energy; such modes are the reason the system is hypostatic at jamming.  The middle frequency band, with extended modes in the center and localized modes at the edges, consists mostly of rotational motion towards the upper side of the band, but mostly translational motion towards the lower side of the band.  The modes in the middle band appear to be quartically constrained exactly at $\phi_J$.
The low frequency band consists of highly localized sliding modes, involving the translation of single spherocylinders parallel to their spine.  It is unclear if such modes are constrained at all, for small displacements, when one is exactly at $\phi_J$.  As the aspect ratio $\alpha$ decreases from unity, we find that the number of modes in the middle band increases, while the number of modes in the lowest frequency band decreases and eventually vanishes.  As  $\alpha$ increases above unity, we find that the number of sliding modes in the lower band increases, while the middle band shrinks and becomes very narrow (and possibly disappears for large enough $\alpha$.).


Our results confirm that small distortions of particles from a perfect circular shape result in hypostatic states at jamming, however we show that behavior for very elongated particles depends in detail on the particle shape:  for ellipses and ellipsoids, particles approach isostaticity at jamming as the aspect ratio increases, while for spherocylinders they remain hypostatic.  We believe that this hypostatic behavior for elongated spherocylinders is a consequence of the long flat sides of the particles, which have a strong effect on the nature of the quadratically unconstrained modes at jamming.
As we were completing this work we learned of similar work being carried out by Vanderwerf et al. \cite{Vanderwerf}.

\section*{Acknowledgements}

This work was supported by  National Science Foundation Grant No.~CBET-1435861.
Computations were carried out at the Center for Integrated Research Computing at the University of Rochester.
We thank S.~V.~Franklin,  P. Olsson and C.~S.~O'Hern for helpful discussions.

\section*{Appendix A}

In this appendix we provide details of our energy minimization procedure, and tests that show how well our procedure results in mechanically stable jammed states. Starting from an initial state obtained by our slow compression algorithm, we energy minimize to obtain a mechanically stable  configuration by using the Polak-Ribiere conjugate gradient method.


With $\boldsymbol{\zeta}=(\boldsymbol{\zeta}_1, \boldsymbol{\zeta}_2, \dots)$ a vector giving the initial position of our configuration in the $N$ particle coordinate space, we compute the steepest descent gradient $\mathbf{v}=-\partial U/\partial\boldsymbol{\zeta}$, take as our initial search direction the unit vector $\mathbf{\hat v}=\mathbf{v}/|\mathbf{v}|$, and perform a line search to find the approximate minimum in this direction.
We begin the line search by choosing a  small step size $\varepsilon$ and finding the energies of the current configuration  $U(\boldsymbol{\zeta})$ as well as  $U(\boldsymbol{\zeta} + \varepsilon \mathbf{\hat v})$ and  $U(\boldsymbol{\zeta} + 2 \varepsilon \mathbf{\hat v})$.
If the energy increases when moving to $\boldsymbol{\zeta} + \varepsilon \mathbf{\hat v}$, i.e., $U(\boldsymbol{\zeta})<U(\boldsymbol{\zeta}+\varepsilon\mathbf{\hat v})$, we take $\varepsilon^{\prime} = \varepsilon/2$ and find the new set of energies with $\varepsilon^\prime$.
If the energy decreases monotonically to $\boldsymbol{\zeta} + 2 \varepsilon \mathbf{\hat v}$, i.e. $U(\boldsymbol{\zeta})>U(\boldsymbol{\zeta}+\varepsilon\mathbf{\hat v})>U(\boldsymbol{\zeta}+2\varepsilon\mathbf{\hat v})$, then we take $\varepsilon^\prime = 2 \varepsilon$ and find the new set of energies.
Once we have a series of three points with the lowest energy at $\boldsymbol{\zeta} + \varepsilon \mathbf{\hat v}$ we make a quadratic fit to the points, and determine the location of the minimum $\boldsymbol{\zeta}_0$ of that quadratic fit.  If $U(\boldsymbol{\zeta}_0)<U(\boldsymbol{\zeta}+\varepsilon\mathbf{\hat v})$ we then move the configuration to $\boldsymbol{\zeta}_0$; otherwise we move it to $\boldsymbol{\zeta}+\varepsilon\mathbf{\hat v}$.  We then
use the Polak-Ribiere method to  define the new, orthogonal, search direction. For our system size and packings near jamming, we find empirically that an initial value of $\varepsilon=10^{-4}$ is a good choice.  Recall, in our units, the smaller spherocylinders have a diameter of unity.

As the algorithm narrows in on a local minimum of $U$, the step size $\varepsilon$ needed to complete a line search gets ever smaller.  Once $\varepsilon < 10^{-16}$ we no longer have sufficient machine precision in the particle coordinates to accurately compute the energy difference $U(\boldsymbol{\zeta})-U(\boldsymbol{\zeta}+\varepsilon\mathbf{\hat v})$, and so we stop the line search, and reinitialize the search using the steepest descent direction at the current configuration coordinates.  When the search in the steepest descent direction similarly fails to find a new minimum with $\varepsilon>10^{-16}$ we terminate the search.


A configuration  in perfect mechanical equilibrium will satisfy the conditions that the net force and net torque on each particle from its elastic contacts vanishes, i.e., $\mathbf{F}_i^\mathrm{el}=0$ and $\tau_i^\mathrm{el}=0$.
In practice, the net force and torque on particles will have some small residual value, due to the finite numerical accuracy of our minimization procedure.  As a measure of how well our minimization procedure is finding the desired mechanically stable states we therefore look at the net residual forces and torques, and see how large they are compared to the average force and torque at individual particle contacts.  For each minimized configuration we therefore compute the average contact force,
\begin{equation}
\bar F_{ij}^\mathrm{el}=\dfrac{2}{zN}\sum_{(i,j)}|\mathbf{F}_{ij}^\mathrm{el}|,
\end{equation}
where the sum is over all particle contacts $(i,j)$, $\mathbf{F}_{ij}^\mathrm{el}$ is the elastic force 
of Eq.~(\ref{eq:Fij}), and $z$ is the average number of contacts per particle.
We similarly compute the average contact torque,
\begin{equation}
\bar\tau_{ij}^\mathrm{el}=\dfrac{1}{zN}\sum_{i=1}^N{\sum_j}^\prime|\mathbf{\hat z}\cdot(\mathbf{s}_{ij}\times\mathbf{F}_{ij}^\mathbf{el})|.
\end{equation}
Here $\mathbf{s}_{ij}$ is the moment arm from the center of mass of spherocylinder $i$ to the point of contact with spherocylinder $j$, the second sum is over all spherocylinders $j$ in  contact with spherocylinder $i$, and  each contact gives rise to two terms, one for the torque on spherocylinder $i$ and one for the torque on spherocylinder $j$.

\begin{figure}
	\centering
	\includegraphics[width=0.8\linewidth]{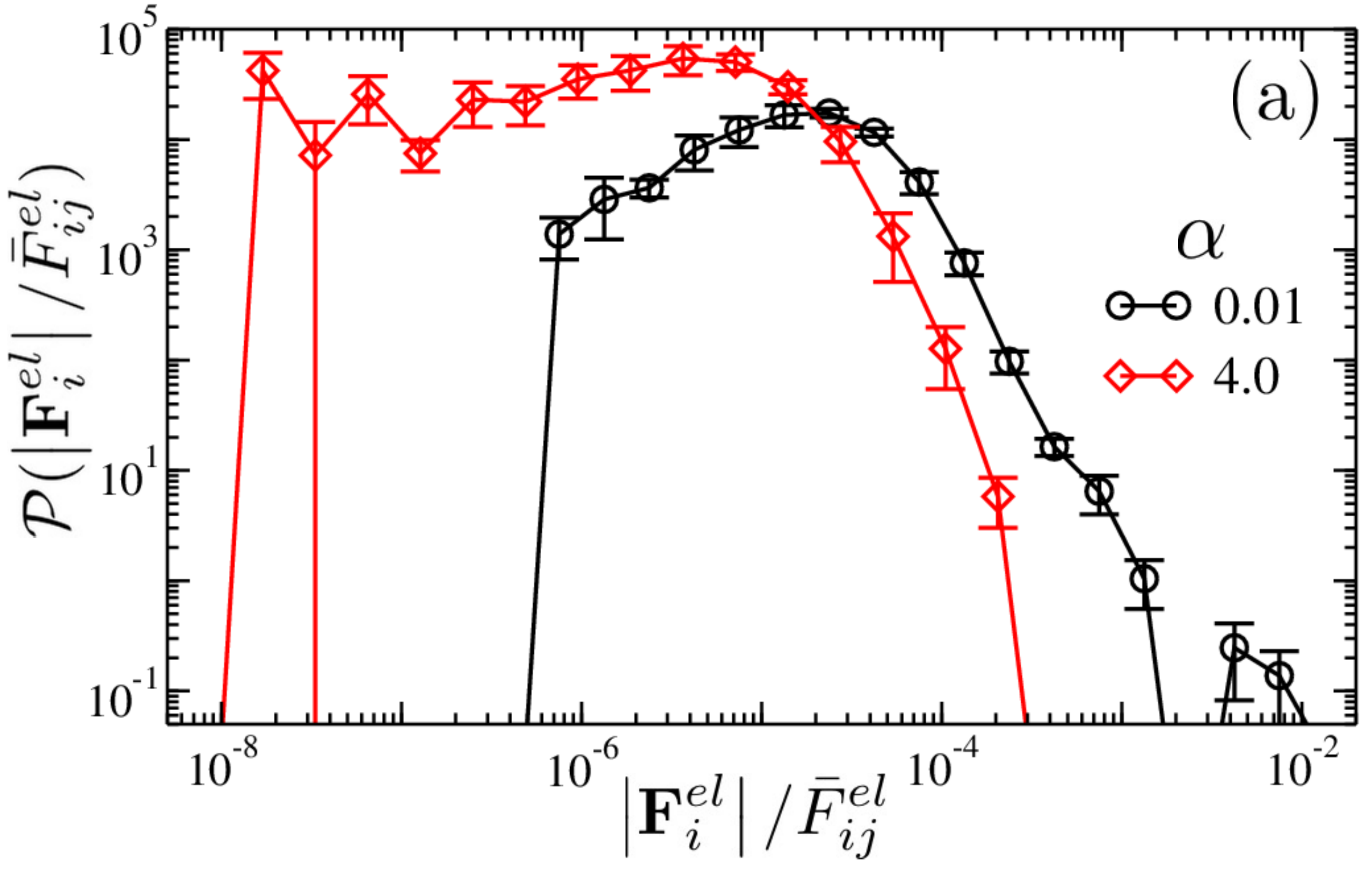}
	\includegraphics[width=0.8\linewidth]{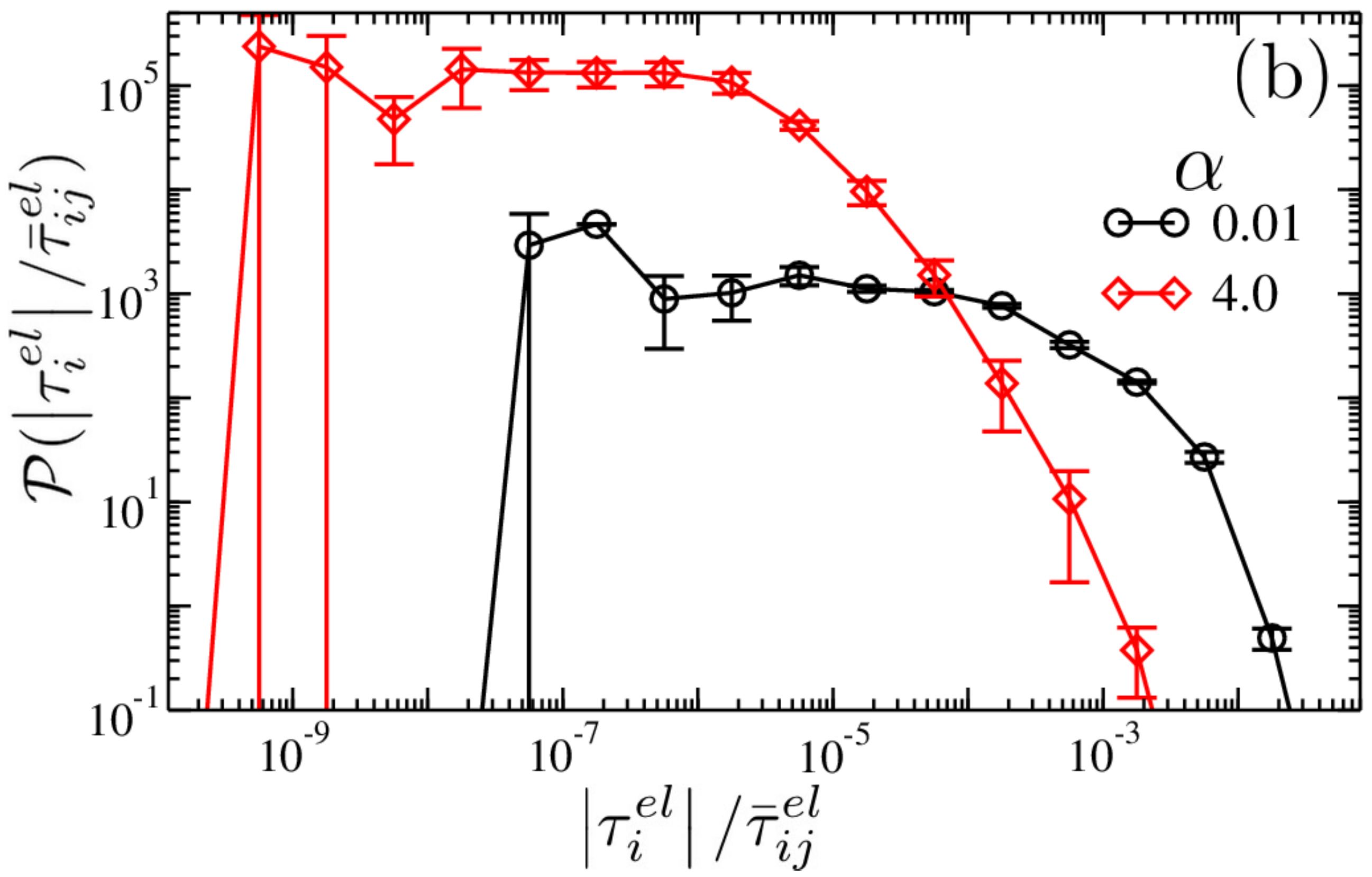}
	\caption{\label{fig:hists} Distribution of (a) net residual forces on spherocylinders $|\mathbf{F}_i^\mathrm{el}|$ relative to the  average contact force $\bar F_{ij}^\mathrm{el}$, and (b) net residual torques on spherocylinders $|\tau_i^\mathrm{el}|$ relative to the average contact torque $\bar\tau_{ij}^\mathrm{el}$, for configurations of $N=1024$ bidisperse spherocylinders at energy $U/L^2=10^{-15}$, very close to jamming, as obtained by our conjugate gradient energy minimization.  In each panel, circles are for $\alpha=0.01$ and diamonds are for $\alpha=4.0$.  Results are averaged over 6 independent samples and the contact averages $\bar F_{ij}^\mathrm{el}$ and $\bar\tau_{ij}^\mathrm{el}$ are computed separately for each sample.
		}
\end{figure}

We then compute the distribution of net residual forces, $\mathcal{P}(|\mathbf{F}_i^\mathrm{el}|/\bar F_{ij}^\mathrm{el})$ and net residual torques $\mathcal{P}(|\tau_i^\mathrm{el}|/\bar\tau_{ij}^\mathrm{el})$, averaging these over our different independent samples.  In Fig.~\ref{fig:hists} we plot these distributions for the two cases $\alpha=0.01$ and $\alpha=4.0$, for configurations minimized to energy $U/L^2=10^{-15}$, very close to the jamming transition, $\phi-\phi_J\lesssim 10^{-7}$.
We see that, on the scale of the contact forces and torques, the net residual forces and torques on the spherocylinders are indeed generally quite small.

\section*{Appendix B}

In this appendix we discuss our method to determine the the eigenmodes and density of states $D(\omega)$ for our system of spherocylinders with aspect ratios $\alpha=1.0$ and 4.0.  Because the eigenmodes in the lowest frequency band have exceedingly small frequencies $\omega_m$, we find that a direct analysis of the dynamical matrix results in large errors in these smallest eigenvalues.  We therefore follow Refs.~\cite{Wyart.EPL.2005,Wyart.PRE.2005,Donev.PRE.2007,Schreck.PRE.2012} and split the dynamical matrix into two pieces,

\begin{equation}
M_{ia,jb} = \dfrac{\partial^2}{\partial \zeta_{ia}\partial\zeta_{jb}}\left[\sum_{(i,j)}V_{ij}(r_{ij})\right] = 
H_{ia,jb} - S_{ia,jb},
\end{equation}
where
\begin{equation}
H_{ia,jb}= \sum_{(i,j)} \dfrac{\partial^2 V_{ij}(r_{ij})}{\partial r_{ij}^2}\left(
\dfrac{\partial r_{ij}}{\partial \zeta_{ia}}\right)\left(\dfrac{\partial r_{ij}}{\partial \zeta_{jb}}\right)
\end{equation}
is known as the stiffness matrix, and
\begin{equation}
S_{ia,jb}= -\sum_{(i,j)}\dfrac{\partial V_{ij}(r_{ij})}{\partial r_{ij}}\left(\dfrac{\partial^2 r_{ij}}{\partial\zeta_{ia}\partial\zeta_{jb}}\right)
\end{equation}
is known as the stress matrix.

From Eq.~(\ref{eVij}), for our harmonic elastic interaction $\partial^2 V_{ij}/\partial r_{ij}^2 = k_e/d_{ij}^2$ at each contact is $O(1)$ in our units (where $k_e=1$ and $d_s=1$), while $\partial V_{ij}/\partial r_{ij}=-k_e(1-r_{ij}/d_{ij})/d_{ij}$ is proportional to the particle overlap and hence very small close to jamming.  Hence we can regard $S_{ia,jb}$ as a small perturbation of $H_{ia,jb}$.  The  eigenvectors $\mathbf{\hat u}_m^{(0)}$ and eigenvalues $\lambda_m^{(0)}$ of the stiffness matrix $H_{ia,jb}$ are therefore the zeroth order approximates to those of $M_{ia,jb}$.  We thus compute these $\mathbf{\hat u}_m^{(0)}$ and find them to include a set of degenerate eigenvectors with $\lambda_m^{(0)}=0$.  These would be the unconstrained (to quadratic order) modes present in a hypostatic system, were the system exactly at the jamming transition where $S_{ia,jb}=0$.  At finite energy $U$ above jamming, we take these as the zeroth order approximates to the modes in the lower frequency bands, while the eigenvectors with  $\lambda_m^{(0)}>0$ are the approximates to the modes in the upper frequency band.

We can then compute the first order corrections to the eigenvalues, due to the non-zero $S_{ia,jb}$ at finite $U$, in the usual way.  For a mode in the upper frequency band we have $\delta \lambda_m = -\mathbf{\hat u}_m\cdot \mathbf{S}\cdot\mathbf{\hat u}_m$.  For such a mode in the upper frequency band we find that $\mathbf{\hat u}_m^{(0)}$ and $\lambda_m^{(0)}+\delta\lambda_m$ differ negligibly from the $\mathbf{\hat u}_m$ and $\lambda_m$ we obtain from a direct analysis of $M_{ia,jb}$.

For the modes in the lower frequency bands, we project $S_{ia,jb}$ onto the subspace spanned by the set of degenerate eigenvectors $\{\mathbf{\hat u}_m^{(0)}\}$ with $\lambda_m^{(0)}=0$, and then diagonalize $-S_{ia,jb}$ on that subspace.  The resulting eigenvectors $\mathbf{\hat u}_m$ and eigenvalues $\lambda_m$ are then the next level approximates to the eigenmodes of the lower frequency bands  of the full dynamical matrix $M_{ia,jb}$, and these values are then used in the construction of the density of states $D(\omega)$.

\newpage

\bibliographystyle{apsrev4-1}
\bibliography{rod_compression7.bib}

\end{document}